\global\def\draftcontrol{0}

   \def\versionno{ W transition }
\catcode`\@=11

\expandafter\ifx\csname draftcontrol\endcsname\relax\global\def\draftcontrol{0} 
\fi 

{\count255=\time\divide\count255 by 60 
\xdef\hourmin{\number\count255} 
\multiply\count255 by-60\advance\count255 by\time 
\xdef\hourmin{\hourmin:\ifnum\count255<10 0\fi\the\count255}} 
\def\draftdate{\number\month/\number\day/\number\year\ \ \ \hourmin } 

\newcommand\makepapertitle{\par

  \begingroup 
    \renewcommand\thefootnote{\@fnsymbol\c@footnote}% 
    \def\@makefnmark{\rlap{\@textsuperscript{\normalfont\@thefnmark}}}% 
    \long\def\@makefntext##1{\parindent 1em\noindent 
            \hb@xt@1.8em{% 
                \hss\@textsuperscript{\normalfont\@thefnmark}}##1}% 
     \newpage 
     \global\@topnum\z@   % Prevents figures from going at top of page. 
     \@makepapertitle 
     \thispagestyle{empty}\@thanks 
  \endgroup 
  \setcounter{footnote}{0}% 
  \global\let\thanks\relax 
  \global\let\makepapertitle\relax 
  \global\let\@makepapertitle\relax 
  \global\let\@thanks\@empty 
  \global\let\@author\@empty 
  \global\let\@date\@empty 
  \global\let\@title\@empty 
  \global\let\title\relax 
  \global\let\author\relax 
  \global\let\date\relax 
  \global\let\and\relax 
  \def\version{\let\version\@version\@gobble} 
} 
\def\@makepapertitle{% 
  \newpage 
   \ifnum\draftcontrol=1 {} 
   \version\versionno 
   \vskip 5em% 
   \else 
   \hfill\hbox to 3cm {\parbox{4cm}{\@pubnum}\hss}% 
   \vskip 5em% 
   \fi 
   \begin{center}% 
   \let \footnote \thanks 
      {\hskip -0\textwidth \hbox to 1\textwidth% 
        {\centerline{\Large\bf{\noindent\@title}}}}% 
     \vskip 2em% 
     {\normalsize%\large 
       \lineskip .5em% 
       \begin{tabular}[t]{c}% 
         \@author 
       \end{tabular}\par}% 
     \vskip 1em% 
     {\@bstract}% 
     \end{center}% 
     \vfill
     \@date%
     \vskip 1.5em%
   \par 
} 

\gdef\@pubnum{} 
\def\pubnum#1{% 
  \gdef\@pubnum{#1}} 

\gdef\@bstract{} 
\def\Abstract#1{% 
  \gdef\@bstract{% 
   \parbox{\textwidth-0pc}{% 
   \centerline{\bf Abstract}\penalty1000
   \noindent%\abstractfont \baselineskip=12pt 
   \renewcommand\baselinestretch{1.0} 
   {#1}}} 
} 

\gdef\@email{}
\def\email#1{%
   \gdef\@email{%
   Email: {\tt #1}}
}

\def\ps@paper{\let\@mkboth\@gobbletwo% 
     \ifnum\draftcontrol=1 
        \def\@oddfoot{\hbox to \textwidth{\tiny \versionno \hfil\tiny\draftdate}% 
        \hskip -\textwidth \hbox to \textwidth{\hfil\rm\thepage\hfil}}% 
     \else\def\@oddfoot{\hbox to \textwidth{\hfil\rm\thepage\hfil}} 
     \fi 
     \let\@evenfoot\@oddfoot 
} 
\def\body{\clearpage 
          \pagestyle{paper} 
        } 
\newenvironment{acknowledgments}{% 
\vskip 3.25ex 
\noindent {\bf Acknowledgments} 
} 

\def\@version#1{\ifnum\draftcontrol=1 
\typeout{}\typeout{#1}\typeout{} 
\vskip3mm\centerline{\hbox{\fbox{\normalsize{\tt DRAFT -- #1 -- } 
                   {\draftdate}}}}\vskip3mm 
\fi} 
\let\version\@version 
\long\def\eqlabel#1{\ifnum\draftcontrol=1 
                    \tag@false  % there are some problems with multline without this 
                    \tag*{(\theequation) \hbox to -0.2cm{\hspace{0cm}\small{#1}\hss}} 
                    \refstepcounter{equation}  
                    \edef\@currentlabel{\theequation} 
                    \ltx@label{#1}          % use old LaTeX \label instead of new definition 
                    \else 
                    \label{#1} 
                    \fi 
                    } 
\let\st@bibitem\@bibitem 
\let\st@lbibitem\@lbibitem 
\ifnum\draftcontrol=1 
  \def\@bibitem#1{% 
    \st@bibitem{#1}\a@@label{#1}\ignorespaces} 
  \def\@lbibitem[#1]#2{% 
    \st@lbibitem[#1]{#2}\a@@label{#2}\ignorespaces} 
  \def\a@@label#1{% 
    \gdef\a@lab{\smash{\normalfont\small#1}} 
    \ifvmode 
      \if@inlabel 
        \global\setbox\@labels\hbox{% 
          \llap{\a@lab\let\a@lab\relax 
                \kern\@totalleftmargin\kern\marginparsep}% 
          \box\@labels}% 
      \fi 
    \fi} 
\fi 
\documentclass[12pt,letterpaper]{article} 
\usepackage[ps,dvips,matrix,arrow,frame,import,curve,color]{xy}
\usepackage{amsmath,bm,amsfonts,amssymb,array,calc,amsthm,rotating}
\usepackage{graphicx,fancybox,epsfig,psfig,color,psfrag} 
\usepackage[nosort]{cite}

\renewcommand\baselinestretch{1.25} 
\renewcommand\section{\@startsection {section}{1}{\z@}% 
                                   {-3.5ex \@plus -1ex \@minus -.2ex}% 
                                   {2.3ex \@plus.2ex}% 
                                   {\normalfont\large\bfseries}} 
\renewcommand\subsection{\@startsection{subsection}{2}{\z@}% 
                                   {-3.25ex\@plus -1ex \@minus -.2ex}% 
                                   {1.5ex \@plus .2ex}% 
                                   {\normalfont\normalsize\bfseries}} 
\renewcommand\subsubsection{\@startsection{subsubsection}{3}{\z@}% 
                                   {-3.25ex\@plus -1ex \@minus -.2ex}% 
                                   {1.5ex \@plus .2ex}% 
                                   {\normalfont\normalsize\it}} 
\renewcommand\paragraph{\@startsection{paragraph}{4}{\z@}% 
                                   {-3.25ex\@plus -1ex \@minus -.2ex}% 
                                   {1.5ex \@plus .2ex}% 
                                   {\normalfont\normalsize\bf}} 
\renewcommand\subparagraph{\@startsection{subparagraph}{5}{\z@}% 
                                   {-1.25ex\@plus -1ex \@minus -.2ex}% 
                                   {0ex \@plus .2ex}% 
                                   {\normalfont\normalsize\it}} 

\numberwithin{equation}{section}

\long\def\@makecaption#1#2{%
  \vskip\abovecaptionskip
  \sbox\@tempboxa{{\bf #1:} #2}%
  \ifdim \wd\@tempboxa >\hsize
    {\small\bf #1:} {\small #2}\par
  \else
    \global \@minipagefalse
    \hb@xt@\hsize{\hfil\box\@tempboxa\hfil}%
  \fi
  \vskip\belowcaptionskip}
\renewcommand*\l@section[2]{%
  \ifnum \c@tocdepth >\z@
    \addpenalty\@secpenalty
    \addvspace{.1em \@plus\p@}%
    \setlength\@tempdima{1.5em}%
    \begingroup
      \parindent \z@ \rightskip \@pnumwidth
      \parfillskip -\@pnumwidth
      \leavevmode \bfseries
      \advance\leftskip\@tempdima
      \hskip -\leftskip
      #1\nobreak\hfil \nobreak\hb@xt@\@pnumwidth{\hss #2}\par
    \endgroup
  \fi}
\renewcommand*\l@subsection{\addvspace{-.15em \@plus\p@}\@dottedtocline{2}{1.5em}{2.3em}}
\renewcommand*\l@subsubsection{\addvspace{-.2em \@plus\p@}\@dottedtocline{3}{3.8em}{3.2em}}

\def\ie{{\it i.e.}}

\def\revise#1       {\raisebox{-0em}{\rule{3pt}{1em}}% 
                     \marginpar{\raisebox{.5em}{\vrule width3pt\ 
                     \vrule width0pt height 0pt depth0.5em 
                     \hbox to 0cm{\hspace{0cm}{% 
                     \parbox[t]{4em}{\raggedright\footnotesize{#1}}}\hss}}}}

\def\call         {{\cal L}} 
\def\calm         {{\cal M}} 
\def\caln         {{\cal N}} 
\def\calo         {{\cal O}}

\def\cals         {{\cal S}}

\def\complex      {{\mathbb C}} 
 
\def\projective   {{\mathbb P}} 
 
\def\reals        {{\mathbb R}} 
\def\zet          {{\mathbb Z}} 
\def\Z            {{\mathbb Z}}

\def\ee           {{\it e}} 
\def\ii           {{\it i}} 
 
\def\tr           {{\rm Tr}} 
\def\Re           {{\rm Re\hskip0.1em}}

\newcommand\topa[2]{\genfrac{}{}{0pt}{2}{\scriptstyle #1}{\scriptstyle #2}}

\def\sqr#1#2{{\vcenter{\vbox{\hrule height.#2pt   
 \hbox{\vrule width.#2pt height#1pt \kern#1pt 
 \vrule width.#2pt}\hrule height.#2pt}}}}

\def\U{{\it U}}
\def\SU{{\it SU}}
\def\SO{{\it SO}}
\def\O{{\it O}}
\def\SP{{\it Sp}}

\def\RP{\reals\projective}

\def\res{{\rm Res}}

\usepackage{psfrag}
\DeclareFontFamily{U}{rsf}{}
\DeclareFontShape{U}{rsf}{m}{n}{
  <5> <6> rsfs5 <7> <8> <9> rsfs7 <10-> rsfs10}{}
\DeclareMathAlphabet\Scr{U}{rsf}{m}{n}

\newcommand{\nn}{\nonumber}
\newcommand{\ts}{\textstyle}
\newcommand{\ds}{\displaystyle}

\newcommand{\MD}{{\mathbb B}_7}
\newcommand{\MF}{{\mathbb D}_7}
\newcommand{\MOD}{{\mathbb A}_7}
\newcommand{\MO}{{\mathbb A}_7/{\mathbb Z}_2}

\newcommand{\mathP}{{\mathbb P}}
\newcommand{\mathZ}{{\mathbb Z}}
\newcommand{\mathC}{{\mathbb C}}
\newcommand{\mathR}{{\mathbb R}}

\catcode`\@=12 

\begin{document} 

%%% 
%%%%%% text starts here 
%%%%%%%%% 

\title{Non-perturbative orientifold transitions at the conifold}

\pubnum{%
hep-th/0506234}
\date{June 2005}

\author{
\makebox[\textwidth][c]{Kentaro Hori$^a$, Kazuo Hosomichi$^a$, 
David C. Page$^a$, Ra\'ul Rabad\'an$^b$ and Johannes Walcher$^b$} \\[0.4cm]
\it $^a$ Department of Physics, University of Toronto,\\
\it Toronto, Ontario, Canada \\[0.2cm]
\it $^b$ School of Natural Sciences, Institute for Advanced Study,\\
\it Princeton, New Jersey, USA 
\vspace{0.5cm}
}

\Abstract{
\\
After orientifold projection, the conifold singularity in
hypermultiplet moduli space of Calabi-Yau compactifications cannot be 
avoided by geometric deformations. We study the non-perturbative
fate of this singularity in a local model involving O6-planes
and D6-branes wrapping the deformed conifold in Type IIA string theory.
We classify
possible A-type orientifolds of the deformed conifold and find that
they cannot all be continued to the small resolution.
When passing through the singularity on the deformed side,
 the O-plane charge generally 
jumps by the class of the vanishing cycle.
To decide which classical 
configurations are dynamically connected, we construct the 
quantum moduli space by lifting the orientifold to M-theory as well 
as by looking at the superpotential. 
We find a rich pattern of 
smooth and phase transitions depending on the total sixbrane charge. 
Non-BPS states from branes wrapped on non-supersymmetric bolts
are responsible for a phase transition.
We also clarify the nature of a $\zet_2$ valued D0-brane charge 
in the 6-brane background. 
Along the way,
we obtain a new metric of $G_2$ holonomy 
corresponding to an O6-plane on the three sphere of the deformed conifold.
}

\makepapertitle

\body

\version\versionno

\tableofcontents

\newpage

\section{Introduction}

The space of all string compactifications with $\caln=1$ supersymmetry in
four dimensions is expected to be quite rich. A poor man's
approach to the problem is to think of $\caln=1$ as $\caln=2$ 
supersymmetry broken by branes, orientifolds, and fluxes.  
$\caln=2$ compactifications have been extensively studied
in the framework of Type II superstrings on
Calabi-Yau manifolds, and
we already have a picture of the variety of vacua.
The most striking aspect of this picture is that the moduli spaces corresponding to
different Calabi-Yau manifolds are connected to each other through
conifold transitions \cite{candelas,conifolds}, which are interpreted as
the $\caln=2$ Higgs mechanism from the viewpoint of the four-dimensional
spacetime physics.

Part of the motivation for our work actually comes from the desire 
to understand
better the stringy interior of $\caln=1$ Calabi-Yau ``moduli'' space,
where we are putting quotation marks to emphasize that these moduli will
generically be lifted by potentials. In particular, one would like 
to know which classical configurations are connected as 
parameters are varied, whether they are connected smoothly
or through phase transitions, and what is the physical nature
of the continuation or the transition.
By a classical configuration, we mean the geometrical data of
internal space, orientifold action, branes and fluxes
in the large volume and weakly coupled regime
where string and non-perturbative corrections 
are small. 
It is expected that the conifolds again play
important roles.

There are various motivations
to study conifolds in $\caln=1$
compactifications.
In the context of Type IIB flux compactifications \cite{sav},
conifolds are the key to explore models with large hierarchy of scales
\cite{GKP}. Also, recent study shows that
a good portion of flux vacua populates a neighborhood of conifolds
\cite{dd,gkt}.
Conifolds are attractive also from cosmology,
see for example \cite{lev,mohsas}.
In this context, conifolds were approached from the ``vector side''
(complex structure in IIB, K\"ahler class in IIA).
We would like to consider also approaching from
the other, ``hyper side''
(complex structure in IIA, K\"ahler class in IIB).
The approach to conifolds from hypermultiplet moduli space
has been studied in $\caln=2$ systems in \cite{gmv}.

\begin{figure}
\centering \epsfxsize=4in \hspace*{0in}\vspace*{.2in}
\epsffile{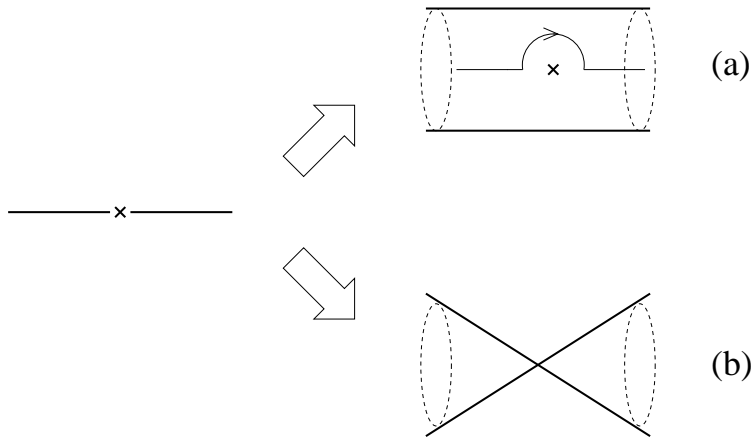}
\caption{\small Is a $\mu$-transition possible? When complexified with 
the RR field, the moduli space can be smooth or it can have 
a real codimension two singularity.}
\label{singu}
\end{figure}

In this paper, we shall study the local behavior of the
$\caln=1$ moduli space around the
conifold loci, in Type II orientifolds.
Before orientifolding,
the conifold locus appears in real codimension two or more
in $\caln=2$ vector multiplet
or hypermultiplet moduli spaces.
The orientifold projects out a part of such closed string moduli fields
(see \cite{BH2} for details).
In particular, it cuts a real slice in the classical geometric
portion of the hypermultiplet moduli space,
 and the singularity appears
in real codimension one. See the left hand side of Figure \ref{singu}.
Since scalar fields 
in $\caln=1$ chiral multiplets are complex,
singularities should only 
be expected in complex co-dimension one. The issue is that the
superpartners of the 
geometric fields are Ramond-Ramond (RR) moduli, and the mixing between the 
two involves non-perturbative effects. It is therefore far from obvious what 
the complexified parameter space will look like in the vicinity
of the classical 
singularity. Two possibilities are sketched on the right of Figure 
\ref{singu}.
%%%%%%%%%%%%%%%%%%%%%%%%
%%%%%%%%%%%%%%%%%%%%%%%%%

For example, it was found in \cite{bhhw} that the tadpole
cancellation conditions are generally different on the two sides of the 
conifold singularity in the real slice of Calabi-Yau moduli space.
In other words, the charge of the orientifold plane is changing,
and any claim of a smooth interpolation between the two classical
limits has to account for this jumping charge.

While these points were raised for compact models, we will here focus
on the singularity and study answers to these questions in the local
model involving just the conifold. Although the results may not
be directly applicable to compact models,
we can test various methods to study the problem.

Since the conifold has been studied from a large number of perspectives
in recent years, it might appear that answers to all these questions
should be known. We therefore explain the basic point 
which we feel has not been addressed in full detail until now. Then we
shall summarize our methods and results.

\subsection{Basic question}

Consider Type IIA orientifolds of the deformed conifold
\begin{equation}
\sum_{i=1}^4 z_i^2 = \mu \,,
\eqlabel{introcon}
\end{equation}
with respect to
the anti-holomorphic involution
\begin{equation}
z_i\to \bar z_i \,.
\eqlabel{introinv}
\end{equation}
Under this projection, the space of complex structures of the conifold,
parameterized by $\mu$, is restricted to the real slice $\mu\in\reals$.
The orientifold 6-plane, given by the fixed point set of \eqref{introinv}
times flat four-dimensional Minkowski space is the locus of
real solutions of \eqref{introcon}. When $\mu>0$,
this leaves an $S^3$ worth, while if $\mu<0$, there is no real solution,
and the O-plane is empty. The point $\mu=0$ is the classical conifold
singularity. We will refer to the transition between $\mu$ positive and 
$\mu$ negative as the {\it $\mu$-transition}.
As we will see, whether or not the $\mu$-transition is possible 
depends on the case.

As alluded to above, the real parameter $\mu$ is complexified by a
RR field, which here arises as the period of the RR three-form
around the vanishing $S^3$ of the deformed conifold. The fundamental 
question is whether or not this complexification allows the two classical 
branches ($\mu>0$ and $\mu<0$) to be connected in the full quantum theory. 
Evidently there are two more classical branches joining in at
$\mu=0$, the resolved conifolds, and the fate of
these branches should be a part of the question.

More generally, we may choose to wrap D-branes on top of the O-plane
on the vanishing $S^3$ of the deformed conifold. These will support
an $\caln=1$ gauge theory at low energies, and one has to make sure
that the proposed quantum dynamics takes account of the vacuum 
structure.

\subsection{Lift to M-theory}

Having exposed the problem, we now explain how we will address it.
The main tool for us will be the lift to M-theory, as studied in 
\cite{bobby,Atiyah:2000zz,AW}, and many other places. Recall that
D6-branes and O6-planes lift in M-theory to purely geometric configurations.
In the local model, the essential idea is to identify all possible
classical geometries with fixed asymptotics. It then turns out that 
with reasonable assumptions about the dynamics, holomorphy essentially 
completely determines the quantum moduli space relating the various 
geometries.

Our second method for analyzing the possible transitions, due to
\cite{VafaSup,Atiyah:2000zz} is to study the critical points of a certain 
superpotential $W$. 
The superpotential is computed on the branch of 
the resolved conifold, and in the orientifold situation is a combination of 
flux and crosscap contributions \cite{Wafa}. We give a careful analysis 
of this superpotential in Section~\ref{analyze}, and reproduce
the component of the moduli space including
the two resolved conifold points.

\subsection{Main results}

Consider wrapping $N$ D6-branes on top of the O6-plane on the $S^3_>$ at 
the bottom of the orientifolded conifold \eqref{introcon}, \eqref{introinv}
for $\mu>0$. Our convention is that the total 6-brane 
charge as measured at infinity is $2N-4$ in the cover ($N-2$ in the quotient), where $-4$ is the contribution
from the O6-plane. Clearly, if a transition to $\mu<0$ is possible, where
there is no O-plane, the charge must be carried by $2N-4$ D6-branes 
wrapped on the $S^3_{<}$ at the bottom of the conifold with $\mu<0$. Now
notice that a D6-brane wrapped on $S^3_>$ preserves the opposite 
combination of supersymmetries to a D6-brane wrapped on $S^3_<$. This
is because the relevant calibration is the real part of the
holomorphic three-form
$\Omega$, which when restricted to the two $S^3$'s leads to opposite
orientation, depending on whether $\mu$ is positive or negative. In
other words, if we fix the supersymmetry preserved at infinity, we
can have D6-branes wrapped on $S^3_>$ or anti-D6-branes\footnote{Here 
and throughout the paper, we refer to objects as branes or anti-branes 
according to the sign of the charge measured at infinity. A more natural 
convention would be to refer to all objects preserving the same 
supersymmetry as branes, but we choose our present convention to 
emphasize that the supersymmetric objects sometimes carry opposite 
charge.} wrapped on $S^3_<$. It is clear, therefore, that we can at best 
expect a transition between $\mu>0$ and $\mu<0$ in the supersymmetric 
parameter space if $N$ is non-negative (so that the D6-branes preserve 
the same susy as the O6-plane), and $2N-4$ is non-positive (so that to 
conserve the charge, we wrap $4-2N$ anti-D6-branes). In other words, 
we can expect a $\mu$-transition if $N=0$, $1$, or $2$. 

In the previous paragraph, we discussed the possibility of wrapping an 
O6$^-$-plane on $S^3_>$ with $N$ D6-branes, which yields an $SO(2N)$ gauge 
group. Alternatively, for $N \geq 4$ we may wrap an O6$^+$ and $N-4$ D6-branes, 
yielding an $Sp(N-4)$ gauge group. For $N <1$ there is a similar choice for 
the action of the free orientifold on the Chan-Paton matrices corresponding 
to the D6-branes on $S^3_<$. This leads to two distinct possibilities for the 
low energy four-dimensional gauge group, $SO(2(2-N))$ or $Sp(2-N)$. 
Finally, for any value of $N$ we have two semi-classical limits corresponding 
to the two resolved conifolds with freely acting orientifold and 
$N-2$ units of RR 2-from flux through $\RP^2$. 

With these observations in mind, we can identify all the possible
semi-classical limits in the IIA description for each value of $N$.
Our results are summarized in the left-hand side of Table \ref{summary}.

\begin{table}[t]
\begin{center}
\begin{tabular}{|c|c|c||c|c|c|}
\hline
\multicolumn{3}{|c||}{O-planes and D-branes wrapping the conifold} & 
\multicolumn{3}{c|}{M-theory geometries} \\
\hline\hline
 blown-up & deformed$(\mu>0)$ & deformed$(\mu<0)$ &  
 blown-up & $\mu>0$ & $\mu<0$      \\
& \multicolumn{2}{c||}{gauge group on$\qquad$} &  
& & \\
\footnotesize flux thru $\RP^2$  & 
\footnotesize $N$ D6-branes &
\footnotesize $4-2N$ $\overline{\rm D6}$-branes & 
& & \\
\hline
\hline
$N-2\ge3$ &\footnotesize $SO(2N)$/$Sp(N-4)$ &---    & $\MF/D_N $& $\MD/D_N $& --- \\
$N-2=3$   &$SO(10)$ or $Sp(1)$ &---    & $\MF/D_5 $& $\MD/D_5 $& --- \\
$N-2=2$   &$SO( 8)$ or none    &---    & $\MF/D_4 $& $\MD/D_4 $& --- \\
$N-2=1$  &$SO( 6)$            &---    & $\MF/D_3 $& $\MD/D_3 $& --- \\
$N-2= 0$   &$SO( 4)$  &none    & 
\multicolumn{3}{c|}{$(\mbox{conifold}\times S^1)/\mathZ_2$} \\
$N-2=-1$  &$SO( 2)$  &$SO(2)$ or $Sp(1)$& $\MF/D'_3$& $\MOD$& $\MD/D'_3 $ \\
$N-2=-2$  &none     &$SO(4)$ or $Sp(2)$& $\MF/D'_4$& $\MO $& $\MD/D'_4 $ \\
$N-2=-3$ &---      &$SO(6)$ or $Sp(3)$& $\MF/D'_5$&---    & $\MD/D'_5 $ \\
$N-2\le-3$ &---      &\footnotesize $SO(2(2-N))$/$Sp(2-N)$& $\MF/D'_{4-N}$
&---    & $\MD/D'_{4-N} $ \\
\hline
\end{tabular}
\caption{M-theory geometries with various fixed values of Kaluza-Klein
flux, their Type IIA interpretations, and low energy gauge group.
The moduli spaces of supersymmetric vacua that link these various
semi-classical limits can be contemplated in Figure \ref{fig:cartoon}
on page \pageref{fig:cartoon}.}
\label{summary}
\end{center}
\end{table}

\def\geqp{\mathrel{{}_{\scriptscriptstyle{(}}\mathord{\geqq}{}_{\scriptscriptstyle{)}}}}

The M-theory lifts of the various semi-classical limits are described in the right-hand 
side of Table \ref{summary}. The problem for $N \geqp 3$ is included in \cite{bobby,AW}. 
For infinite string coupling (the size of the M-theory circle growing without 
bounds asymptotically), the M-theory geometries are quotients
of  smooth $G_2$ holonomy manifolds,  $X_i$,  by the dihedral group $D_{N}$. The $X_i$ are 
all isomorphic to the spin bundle over $S^3$, whose $G_2$ holonomy metric was found 
in \cite{Gibbons:1989er}. They differ in the breaking pattern of the asymptotic 
discrete symmetries, and discrete fluxes at the singularities. In the $D_{N}$ case, 
there are four semi-classical limits corresponding to the four IIA limits described 
above. 
 
For $N=3$, the dihedral group $D_3$ is isomorphic to
$A_3=\zet_4$, and the problem 
is equivalent, from the M-theory perspective, to a case without
orientifold. But for the appropriate identification of the M-theory
circle, we still end up with an orientifold in Type IIA.

The extension to lower values of $N$, as well as the generalization to
finite values of the string coupling, requires more complicated 
$G_2$ holonomy metrics with reduced symmetry. These metrics have been 
partially constructed in \cite{bggg,cvetic,ab}, and we now describe
their relevance to our problem.

For $N>2$, if one wishes to keep the asymptotic IIA string coupling
finite, the M-theory lift of the deformed conifold geometry involves a 
$G_2$ metric called $\MD$ in \cite{cvetic}. Roughly, the manifold $\MD$ 
is a Taub-NUT manifold fibered over an $S^3$. Quotienting $\MD$
by the dihedral group leaves a $D_N$ singularity supporting an
$SO(2N)$ or $Sp(N-4)$ gauge group, depending on a discrete
flux. The resolved conifold with flux (and finite string coupling) has 
an M-theory lift called $\MF$ in \cite{cvetic}. This $G_2$ metric is 
smooth and $D_N$ acts freely on $\MF$.

For $N=2$, the M-theory lift will be
\[
 \frac{\mbox{(conifold)}\times S^1}{\mathZ_2},
\]
where $\mathZ_2$ acts as an antiholomorphic involution on the conifold
and reverses the M-theory $S^1$. As such, we know the classical M-theory 
geometry exactly.

For $N=0$ or $N=1$, one can guess that the M-theory lift of the deformed
conifold with O-plane (namely, $\mu>0$) will look like an Atiyah-Hitchin
manifold \cite{AH} or Dancer's manifold \cite{Dancer} fibered over an $S^3$.
Such $G_2$ metrics were not previously known but we find that
they do indeed exist. We will call these manifolds $\MO$ and $\MOD$,
respectively. 

In all cases with $N<2$, the deformed conifold with $\mu<0$ (which is 
wrapped by $4-2N$ anti-D6-branes) and the resolved conifold with flux 
again lift to the quotient of $\MD$ and $\MF$, respectively, by the 
dihedral group $D_{4-N}$. The difference to the $N>2$ case is in the
action of $D_{4-N}$ on the space, as we will explain in more detail 
later. The gauge group living on the $4-2N$ anti-D6-branes in the freely 
acting orientifold can be $\SO(2(2-N))$ or $\SP(2-N)$, depending on the
action on the Chan-Paton factors. This corresponds to the value of a 
discrete torsion. Thus, the $\MD/D_{4-N}$ geometries yield two semi-classical 
limits with $\mu<0$ in each case.

No analytic expressions are known for any of the $G_2$ metrics $\MOD$,
$\MD$, $\MF$ except one special point on the parameter space of $\MD$ 
found in \cite{bggg}, as well as the limits of zero or infinite string 
coupling. From symmetry requirements, one can determine the
metrics up to a small number of unknown functions (of a radial coordinate)
and derive a set of differential equations for the unknowns. It is not 
difficult to verify numerically the existence of such solutions.
One can also see that they depend on two parameters, one corresponding to
the radius of the M-theory circle at infinity and the other to the
volume of $S^3$ at the center. 

The next question is how these semi-classical limits with the same asymptotic
flux, $N-2$, and the same supersymmetry fit together into a complex space
parameterizing supersymmetric vacua. For $N\geqp 3$, which is the case 
discussed in \cite{AW}, the parameter space is a copy of $\projective^1$
with four (three for $N=3$) marked points corresponding to the semi-classical 
limits we have discussed above. Good local coordinates around each of these 
points correspond to the volume deficits of certain three-cycles in the 
asymptotic geometries together with the period of the M-theory three-form 
around the same three-cycles. These complex parameters
are the instanton coefficients in the four-dimensional 
low-energy gauge theory. 

The case $N=2$ was discussed in \cite{kamc}, where it was argued that a
$\mu$-transition should be possible. Our analysis will confirm the
expectation in this case.
In addition, we will find a second branch of moduli space containing two
vacua of $SO(4)$ gauge theory as well as the two resolved conifolds
that were not treated in \cite{kamc}.

In the case $N=1$, there are five semi-classical limits. As can be seen 
from Table \ref{summary} two of them have an $\SO(2)\cong U(1)$ gauge
theory at low energies, one has an $\SP(1)\cong \SU(2)$, while the
two remaining ones have no gauge theory at all. The first two have
free massless gauge bosons while the latter three points do not.
Since the massless spectrum is different, one must
pass through a phase transition when interpolating between the various 
limits. This situation is very similar to ones studied recently in 
\cite{hp2005}, and we will be able to use these methods to deduce
the structure of the quantum parameter space, confirming the naive picture
we have just sketched.

When $N=0$, it appears at first sight that we have also five semi-classical
limits: two on the resolved conifold, two from the deformed conifold with
$\mu<0$, and one from the deformed conifold with $\mu>0$. However, as we will 
see, there are in fact two distinct semi-classical limits corresponding
to just the O-plane wrapping the $S^3$. In M-theory, this can be simply 
seen from the existence of an asymptotic discrete $\zet_2'$ symmetry that is 
spontaneously broken in the interior of $\MO$. In Type IIA string 
theory, this symmetry corresponds to D0-brane charge modulo $2$, which 
is broken only at the bare orientifold plane, but is preserved in the
presence of just a single D6-brane on top. (We will explain why this
statement is not in conflict with the K-theory classification of D-brane
charge.) There are therefore six semi-classical limits to consider.
We will argue that the quantum parameter space consists of two disconnected
branches, with a certain distribution of vacua consistent with the
discrete symmetries of the problem.

For $N<0$, we again have four semi-classical limits, each of which has a
mass gap at low energies. From the analysis of the holomorphic parameters
associated with the gauge theories, as well as our later superpotential
analysis, we will deduce that the space on which these four limits sit
is again a copy of $\projective^1$. In fact, we will see that the curve 
for $N<0$ is isomorphic to the curve for $N'=4-N> 4$, with the only
difference being the association between two of the points and $\SO/\SP$
gauge group!

\subsection{Summary}

In addition to the literature that we have cited already, aspects 
of the problem have also been discussed elsewhere. The basic question whose
solution we have presented in the previous subsection has been broached,
for example, in \cite{kamc,giveon}, with a restriction to the locally 
tadpole canceling case with exactly two D6-branes on top of the O-plane. 
More recently, similar transitions have been found in Type IIB 
compactifications in \cite{ddfgk}. The main result of our paper 
is to explain under which conditions in the local model we can actually 
expect a transition, and to determine the quantum parameter space, whenever
possible. Our main tool of analysis is the lift to M-theory on $G_2$ 
manifolds which are described as quotients by finite groups. Similar
system have been analyzed in the past also in \cite{g2ref}, for example.
Conifold transitions in the $G_2$ context have also been studied
in \cite{papi}.

We also have a number of subsidiary and complementary results to
offer, as we now summarize. We will start out with a classification of 
A-type orientifolds \footnote{The construction of several orientifolds of the 
conifold in Type IIB has been analyzed in \cite{pru}. These orientifolds were 
constructed by partially blowing up orientifolds of orbifold singularities.} 
of the deformed conifold in Section~\ref{oricon}. In this section, we also show 
that when passing through a ``$\mu$-transition'' between $\mu>0$ and $\mu<0$ 
(independent of whether this is possible dynamically or not), the orientifold 
charge changes by the class of the vanishing cycle. We also study how the 
previous orientifolds act on the resolved conifold. We then return to the
main case of interest, the orientifold \eqref{introinv}.
In Section~\ref{symmetries}, 
we explain generalities about the symmetries of the 
underlying $G_2$ holonomy metrics. The nuts and bolts of these space are
assembled in Section~\ref{geometry}. We find a new class of $G_2$ holonomy 
metrics which we label $\MOD$. In Section~\ref{analyze}, we derive
the quantum parameter spaces. In particular, we show how a careful analysis
of the Vafa superpotential produces most of the structure of the parameter 
space. For the cases $N=2,1$ and $0$,
we need to invoke some additional 
information, partly from \cite{hp2005}. In Section~\ref{other}, we briefly
discuss the problem of $\mu$-transition for the other classes of orientifolds.
Since they break more of the geometrical symmetries, we are unable to write 
down explicit metrics. The Vafa superpotential gives a prediction for one 
other class of models. However, when the orientifold does not admit the
resolved conifold, the superpotential method is not applicable. In these
cases, we describe our best educated guesses for possible $\mu$-transitions.
We finally conclude in Section~\ref{conclusions}.

\section{Orientifolds of the conifold}
\label{oricon}

In this section, we describe the possible A-type orientifolds of the deformed 
conifold. In particular, we will see that when crossing
the conifold the class of 
the orientifold locus changes by the class of the
vanishing cycle. We then analyze 
how these orientifolds act on the resolved conifold. 

\subsection{Deformed conifold}

An orientifold of the deformed conifold
\begin{equation}
z_1^2+z_2^2+z_3^2+z_4^2 = \mu
\eqlabel{coni}
\end{equation}
in Type IIA string theory can be obtained from an anti-holomorphic involution, 
which acts on the complex coordinates $z_i$ by complex conjugation followed 
by a symmetry of the quadric \eqref{coni},
\begin{equation}
z_i \to \ee^{\ii\alpha} M_i^j \bar z_j.
\eqlabel{ori}
\end{equation}
Here $M$ is an orthogonal matrix with $M^2=1$, and $\alpha$ is
a phase, which if we assume that $\mu$ is real can be set to zero.

Since all such orthogonal matrices can be diagonalized with $\pm 1$ on 
the diagonal, inequivalent anti-holomorphic involutions of the deformed 
conifold are classified by the number of $+1$ and $-1$ eigenvalues of
$M$.

In each of these five possibilities, the fixed point set is described 
by setting $z_i=x_i$ or $z_i=\ii y_i$, with $x_i$, $y_i$ real, depending 
on the corresponding sign. The fixed point locus are the orientifold 
6-planes. Let us take $\mu>0$ (the case with $\mu$ negative can easily 
be obtained by interchanging the real and imaginary components of the $z_i$). 
We have the following inequivalent cases:

\newenvironment{rrlist}{%
  \begin{list}{}{%
      \setlength{\labelwidth}{.5cm}%
      \setlength{\leftmargin}{.5cm}%
      \setlength{\rightmargin}{0cm}%
      \setlength{\listparindent}{0cm}%
      \setlength{\parskip}{0cm}%
      \setlength{\topsep}{0.2cm}
      \setlength{\itemsep}{0.2cm}%
      }%
    }%
  {\end{list}}

\begin{rrlist}

\item[(0)] For $(z_1,z_2,z_3,z_4)\to (\bar z_1,\bar z_2,\bar z_3,\bar z_4)$ the 
O6-plane is described by 
\begin{equation}
x_1^2+x_2^2+x_3^2+x_4^2 = \mu.
\end{equation}
Since $\mu>0$, this is an $S^3$.

\item[(1)] If the involution takes $(z_1,z_2,z_3,z_4)\to (-\bar z_1,\bar z_2,\bar 
z_3,\bar z_4)$ the orientifold set is non-compact,
\begin{equation}
-y_1^2+x_2^2+x_3^2+x_4^2 = \mu ,
\end{equation}
and isomorphic to $S^2\times \reals$.

\item[(2)] When $(z_1,z_2,z_3,z_4)\to (-\bar z_1,-\bar z_2,\bar z_3,\bar z_4)$ 
the O-plane is at 
\begin{equation}
-y_1^2-y_2^2+y_3^2+y_4^2 = \mu,
\end{equation}
which describes $S^1\times \reals^2$.

\item[(3)] If $(z_1,z_2,z_3,z_4)\to (-\bar z_1,-\bar z_2,-\bar z_3,\bar z_4)$ 
the orientifold set is not connected: The equation
\begin{equation}
-y_1^2-y_2^2-y_3^2+x_4^2 = \mu
\end{equation}
is solved by two copies of $\reals^3$.

\item[(4)] Finally, when $(z_1,z_2,z_3,z_4)\to (-\bar z_1,-\bar z_2,-\bar z_3,
-\bar z_4)$, the orientifold set is empty:
\begin{equation}
-y_1^2 - y_2^2-y_3^2-y_4^2 = \mu
\end{equation}
has no real solutions.
\end{rrlist}

Generalizing the relation between (0) and (4) discussed in the introduction,
flipping the sign of $\mu$ maps (1) to (3) and (2) to itself.

%%%%%%%%%%%%%%%%%%%%%%%%
\begin{figure}
\begin{center}
\input{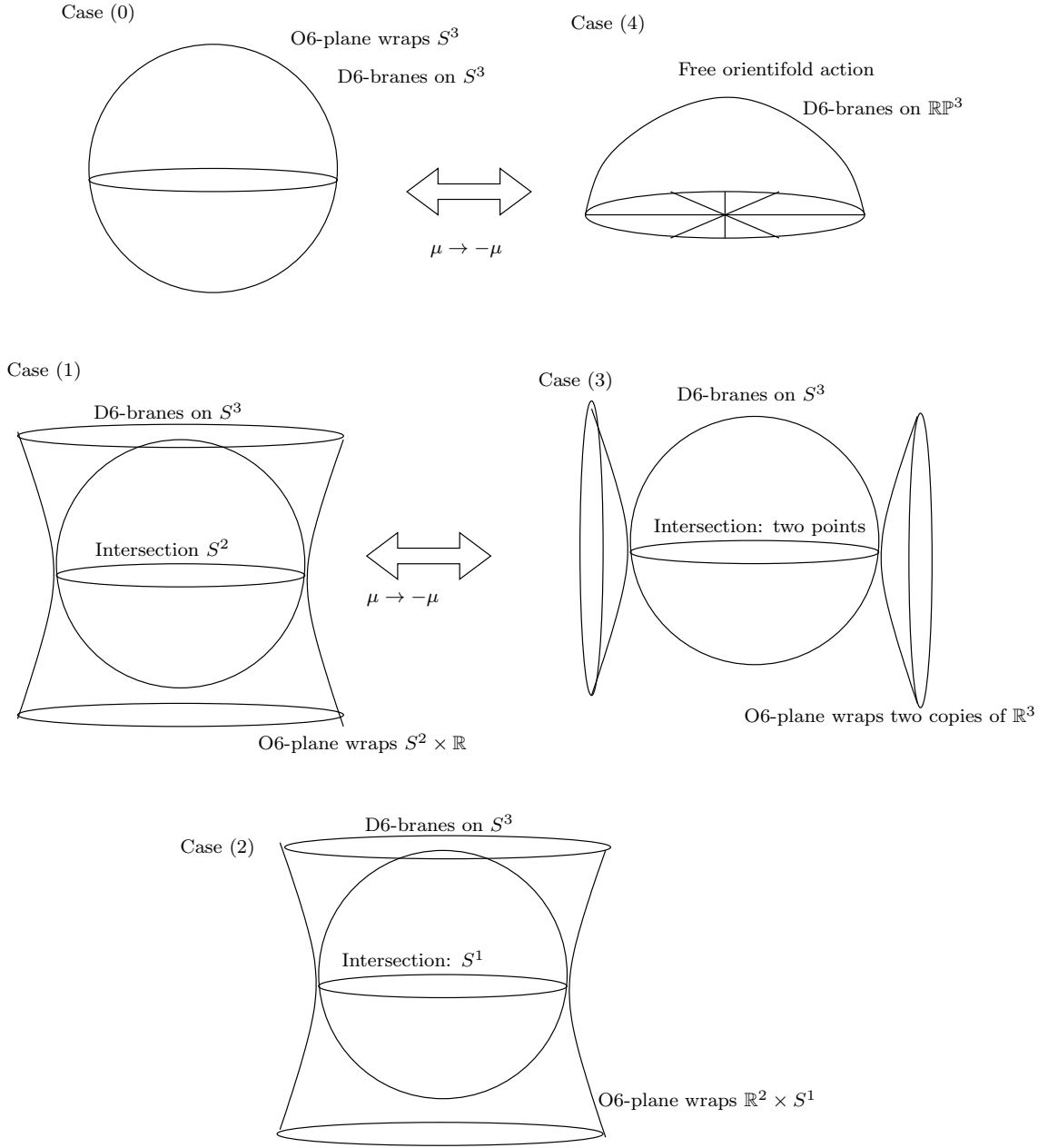}
\end{center}
\caption{Representations of the 5 orientifolds of the deformed conifold 
in Type IIA theory. The orientifold loci are the fixed points of the 
anti-holomorphic involution. The cases (1) and (3) break supersymmetry when 
D6-branes are wrapped on $S^3$ and they are related by taking $\mu 
\rightarrow -\mu$.}
\label{classification} 
\end{figure}
%%%%%%%%%%%%%%%%%%%%%%%%%

We can introduce D6-branes wrapping supersymmetric cycles of the 
deformed conifold. In order to have dynamical gauge symmetry in 
four dimensions, the cycle must be compact, and the only possibility is 
the three-dimensional sphere at the center of the conifold. In the 
cases (1) and (3), the involution reverses the orientation of the $S^3$, 
\ie, maps branes to anti-branes. Therefore, wrapping branes on $S^3$ in 
these cases will break supersymmetry. In fact, in those cases, there is 
a RR tadpole which originates from the non-trivial intersection of the 
O-plane with the compact $S^3$. To get a stable configuration, we are 
forced to wrap branes on some other non-compact cycles. This yields quite
an interesting class of models, which is briefly analyzed in
Section~\ref{other}. The case (0) is supersymmetric: the D6-branes 
wrap the same $S^3$ as the orientifold. In the case (4) the action 
of the orientifold is free, the D6-branes are then wrapping an $\RP^3$. In
case (2) the orientifold plane intersects the $S^3$ where the D6-branes 
are wrapping in an $S^1$. We have depicted these circumstances in
Figure \ref{classification}.

Compact orientifold models including invariant conifolds must be one
of these five types. For example,
let us consider the Type IIA orientifold
on the mirror $X$ of the Fermat quintic, which is a resolution of the
orbifold of
$$
z_1^5+z_2^5+z_3^5+z_4^5+z_5^5-5\psi z_1z_2z_3z_4z_5=0
$$
by the $\Z_5^3$ action, $z_i\to \omega_i z_i$,
$\omega_i^5=\prod\omega_i=1$.
For $\psi=1$, $X$ has a single conifold singularity at $z_1=\cdots =z_5=1$.
We consider the orientifold with respect to the involution
$\tau_{\sigma}: z_i\to\overline{z}_{\sigma(i)}$, where $\sigma$
is an order two permutation (exchange), that fixes the conifold singularity
when $\psi=1$.
The orientifold is allowed only when $\psi$ is real.
There are three distinct cases --- $\sigma$ is identity,
an exchange of a pair, and an exchange of two pairs.
Depending on the sign of $\epsilon=\psi-1$, these cases are one of the five
possibilities:
Without exchange, $z_i\to \overline{z}_i$, it is case (0) if
$\epsilon>0$ and case (4) if $\epsilon<0$.
With an exchange of one pair, such as
$(z_1,z_2,z_3,z_4,z_5)\to
(\overline{z}_2,\overline{z}_1,\overline{z}_3,\overline{z}_4,\overline{z}_5)$,
it is case (1) if $\epsilon>0$ and case (3) if $\epsilon<0$.
With an exchange of two pairs, such as
$(z_1,z_2,z_3,z_4,z_5)\to
(\overline{z}_2,\overline{z}_1,\overline{z}_4,\overline{z}_3,\overline{z}_5)$,
it is case (2) for both signs of $\epsilon$.
Note that cases (0) and (4) are mirror to Type IIB orientifold with
an O9-plane (Type I), case (1) and (3) are mirror to IIB orientifold with
O3/O7 planes, and case (2) is mirror to IIB orientifold with
O5-planes \cite{BH2}.

\subsection{Homology of O-planes}
\label{Ohomology}

We can study the homology classes of the Lagrangian manifolds of the 
orientifold loci by computing the integral of the holomorphic three-form 
$\Omega$ around them, where
\begin{equation}
\Omega = \frac{dz_1\wedge dz_2\wedge dz_3}{z_4}
\end{equation}
on the sheet $z_4=\sqrt{\mu-z_1^2-z_2^2-z_3^2}$. 

We find that
\begin{rrlist}

\item[(0)] For the case (0) the orientifold wraps a compact manifold with period: 
\begin{equation}
\varpi^{(0)}(\mu) =
\int_{S^3} \Omega = 2 \int_{x_1^2+x_2^2+x_3^2\le \mu} 
\frac{dx_1dx_2dx_3}{\sqrt{\mu-x_1^2-x_2^2-x_3^2}}
= 8\pi\mu\int_0^1 \frac{r^2 dr}{\sqrt{1-r^2}}
= 2\pi^2\mu.
\end{equation}

\item[(1)] The orientifold locus is non-compact, and we need to introduce a 
cutoff, which we put at $|z|^2= x^2+y^2=\Lambda$. 
\begin{equation}
\begin{split}
\varpi^{(1)}(\mu) &= \int_{S^2\times\reals} \Omega = 
2 \ii \int_{\mu\le x_2^2+x_3^2+x_4^2\le \frac{\Lambda+\mu}2} 
\frac{dx_2dx_3dx_4}{\sqrt{x_2^2+x_3^2+x_3^2-\mu}} \\
&= 8\pi\ii \mu \int_{1}^{\sqrt{\frac{\Lambda+\mu}{2\mu}}}
\frac{r^2 dr}{\sqrt{r^2-1}}
= 4\pi\ii\mu\Bigl[
\frac{\sqrt{\Lambda^2-\mu^2}}{2\mu} + 
\log\frac{\sqrt{\Lambda+\mu}+\sqrt{\Lambda-\mu}}{\sqrt{2\mu}}
\Bigr].
\end{split}
\end{equation}

\item[(2)] As in the previous case, the orientifold is non-compact.
\begin{equation}
\begin{split}
\varpi^{(2)}(\mu) &= \int_{S^1\times\reals^2} \Omega =
-2 \int_{\topa{\mu+y_1^2+y_2^2-x_3^2\ge 0}{y_1^2+y_2^2\le \frac{\Lambda-\mu}2}}
\frac{dy_1dy_2dx_3}{\sqrt{\mu+y_1^2+y_2^2-x_3^2}} \\
&= - \pi^2 (\Lambda - \mu). 
\qquad\qquad\qquad\qquad\qquad\qquad\qquad\qquad\qquad\qquad\qquad\qquad
\end{split}
\end{equation}

\item[(3)] We have:
\begin{equation}
\begin{split}
\varpi^{(3)}(\mu) &=
\int_{\reals^3}\Omega = -2\ii \int_{y_1^2+y_2^2+y_3^2\le \frac{\Lambda-\mu}2}
\frac{dy_1dy_2dy_3}{\sqrt{\mu + y_1^2+y_2^2+y_3^2}}  \\
&= -8\pi\ii \mu \int_0^{\sqrt{\frac{\Lambda-\mu}{2\mu}}} \frac{r^2 dr}{\sqrt{1+r^2}} 
= -4\pi\ii\mu\Bigl[ 
\frac{\sqrt{\Lambda^2-\mu^2}}{2\mu} - 
\log\frac{\sqrt{\Lambda+\mu}+\sqrt{\Lambda-\mu}}{\sqrt{2\mu}}\Bigr].
\end{split}
\end{equation}

\item[(4)] This is empty.
\end{rrlist}

If we evaluate case (1) for $\mu<0$ (following the computation in case
(3)), and subtract the result from the direct analytic continuation of (1), the
difference is
\begin{equation}
\varpi^{(1)}(\mu) - \varpi^{(1)}(-\mu) = 4\pi\ii\mu \log\sqrt{-1} = 2\pi^2\mu,
\end{equation}
which is exactly the period of the vanishing $S^3$. The same result also holds
for case (2)
\begin{equation}
\varpi^{(2)}(\mu)-\varpi^{(2)}(-\mu) = 
- \pi^2 (\Lambda-\mu) - (-\pi^2(\Lambda+\mu)) = 2\pi^2\mu,
\eqlabel{delta2}
\end{equation}
and trivially for (0)/(4). 

We can understand this by noting that the transition
does not affect the boundary of the O-plane and so we may glue the O-planes for $\mu<0$ and $\mu>0$ along their common boundary to form a compact three-cycle which must then be homologous to an integer multiple of the minimal $S^3$. The calculation shows that this integer is one. 

It is natural to propose that this holds as
a universal result, and not just for the simple conifold singularity we
have studied here. We conjecture:

\begin{center}
\shadowbox{\parbox{\textwidth-1cm}{When crossing the conifold locus of real 
co-dimension one in the geometric moduli space of an orientifold model, the 
class of the O-plane changes by the class of the vanishing cycle.}}
\end{center}

The non-compact part of the homology of the O-planes (which contributes the
$\mu\log\mu$ part in the expressions above) can be understood from the 
intersection with the compact three-cycle. It is easy to see that when we 
make the intersection between O-plane and the $S^3$ transversal, the $S^1$ 
in case $(2)$ disappears completely, while the $S^2$ leaves two
intersection points. This follows from the fact that for a Lagrangian
submanifold such as our three-cycles, the normal bundle is isomorphic to 
the tangent bundle via contraction with the K\"ahler form. The number of
intersection points is then simply the Euler characteristic of the
(non-transversal) intersection locus.

\subsection{Orientifolds of the Resolved Conifold}

Let us now discover what these involutions look like for the blown-up conifold.
This space, also known as $\calo_{\projective^1}(-1)\oplus\calo_{\projective^1}(-1)$,
can be written as
\begin{equation}
\begin{pmatrix} x & u \\ v & y \end{pmatrix} 
\begin{pmatrix} \lambda_1 \\ \lambda_2 \end{pmatrix} = 0,
\end{equation}
with $(\lambda_1,\lambda_2)\in\projective^1$ and
\begin{equation}
\begin{split}
x &= z_1 + \ii z_2, \qquad\qquad y = z_1 - \ii z_2, \\
u &= z_3 + \ii z_4, \qquad\qquad v = -z_3 + \ii z_4.
\end{split}
\end{equation}
The blow up breaks the $\O(4)$ symmetry we have used above to $\SO(4)$,
while at the same time restoring the $\U(1)$ phase symmetry. Thus, in
\eqref{ori}, $M$ must have determinant $+1$, while $\alpha$ is a priori
arbitrary. But note that by a change of coordinates, $\alpha$ can be 
conjugated to $0$. The condition that the number of $-1$ eigenvalues
of $M$ must be even eliminates cases (1) and (3) discussed previously for
the deformed conifold, while (0) and (4) become equivalent on the
blown-up side. In cases (1) and (3) the K\"ahler parameter is projected out by 
the orientifold action and it is not possible to blow up the singularity.
This leaves two cases:

\begin{rrlist}
\item[(0)] (equivalent to (4)) maps $(x,y,u,v)\to (\bar y, \bar x,-\bar v,
-\bar u)$. In terms of the inhomogeneous coordinate $z=\lambda_1/\lambda_2 
= - u/x = -y/v$ the $\projective^1$ is mapped as
\begin{equation}
z\to - \frac{1}{\bar z}\,\,.
\label{inv1}
\end{equation}
This is a freely acting orientifold.

\item[(2)] The action $(x,y,u,v) \to (-\bar y,-\bar x,-\bar v,-\bar u)$ must
be accompanied by
\begin{equation}
z \to \frac{1}{\bar z}\,\,.
\eqlabel{accompa}
\end{equation}
The fixed point set is an $S^1\times \reals^2$.
\end{rrlist}

We note that these cases coincide with the cases where the orientifold action 
preserves the same supersymmetry as the D6-branes wrapping the $S^3$, and are
precisely the cases discussed by Acharya, Aganagic, Hori and Vafa in \cite{Wafa}.

\subsection{The gauge group}
\label{gauge}

For future reference, it is useful to describe here which gauge theories will
be living on the worldvolume of D6-branes that are wrapping this geometry.

In flat space $N$ dynamical D6-branes on the top of an O6$^-$-plane yield an 
$\SO(2N)$ gauge group on the worldvolume. The O6$^-$-plane has a RR charge 
$-2$ in D6-brane units, so the total charge of the system is $N-2$. When 
wrapping an $S^3$ the D6-branes cannot be higgsed away. 
The four-dimensional gauge theory is pure  ${\cal N}=1$ $\SO(2N)$
super Yang-Mills. This theory is confining with $h= 2N-2$ different vacua.

A similar classical configuration is a system of $N-4$ D6-branes on the top 
of an O6$^+$-plane. The  O6$^+$-plane has RR charge $+2$, so the whole system 
also has charge $N-2$. The low energy theory is pure ${\cal N}=1$ $\SP(N-4)$ 
super Yang-Mills. The theory is confining and has $h=N-3$ different vacua.

This discussion was of course standard. Slightly less familiar are D-branes 
wrapping in freely acting orientifolds (such as case (4) above), but it is also 
clear what will result. The involution acts on the $S^3$ as the antipodal map,
$x\to -x$. Locally, this simply identifies excitations at antipodal points 
on the sphere, via an anti-unitary transformation on the D6-brane degrees of 
freedom. That can be understood as an action on the Chan-Paton matrices:
\begin{equation}
\lambda(x) \rightarrow  - \gamma_{\Omega} \lambda(-x)^T \gamma_{\Omega}^{-1},
\end{equation}
where $\lambda$ are $2M\times 2M$
hermitian matrices for $2M$ D6-branes wrapping
the covering $S^3\ni x$. Locally on $S^3$, this simply yields a $\U(2M)$
gauge group. The zero modes, however, suffer a slightly different projection.
The orientifold action is an involution if $\gamma_{\Omega} (\gamma_{
\Omega}^T)^{-1} = \epsilon$, with $\epsilon = \pm 1$, i.e. $\gamma_{\Omega}$ 
is symmetric or antisymmetric. Depending on the sign the four-dimensional
theory will be pure super Yang-Mills with gauge group $\SO(2M)$ or $\SP(M)$. 
As before the system preserves  ${\cal N}=1$.

\section{Symmetries of $G_2$ holonomy metrics}
\label{symmetries}

In this section and the next, we discuss aspects of the $G_2$ lift of the 
deformed and resolved conifold with branes and fluxes. Most of this section is
review \cite{cande,bggg,cvetic}, but the careful discussion of the symmetry breaking
pattern will be crucial in our subsequent analysis. 

\subsection{Deformed and resolved conifold}

We begin by recording the symmetry group of the conical Calabi-Yau metric
on the (singular) conifold,
\begin{equation}
z_1^2+z_2^2+z_3^2+z_4^2 = 0.
\eqlabel{conii}
\end{equation}
These isometries must preserve \eqref{conii} together with 
\begin{equation}
r^2 = \sum |z_i|^2.
\eqlabel{unit}
\end{equation}
In full glory, the symmetry group is
\begin{equation}
\bigl(\SU(2)\times\SU(2)\ltimes \widetilde\zet_2 \times 
\U(1)^{\rm phase}\ltimes \zet_2^{\rm cc}\bigr) 
/_{\zet_2\times\zet_2}.
\eqlabel{group}
\end{equation}
Here, the $\SO(4)\cong \SU(2)\times\SU(2)/\zet_2$ is extended
by $\widetilde\zet_2$ to the $\O(4)$ leaving the quadric invariant.
$\widetilde \zet_2$ acts as
$$
(z_1,z_2,z_3,z_4) \mapsto (-z_1,z_2,z_3,z_4).
$$
The $\U(1)^{\rm phase}$ contains the rotations
$$
R_\alpha: z_i\mapsto \ee^{\ii\alpha/2} z_i
$$
with $\alpha\in[0,4\pi]$, and $\alpha=2\pi$ corresponding to an element
of $\SO(4)$. Finally, $\zet_2^{\rm cc}$ is complex conjugation, represented
by
\begin{equation}
c_0: z_i \mapsto \bar z_i.
\eqlabel{cc}
\end{equation}
When conjugated by elements of $\U(1)^{\rm phase}$, $c_0$ becomes
\begin{equation}
c_\alpha = R_\alpha c_0 R_\alpha^{-1}: z_i \mapsto 
\ee^{\ii\alpha} \bar z_i.
\eqlabel{anti}
\end{equation}

When the singular conifold is smoothed out, some of these symmetries
are broken. The {\it deformation} which replaces \eqref{conii} by
\begin{equation}
\sum z_i^2 = \mu
\eqlabel{deform}
\end{equation}
breaks $\U(1)^{\rm phase}$ to the $\zet_2$ which is already part of 
$\O(4)$ (namely $\alpha=2\pi$). Since \eqref{anti} is a symmetry of 
\eqref{deform} for both $\alpha= \arg(\mu)$ and $\alpha=\arg(\mu)+
\pi$, we have a choice of orientifold $c_0: z_i\mapsto \bar z_i$ or 
$c_\pi: z_i \mapsto -\bar z_i$.

The blowup of the conifold is obtained by rewriting \eqref{conii} as
\begin{equation}
x y - u v = 0,
\end{equation}
and then replacing it with the two equations
\begin{equation}
\begin{pmatrix} x & u \\ v & y \end{pmatrix}
\begin{pmatrix} \lambda_1 \\ \lambda_2 
\end{pmatrix} = 0
\end{equation}
in $\complex^4\times \projective^1$. The blowup clearly 
preserves $\U(1)^{\rm phase}$, but breaks $\widetilde\zet_2$. 
Indeed, transposition of $z$ amounts to exchanging $u$ and 
$v$ and is equivalent to flopping the $\projective^1$.

As equation \eqref{anti} shows, orientifolding breaks $U(1)^{\rm phase}$ down 
to the $\mathbb Z_2$ subgroup which is in $SO(4)$. The addition of RR 2-form 
flux also breaks this $U(1)^{\rm phase}$ as will become clear after lifting 
to M-theory. Thus the symmetry group of the asymptotic IIA geometry relevant 
to our problem is:
\begin{equation}
\bigl(\SO(4)\ltimes \widetilde\zet_2 \times \zet_2^{\rm cc}\bigr).
\eqlabel{group2}
\end{equation}
Deforming the conifold  leaves these symmetries intact whilst resolving breaks 
$\widetilde\zet_2$. 

\subsection{Lift}

Next we would like to lift to M-theory and understand the action of the 
symmetry group \eqref{group2} there. The asymptotic boundary of the conifold 
is $T^{1,1}=\SU(2)\times \SU(2)/\U(1)$ and in the presence of RR 2-form flux 
the boundary of the M-theory lift is an orbifold of $\SU(2)\times\SU(2)$. 

This can be described a little more explicitly by introducing
some coordinates on the conifold. We write quaternionically
\begin{equation}
z = 
\begin{pmatrix} z_1+\ii z_2 & z_3+\ii z_4 \\
-z_3 +\ii z_4 & z_1-\ii z_2 \end{pmatrix}
= x + \ii y,
\end{equation}
with 
\begin{equation}
\begin{split}
x &= \begin{pmatrix} x_1 + \ii x_2 & x_3+\ii x_4  \\
-x_3 +\ii x_4 & x_1-\ii x_2\end{pmatrix}
= X \tilde{X}^\dagger, \\
y &= \begin{pmatrix} y_1 + \ii y_2 & y_3 + \ii y_4\\
-y_3 + \ii y_4 & y_1 -\ii y_2 \end{pmatrix}
= X \sigma \tilde{X}^\dagger
\end{split}
\end{equation}
and $\sigma = \bigl(\begin{smallmatrix} \ii & 0 \\ 0 & 
-\ii\end{smallmatrix}\bigr)$. Here we have used that 
\eqref{conii} and \eqref{unit} imply that (for $r=\sqrt{2}$) 
$x$, $y$ are orthogonal unit quaternions
\begin{equation}
\det x = \det y = 1\qquad \tr\; y^{\dagger} x = 0.
\eqlabel{base}
\end{equation}
Choosing our standard traceless $\SU(2)$ matrix $\sigma$, 
this equation is solved in terms of two $\SU(2)$ matrices 
$X$ and $\tilde{X}$, modulo the relation $(X,\tilde{X})\equiv (X\Theta,
\tilde{X}\Theta)$ with $\Theta=\Bigl(\begin{smallmatrix} \ee^{\ii\psi/2}
& 0 \\ 0 & \ee^{-\ii\psi/2}\end{smallmatrix}\Bigr)$. This 
exhibits the base of the conifold as either a (topologically
trivial) $S^2$ bundle over $S^3$ or as $T^{1,1}=\SU(2)\times
\SU(2)/\U(1)$. (We won't need its presentation as $T^{1,0}$, 
which uses $x$ and $X$ instead.)

$X$ and $\tilde{X}$ are the $S^3 \times S^3$ of the M-theory lift.  
The action of $SO(4)$ is by left multiplication of $SU(2) \times SU(2)$ on
$X$ and $\tilde{X}$. We can choose $\widetilde\zet_2$ to be $z_1 
\rightarrow -z_1$ which corresponds to exchanging $X$ and $\tilde{X} \beta$, 
where $\beta=\bigl(\begin{smallmatrix} 0 & 1 \\ -1 & 0\end{smallmatrix}
\bigr)$. The action of complex conjugation  $c_0$ is given by
\begin{equation}
X \mapsto X \beta,\qquad \tilde{X} \mapsto \tilde{X} \beta.
\eqlabel{beta}
\end{equation}
 
Adding the M-theory circle adds to the asymptotic symmetry group a factor of
$U(1)^M$ which is given by right multiplication of $X$ and $\tilde{X}$ by 
$\Theta$ as described above. Note that the orientifold action \eqref{beta} 
inverts the M-theory circle as it should. Also note that the right action 
of $\Theta$ and $\beta$ on $(X,\tilde{X})$ generates a larger group of 
symmetries which in particular contains the dihedral group $D_N$ for any 
$N$. We denote the group generated in this way by $G$.
 
A further useful set of coordinates is given by writing $S^3\times S^3$
as the quotient $\SU(2)^3/\SU(2)$,
\begin{equation}
(g_1,g_2,g_3)\sim (g_1g',g_2g',g_3g') \in \SU(2)^3/\SU(2).
\eqlabel{su2c}
\end{equation}
The base of the conifold is obtained by reduction along the maximal
torus of $g_1$. The explicit identification is
\begin{equation}
\begin{split}
x &= g_2 g_3^{-1}, \\
\tilde{X} &= g_3 g_1^{-1}, \\
X &= x \tilde{X} = g_2 g_1^{-1}.
\end{split}
\eqlabel{iden}
\end{equation}
Note that in these coordinates the flop is described by 
\begin{equation}
(g_1,g_2,g_3) \rightarrow (-\beta g_1,g_3, - g_2)\, .
\label{flop}
\end{equation}

Topologically then, the boundary of our $G_2$-holonomy manifolds will be an 
orbifold of $S^3\times S^3$. One interesting metric on $S^3\times S^3$, which 
is the one underlying the $G_2$ metric on the spin bundle on $S^3$ 
\cite{Bryand:1989mv,Gibbons:1989er}, has $\SU(2)^3\times \Sigma_3$ 
symmetry, where $\SU(2)^3$ acts on the left in \eqref{su2c}, and $\Sigma_3$ 
is the permutation of the three $\SU(2)$ factors \cite{AW}. $\Sigma_3$ is 
``spontaneously'' broken in the interior, and only $\SU(2)^3\times\zet_2$
are isometries of the full $G_2$ holonomy metrics. These metrics are 
relevant to the problem at infinite IIA coupling.

The metrics that are relevant for our discussion (at finite string coupling) 
have asymptotic symmetry group $(\SU(2) \times \SU(2) \ltimes\widetilde\zet_2 
\times G)/_{\zet_2}$, where $G$ was defined above as the group generated
by $U(1)^M$ and the orientifold action \eqref{beta}. The trivial $\zet_2$ 
is generated by the element $(-1,-1,-1) \in \SU(2) \times \SU(2) \times 
U(1)^M$.

In the interior, various symmetry breaking patterns are possible. Moreover, 
in certain cases it happens that the symmetry group is enhanced
to $\SU(2)^3$ in the deep interior. 

\subsection{Orientifolding}
\label{symori}

As in \cite{bobby,Atiyah:2000zz, AW}, we can consider dividing out by the action 
of a discrete group $\Gamma$ preserving the $G_2$ metric. Of interest to us 
is the case that $\Gamma$ is the (binary) dihedral group $D_N$. This group
has generators $a$, and $b$ satisfying the relations
\begin{equation}
a^{2N-4} = 1, \qquad b^2 = a^{N-2}, \qquad b a b^{-1} = a^{-1}.
\end{equation}
The group has a presentation
\begin{equation}
\zet_{2N-4} \longrightarrow D_N \longrightarrow \zet_2.
\eqlabel{present}
\end{equation}
The dihedral group has a standard action on the three sphere coming
from its embedding as a discrete subgroup of $\SU(2)$, namely
\begin{equation}
a = \begin{pmatrix} \ee^{\pi\ii/(N-2)} & 0 \\ 0 & \ee^{-\pi\ii/(N-2)}
\end{pmatrix},
\qquad
b = \begin{pmatrix} 0 & 1 \\ -1 & 0 \end{pmatrix}.
\eqlabel{standard}
\end{equation}
We will call the action of $D_N$ on $S^3$ in which $a$ and $b$ are 
represented in this way as $\rho$. The quotient of our interest is
obtained by letting $D_N$ act as $\rho$ on the right on both $X$
and $\tilde{X}$, in the variables \eqref{iden}. The effect of the $\zet_{2N-4}$
factor in \eqref{present} is to reduce the length of the M-theory
circle, thereby increasing the flux to $2N-4$ units.
After reduction on $\U(1)^M$, the remaining $\zet_2$ sends
$(x,y)\to (x,-y)$, so is indeed complex conjugation $c_0$.
In terms of the variables $(g_1,g_2,g_3)\in\SU(2)^3/\SU(2)$,
$D_N$ acts as $\rho$ on $g_1$ and trivially on $g_2$ and $g_3$.

We have discussed this action in detail in order to make the following
point. There is another action, call it $\tilde \rho$, of $D_N$ on $S^3$ 
in which $b$ is represented by $\left(\begin{smallmatrix}0 & -1 \\ 1 & 0 
\end{smallmatrix}\right)$. Acting with $D_N$ as $\tilde\rho$ on the right of
$X$ and as $\rho$ on the right of $\tilde{X}$ is equivalent to acting on 
$g_1$ as $\rho$ and on $g_2$ via the central action, $a=1$, $b=-1$. After 
reduction to $T^{1,1}$, the action is $(x,y)\to (-x,y)$, corresponding to
$c_\pi$. We shall sometimes refer to this action as $D_N'$. 

Note that the actions of $D_N$ and $D_N'$ can be conjugated into each 
other by a diffeomorphism of $S^3\times S^3$, sending $(X,\tilde{X})\to 
(X \ii\sigma_3 , \tilde{X})$. However, this diffeomorphism is {\it not} 
an isometry of the boundary metric since $U(1)^{\rm phase}$ is explicitly 
broken by the addition of flux.  

We should now ask, what are the symmetries of the boundary metric which 
also preserve the orbifold group $D_N$ or $D_N'$? In each case, $(\SU(2) 
\times \SU(2) \ltimes\widetilde\zet_2 \times G)/_{\zet_2}$ is broken to 
$\SO(4) \ltimes\widetilde\zet_2 \times \zet_2'$ where $\zet_2'$ is the 
centralizer of $D_N$ or $D_N'$ in $G$ and is generated by 
\begin{equation}
\sqrt{a} = \begin{pmatrix} \ee^{\pi\ii/2(N-2)} & 0 \\ 0 & \ee^{-\pi\ii/2(N-2)}
\end{pmatrix}.
\end{equation}

We can now make a few comments about the pattern of symmetry breaking by 
the various geometries in the interior. As we have already commented, we 
expect that $\widetilde\zet_2$ will be broken for the resolved conifold 
and unbroken for the deformed conifold. The two resolved conifolds
are interchanged by the broken  $\widetilde\zet_2$. On the other 
hand $\zet_2'$ is a subgroup of $U(1)^M$ and will be unbroken whenever 
translation  along the M-theory circle is a symmetry. This will fail to 
be true only for the deformed conifold with O6-plane on $S^3_>$ and $N=0$ 
or $1$ D6-branes. Near the O6-plane we expect the geometry to look like 
Atiyah-Hitchin space or its double cover, wrapped on $S^3_>$. The M-theory 
circle is no longer an isometry direction, but for $N=1$, corresponding to 
the double cover, $\zet_2'$ is unbroken whilst for $N=0$, $\zet_2'$ is 
broken. For $N=0$, this leads to two distinct geometries interchanged by 
the broken symmetry generator. We shall return to this point and its 
interpretation in Type IIA later.

\section{M-theory geometry}
\label{geometry}

In this section, we will flesh out our discussion of the M-theory geometries 
with some details of the explicit $G_2$ holonomy metrics. We shall describe 
in turn the three distinct classes of $G_2$ holonomy metrics relevant to 
our discussion. These are labeled $\MD, \MOD$ and $\MF$ and their various 
orbifolds correspond respectively to orientifolds of the deformed conifold 
with D6-branes, an O6-plane on $S^3_>$ with $N=0$ or 1 D6-branes and
orientifolds of the resolved conifold with flux. 

We present the details of these metrics in order to confirm the details of 
the symmetry breaking discussion of the previous section and also because 
the existence of the $\MOD$ metrics were not previously known in the 
literature. Furthermore, we show that there exists  a normalizable harmonic 
two-form on $\MOD$ which leads to a massless $U(1)$ gauge field in the IR. 
We can use this to rule out a smooth  transition between these backgrounds 
and others  with mass gap in the IR.

The reader who is not interested in the details of the metrics may wish to 
skip  to the discussion of the topology of the various spaces in
Section~\ref{topology}.

\subsection{Preliminaries}

We recall the Euler angle representation of $SU(2)$
matrices:
\begin{equation}
  X = \left(\begin{array}{rr}
      \cos\frac\theta2e^{ \frac{i}{2}(\psi+\phi)} &
     -\sin\frac\theta2e^{-\frac{i}{2}(\psi-\phi)} \\
      \sin\frac\theta2e^{ \frac{i}{2}(\psi-\phi)} &
      \cos\frac\theta2e^{-\frac{i}{2}(\psi+\phi)}
      \end{array}\right),~~~
\end{equation}
where the coordinates take values
\begin{equation}
  0\le \psi < 4\pi,~~~
  0\le \phi < 2\pi,~~~
  0\le \theta < \pi.
\end{equation}
The associated Maurer-Cartan one-forms $\sigma_a$ are defined
by $X^{-1}dX = \frac i2\tau_a\sigma_a$ where $\tau_a$ are Pauli's
matrices, and they satisfy
$d\sigma_a = \frac{1}{2}\epsilon_{abc}\sigma_b\sigma_c$.
They read
\begin{equation}
 \sigma_1 = -\cos\psi\sin\theta d\phi +\sin\psi d\theta,~~~
 \sigma_2 = -\sin\psi\sin\theta d\phi -\cos\psi d\theta,~~~
 \sigma_3 = d\psi +\cos\theta d\phi.
\end{equation}
One can write down the metric on $\mathR^4$ using polar coordinates
$r,\theta,\phi,\psi$ simply by relating the Cartesian
and polar coordinates as follows:
\begin{equation}
 W = \left(\begin{array}{rr}
      w_1+iw_2 & -w_3+iw_4 \\ w_3+iw_4 & w_1-iw_2\end{array}\right)
   = r\cdot X(\theta,\phi,\psi),~~~
 ds^2 = \frac12{\rm Tr}(dWdW^\dag)
      = dr^2 + \frac{r^2}{4}\sigma_a\sigma_a.
\label{R4}
\end{equation}

\subsection{Deformed conifold with D6-branes and orientifold}

Let us consider the M-theory lift of orientifolds of deformed conifold 
with D6-branes. The metrics should at least preserve the $\SU(2)\times 
\SU(2)\ltimes\widetilde\zet_2$ isometry of the deformed conifold, so we 
write the metrics in terms of $SU(2)$ matrices $X,\tilde X$ and the 
associated Maurer-Cartan forms $\sigma_a,\tilde\sigma_a$. The following ansatz 
for a $\widetilde\mathZ_2$-symmetric metric was considered in \cite{bggg}:
\begin{equation}
  ds^2 ~=~ dr^2 +\sum_{i=1}^3A_i^2(\sigma_i-\tilde\sigma_i)^2
                +\sum_{i=1}^3B_i^2(\sigma_i+\tilde\sigma_i)^2.
\label{ansdef}
\end{equation}
This metric is of $G_2$ holonomy provided the metric components obey
\begin{eqnarray}
 \ds\frac{dA_1}{dr} &=& 
 -\frac14\left[\frac{A_1^2-A_3^2-B_2^2}{A_3B_2}
              +\frac{A_1^2-A_2^2-B_3^2}{A_2B_3}\right], \nn\\
 \ds\frac{dA_2}{dr} &=& 
 -\frac14\left[\frac{A_2^2-A_1^2-B_3^2}{A_1B_3}
              +\frac{A_2^2-A_3^2-B_1^2}{A_3B_1}\right], \nn\\
 \ds\frac{dA_3}{dr} &=& 
 -\frac14\left[\frac{A_3^2-A_2^2-B_1^2}{A_2B_1}
              +\frac{A_3^2-A_1^2-B_2^2}{A_1B_2}\right], \nn\\
 \ds\frac{dB_1}{dr} &=&
  \frac14\left[ \frac{A_2^2+A_3^2-B_1^2}{A_3A_2}
               +\frac{B_1^2-B_2^2-B_3^2}{B_2B_3}\right], \nn\\
 \ds\frac{dB_2}{dr} &=&
  \frac14\left[ \frac{A_3^2+A_1^2-B_2^2}{A_1A_3}
               +\frac{B_2^2-B_3^2-B_1^2}{B_3B_1}\right], \nn\\
 \ds\frac{dB_3}{dr} &=&
  \frac14\left[ \frac{A_1^2+A_2^2-B_3^2}{A_2A_1}
               +\frac{B_3^2-B_1^2-B_2^2}{B_1B_2}\right].
\label{diffeq}
\end{eqnarray}
and the $G_2$ three-form is given by
\begin{eqnarray}
 \Phi &=& -e_1e_2e_3
        + e_0e_ie_{\tilde i}
        + \frac12\epsilon_{ijk}e_ie_{\tilde j}e_{\tilde k}
 \nn\\
  &=& -e_1e_2e_3 
     +e_0e_1e_{\tilde 1}
     +e_0e_2e_{\tilde 2}
     +e_0e_3e_{\tilde 3}
     +e_1e_{\tilde 2}e_{\tilde 3}
     +e_2e_{\tilde 3}e_{\tilde 1}
     +e_3e_{\tilde 1}e_{\tilde 2}.
\end{eqnarray}
The M-theory geometries $\MD$ and $\MOD$ both take the above form.
The metric for $\MD$ has an additional $U(1)^M$ symmetry corresponding
to translation along M-theory circle, while $\MOD$
has no such $U(1)$ symmetry.

Numerical analysis of the differential equations proceeds in the
following way.
We first perform a power series analysis at the origin $r=0$ to find
correct initial values for $A_i,B_i$ that make the solution smooth there.
We then let them evolve according to (\ref{diffeq}) and see if the
metric asymptotes  to that of a $(\mbox{conifold})\times S^1$ with flux,
\begin{equation}
 (A_1,A_2,A_3,B_1,B_2,B_3) ~\propto~
 \frac{r}{6} \cdot (\sqrt3,\sqrt3,~2,\sqrt3,\sqrt3,~0),
\label{asconi}
\end{equation}
or that of a $G_2$ cone over $S^3\times S^3$ (which describes the IIA 
theory at infinite string coupling),
\begin{equation}
 (A_1,A_2,A_3,B_1,B_2,B_3) ~\propto~
 \frac{r}{6} \cdot (\sqrt3,\sqrt3,\sqrt3,~1,~1,~1).
\label{ascone}
\end{equation}
We will find that initial conditions which are regular at the origin
do not always lead to a sensible asymptotic behavior.

\subsubsection{Numerical analysis for $\MD$}

The correct initial data for smooth metric with an $S^3$ of unit
size is given by the following power series
\begin{equation}
 A_i=1+\frac1{16}r^2+{\cal O}(r^3),~~~
 B_i=\frac{r}{4}+\frac{b_i}{192}r^3+{\cal O}(r^4),~~~~
 b_1+b_2+b_3=-3.
\label{init1}
\end{equation}
This leading-order behavior uniquely determines
the solution as a power series at $r=0$.

\begin{figure}[htb]
{\scriptsize
\psfrag{b1}{$b_1$}
\psfrag{b2}{$b_2$}
\psfrag{C1}{$C_1$}
\psfrag{C2}{$C_2$}
\psfrag{C3}{$C_3$}
\psfrag{A}{$-3$}
\psfrag{B}{$-1$}
\psfrag{C}{$-1$}
\psfrag{D}{$-3$}
\psfrag{D1}{(I)}
\psfrag{D2}{(II)}
\psfrag{D3}{(III)}
\psfrag{E}{$P$}
\centerline{\epsfig{width=6cm,file=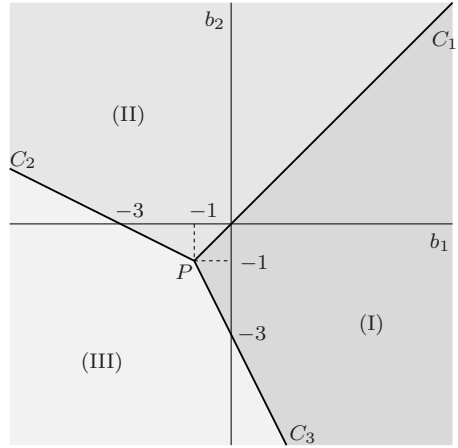}}
}
\caption{Parameter space of smooth initial data for $\MD$.
 The metric has sensible asymptotics only when the initial data is
 on one of the three half lines $C_{1,2,3}$ bounding the regions (I),
 (II), (III).}
\label{fig:init1}
\end{figure}

The family of smooth initial data is parameterized by $b_1,b_2$.
However, the metric asymptotes to (\ref{asconi}) only when
the initial data is on a half-line $C_1:b_1=b_2\ge-1$ in the parameter
space. (Owing to the permutation symmetry of $1,2,3$, there are three 
related families of initial conditions leading to metrics with sensible 
asymptotics. They are three half-lines $C_1,C_2,C_3$ meeting at 
$P:b_1=b_2=b_3=-1$ as depicted in Figure \ref{fig:init1}.)

Let us focus on the family of solutions $C_1$. Since we have fixed the 
size of the minimal $S^3$, the value of $b_1=b_2$ determines the 
radius of the M-theory circle at infinity, which is roughly 
$\ds\lim_{r\to\infty}B_3$. The radius blows up as $b_1=b_2$ approach 
$-1$, and the solution has $SU(2)$ enhanced isometry and asymptotes 
to (\ref{ascone}) at infinity. The solution is nothing but the familiar 
asymptotically conical $G_2$ holonomy metric on the spin bundle over 
$S^3$ \cite{Bryand:1989mv,Gibbons:1989er}. Alternatively, if we consider 
the family $C_1$ of rescaled solutions having a fixed radius of 
M-theory circle, then the limit of approaching $P$ is the limit of 
vanishing $S^3$ at the center.

It is easy to see that, for the solutions on $C_1$, the equalities
$A_1=A_2, B_1=B_2$ hold all the way along the radial evolution.
So the manifold $\MD$ has an additional $U(1)^M$ isometry corresponding
to translation along the M-theory circle.
At the point $P$ the solution satisfies $A_1=A_2=A_3, B_1=B_2=B_3$
along the radial evolution, so the $U(1)^M$ is enhanced further to an $SU(2)$.

The solutions at $r=0$ take the form
\begin{eqnarray}
 ds^2 &\simeq&
 dr^2 + \sum_{i=1}^3(\sigma_i-\tilde\sigma_i)^2
      + \frac{r^2}{16}\sum_{i=1}^3(\sigma_i+\tilde\sigma_i)^2
 \nn\\ &=&
 dr^2 + \sum_{i=1}^3\tilde\Sigma_i^2
      + \frac{r^2}{16}\sum_{i=1}^3(2\Sigma_i-\tilde\Sigma_i)^2,
\end{eqnarray}
where we introduced $Y=\tilde XX^{-1}$ and
$\Sigma=XdX^{-1},\tilde\Sigma=Y^{-1}dY$.
Recalling the metric (\ref{R4}) on $\mathR^4$ one finds that
the geometry is a smooth $\mathR^4$ bundle over $S^3$,
where $Y$ gives the base $S^3$ and $(r,X)$ parameterize
the fiber $\mathR^4$.

\subsubsection{Orbifolding}

The M-theory lifts of $N$ D6-branes or (O6 + $N$D6)  are
orbifolds of $\MD$ by $A_{N-1}$ or $D_N$ groups $\Gamma$.
The orbifold group acts diagonally on $X$ and $\tilde X$ from the right.
\begin{equation}
 g\in\Gamma~:~~(X,\tilde X) \mapsto (Xg,\tilde Xg).
\end{equation}
As we have seen, near the origin the coordinate $Y=\tilde XX^{-1}$
describes the base $S^3$ of finite volume, and $X^{-1}$ describes
the shrinking $S^3$.
$\Gamma$ acts on them as
\begin{equation}
 g\in\Gamma~:~~(Y,X^{-1}) \mapsto (Y,g^{-1}X^{-1}).
\end{equation}
Thus orbifolding gives a geometry which is $\mathC^2/\Gamma$
fibered over $S^3$.

The M-theory circle corresponds to the shift
$\psi\to\psi+\alpha,\tilde\psi\to\tilde\psi+\alpha$.
Its radius can be read off from the metric at infinity
\begin{eqnarray}
 ds^2 &=& \cdots + A_3^2(\sigma_3-\tilde\sigma_3)^2
                 + B_3^2(\sigma_3+\tilde\sigma_3)^2,
 \nn\\&=& \cdots + A_3^2(d\check\psi+\cos\theta d\phi
                         -\cos\tilde\theta d\tilde\phi)^2
                 + B_3^2(d\hat\psi+\cos\theta d\phi
                         +\cos\tilde\theta d\tilde\phi)^2.
\end{eqnarray}
Here $\hat\psi=\psi+\tilde\psi$ is the coordinate on M-theory circle,
and $\check\psi=\psi-\tilde\psi$ is one of the coordinates of $T^{1,1}$.
From the periodicity of $\psi,\tilde\psi$ it follows that
\begin{equation}
 0\le \check\psi < 4\pi,~~~~
 0\le \hat\psi < 8\pi.
\end{equation}
The RR charge is an integral of the field strength of
the one-form potential $A$,
\begin{equation}
 A = \frac14(\cos\theta d\phi + \cos\tilde\theta d\tilde\phi)
\end{equation}
over the $S^2$ defined by $\theta=\tilde\theta, \phi=\tilde\phi$
(One can check that its volume becomes zero at $r=0$).
We normalized $A$ so that it appears in the
M-theory metric as $(d\theta+A)^2$, where $\theta$ is of period $2\pi$.
The D6-brane charge is defined by
\begin{equation}
  Q_{\rm D6} ~=~ \int_{S^2}\frac{dA}{2\pi},
\end{equation}
which is unity for $\MD$ without orbifolding.

The $A_{N-1}$ orbifold group is generated by
$a=\exp(\frac{2\pi i\tau_3}{N})$.
This simply shortens the period of M-theory circle to $1/N$,
and therefore increases the D6-brane charge.
For $D_N$ orbifolds, $\Gamma$ is generated by
$a=\exp(\frac{i\pi\tau_3}{N-2})$ and $b=i\tau_2$.
The element $b$ acts as an antipodal map on $S^2$, and the D6-brane charge
as counted in the covering space becomes
\begin{equation}
  Q_{\rm D6} ~=~ 2\times\int_{{\mathR\mathP}^2}\frac{dA}{2\pi} ~=~ 2N-4.
\end{equation}
The action of $b$ on the coordinates of the conifold was obtained in the 
previous section and is the complex conjugation $c_0$: 
\begin{equation} z_i \rightarrow \bar{z}_i. 
\end{equation}

The M-theory lift of the free orientifold of the deformed conifold with $2N-4$
D6-branes wrapping the $S^3_<$ uplifts to a different orbifold $\MD/D'_N$.
The generators $a, b$ of $D'_N$ act on $(X,\tilde X)$ as
\begin{equation}
\begin{array}{rcl}
  a&:&(X,\tilde X)\mapsto
  \ts(~X\exp(\frac{i\pi\tau_3}{N-2}),~\tilde X\exp(\frac{i\pi\tau_3}{N-2})~),\\
  b&:&(X,\tilde X)\mapsto
  \ts(~X\cdot i\tau_2,~-\tilde X\cdot i\tau_2~).
\end{array}
\label{D'_N}
\end{equation}
The generator $b$ maps $z_i\mapsto -\bar z_i$, and in particular
acts on the minimal $S^3_<$ as the antipodal map.

\subsubsection{Numerical Analysis for $\MOD$}

Next we would like to describe the M-theory metrics corresponding to
O6$^-$ and $N=0$ or $1$ D6-branes on $S^3_>$ of the deformed conifold.
In this case the net flux from the O6$^-$ and D6 is negative and the M-theory
geometry is expected to look locally like Atiyah-Hitchin space ($N=0$)
or its double cover ($N=1$), wrapped on $S^3_>$. 

We are looking for a $G_2$ metric which is asymptotic to  \eqref{asconi} or 
\eqref{ascone}. Furthermore, we expect the orbifold at infinity to be $D'_4$ 
for $N=0$ and $D'_3$ for $N=1$. Recall that $D'_3$ is generated by $b$ which 
acts on $(X, \tilde{X})$ as in \eqref{D'_N}.

By analogy with the construction of Atiyah-Hitchin space\cite{AH},
we expect that 
orbifolding by $D'_3$ is necessary in order to remove a conical singularity 
at the origin. This will be the case if $A_2 = r + ...$ near $r=0$ so that 
the metric becomes
\begin{eqnarray}
 ds^2 &=&
   A_1^2 (\sigma_1-\tilde\sigma_1)^2+A_3^2(\sigma_3-\tilde\sigma_3)^2
  +B_1^2 (\sigma_1+\tilde\sigma_1)^2+B_2^2(\sigma_2+\tilde\sigma_2)^2
  +B_3^2(\sigma_3+\tilde\sigma_3)^2
 \nn\\ &&
  +dr^2+r^2(\sigma_2-\tilde\sigma_2)^2 +\cdots.
\end{eqnarray}
Here $A_1, A_3, B_1, B_2, B_3$ should be regarded as constants near $r=0$.
The first line gives the metric of a five-dimensional bolt which is
topologically $S^2 \times S^3$. The second line yields a conical singularity 
which is removed by orbifolding by $D'_3$ as in \eqref{D'_N}. Note that in 
order for the shrinking $S^1$ to become an isometry direction at the origin 
(which is needed to avoid a singularity) we require $A_1 = B_3$ and $B_1 = 
A_3$ at $r=0$.

The initial data for a smooth metric at $r=0$ with five-dimensional bolt is 
\begin{equation}
A_1=B_3,~ B_1=A_3,~A_2=0.
\label{initMO}
\end{equation}
Interestingly, if we set $(A_3, B_1, B_2)$ all equal and much greater than 
$A_1 (= B_3)$, then for small $r$, $(A_3,B_1,B_2)$ stay almost constant while 
$(A_1,A_2,B_3)$ approximately obey the equations for the Atiyah-Hitchin 
manifold\cite{AH}.
However, the numerical analysis shows that such solutions do not
extend toward large $r$ and we should not try to impose these extra conditions.

\begin{figure}[h]
{\scriptsize
\psfrag{D1}{(I)}
\psfrag{D2}{(II)}
\psfrag{D3}{(III)}
\psfrag{P}{$P$}
\psfrag{C1}{$C_1$}
\psfrag{C2}{$C_2$}
\psfrag{C3}{$C_3$}
\psfrag{A1}{$B_1=A_3$}
\psfrag{A3}{$B_2$}
\psfrag{B1}{$A_1=B_3$}
\centerline{\epsfig{width=6cm,file=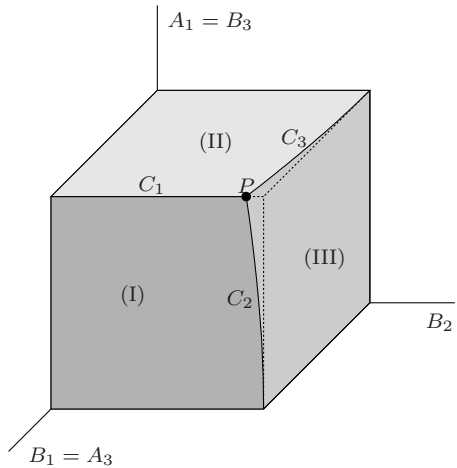}}
}
\caption{
  Parameter space of smooth initial data for the space $\MO,\MOD$
  with fixed scale.
  The metric behaves nicely when the initial data is chosen from one of
  the curves $C_{1,2,3}$ separating the three regions (I), (II), (III).
  The special point $P$ corresponds to
  $(A_1,A_2,A_3,B_1,B_2,B_3)=(1,0,1,1,.9171,1)$.}
\label{fig:init2}
\end{figure}

The power series analysis shows that any initial values for $(A_1,B_1,B_2)$ 
uniquely determine a solution, but this will have sensible asymptotics at large 
$r$ only for some particular fine-tuned initial data. After fixing an overall 
scale, one gets a two-dimensional parameter space of initial conditions.
We choose this to be three faces of a cube,
\[
 \{B_1=1,~~ B_2,A_1\in[0,1]\},~~~~
 \{B_2=1,~~ B_1,A_1\in[0,1]\},~~~~
 \{A_1=1,~~ B_1,B_3\in[0,1]\}
\]
as depicted in Figure \ref{fig:init2}.
One can numerically see that the solutions for generic initial conditions
do not behave nicely at infinity: either $B_1, B_2$ or $B_3$ blows up
much faster than the others.
The generic initial conditions are grouped into
regions (I), (II), (III) shown in Figure \ref{fig:init2} according to
the asymptotics.
The solution behaves nicely at large $r$ if the initial
condition is chosen on the curves $C_{1,2,3}$ separating three regions.

The curve $C_1$ is a straight line segment corresponding to the initial data
\[
 A_1=B_3=B_1=A_3=1,~~0<B_2\le0.917,~~A_2=0.
\]
Generic solutions with this initial condition asymptote locally
to the conifold times an $S^1$ with flux,
\begin{equation}
(A_1,A_2,A_3,B_1,B_2,B_3)~\propto~(\sqrt{3},2,\sqrt{3},\sqrt{3},0,\sqrt{3}).
\end{equation}
These solutions were discovered in \cite{cvetic} and named $\mathC_7$.
They do not have the expected asymptotics \eqref{asconi} since the M-theory 
circle is in the $B_2$ direction rather than $B_3$. With these asymptotics, 
the orbifolding by $D'_3$ acts as an orbifold on the base rather than as an 
orientifold. In addition, the solution has a $U(1)$ isometry 
$A_1=A_3, B_1=B_3$ 
all along the flow, corresponding to translation along the M-theory circle. 
The 
solutions which we seek with O6$^-$-plane and $N=0$ or 1 D6 are not expected 
to have 
such a $U(1)^M$ isometry. As explained in \cite{hp2005} the solution
$\mathC_7$ should be interpreted as the uplift of IIA on a manifold with
local $\mathP^1\times\mathP^1$ and unit flux through each $\mathP^1$. This 
solution will have no part to play in our current analysis.

The other two curves $C_2, C_3$ are related by the symmetry of the
differential equation (\ref{diffeq}).
\begin{equation}
 (A_1,A_2,A_3,B_1,B_2,B_3)\to
 (A_3,A_2,A_1,B_3,B_2,B_1).
\label{sym1}
\end{equation}
The solutions on $C_2$ flow to the correct asymptotic boundary conditions 
\eqref{asconi} and these turn out the solutions which we label $\MOD$
after an appropriate orbifolding.
The solutions 
on $C_3$ describe the same set of solutions in different coordinates. 
We do not 
need to take account of them separately since the asymptotic boundary 
conditions 
\eqref{asconi} partially fix our choice of coordinates and the solutions 
on $C_3$ 
do not respect these asymptotics.

As we have discussed, it is necessary to orbifold the solutions 
by $D'_3$ in order to remove a conical singularity at the origin.
The resulting family of solutions $\MOD$ is asymptotically an
orientifold of the conifold with RR charge $(-1)$ and 
one can identify them with the M-theory lift of the O6$^-$+D6 system.

Further orbifolding by the right action of $(i\tau_3)\otimes(i\tau_3)$
on $(X,\tilde X)$ halves the radius of the M-theory circle and thereby
doubles the RR charge. Note that this does not lead to any orbifold 
singularities 
in the M-theory geometry $\MO$. The enlarged orbifold group acts as $D'_4$ 
on the 
boundary and this is the solution corresponding to an O6$^-$ and no D6 branes.

\subsubsection{Normalizable harmonic two-form?}

If there is such a two-form $\omega$, it will lead to the presence of $U(1)$ 
gauge dynamics in the IR. We expect that this should be the case for the 
M-theory lift of O6$^-$+D6 since this will have $SO(2)=U(1)$ gauge group.

The two-form can be written locally as a derivative of a one-form.
We assume that it takes the form
\[
 \omega = \sum_i d\left\{
    f_i(r)\cdot(\sigma_i-\tilde\sigma_i)
   +g_i(r)\cdot(\sigma_i+\tilde\sigma_i)
                   \right\}.
\]
Instead of trying to solve $d\omega=d\ast\omega=0$, we try to solve
the simpler equation
\begin{equation}
  \ast\omega = \alpha\Phi\wedge\omega,
\end{equation}
where $\alpha$ is a real constant, requiring $\omega$ to fall into an
irreducible representation of $G_2$.
A little calculation shows that $\alpha=1$ and $f_i,g_i$ have to satisfy
\begin{equation}
 \frac{1}{f_1}\frac{df_1}{dr} +\frac12\left(
    \frac{A_1}{A_2B_3}+\frac{A_1}{B_2A_3} \right) ~=~ 0,~~~~~
 \frac{1}{g_1}\frac{dg_1}{dr} +\frac12\left(
    \frac{B_1}{A_2A_3}-\frac{B_1}{B_2B_3} \right) ~=~ 0,~~{\rm etc.}
\end{equation}
The equation shows the existence of six independent
harmonic two-forms corresponding to $f_i, g_i$.
The normalizability of each mode can be analyzed using the form of
$(A_i, B_i)$ at the origin and infinity.

For $\MOD$, the mode proportional to $f_2$ behaves at $r=0$ and
$r=\infty$ as
\begin{equation}
f_2 \stackrel{r\to0}\sim 
       1-\frac{1+A_1^{-2}}{4}r^2+{\cal O}(r^3),~~~~~
f_2 \stackrel{r\to\infty}\sim 
       \exp\left(-\frac{r}{2 B_3}\right),
\end{equation}
and turns out to be normalizable.
Note that this mode is projected out upon orbifolding further
to get $\MO$.
For $\MD$, none of these two-forms is normalizable.
The absence of a normalizable harmonic two-form for $\MD$ is puzzling,
as discussed in \cite{bggg}, since one would expect
the gauge group on $N$ coincident D6-branes should be $U(N)$
and its $U(1)$ part should remain in the infrared limit.

\subsection{Resolved conifold with flux and orientifold}

For the M-theory lift of the resolved conifold with flux and orientifold,
one expects the symmetry $SU(2)\times SU(2)\times U(1)^M$ but no
$\widetilde\zet_2$ exchanging the two sets of Maurer-Cartan forms.
The following ansatz for the metric was considered
in \cite{cvetic}
\begin{equation}
 ds^2 ~=~ dr^2
 + a^2\{(\sigma_1+g\tilde\sigma_1)^2+(\sigma_2+g\tilde\sigma_2)^2\} 
 + b^2(\tilde\sigma_1^2+\tilde\sigma_2^2) + c^2(\sigma_3+g_3\tilde\sigma_3)^2
 + f^2\tilde\sigma_3^2
\label{ansres}
\end{equation}
The metric is of $G_2$ holonomy provided $g=\frac{af}{2bc}, g_3=-1+2g^2$ and
\begin{equation}
\begin{array}{rcl}
 \dot a &=& \ds-\frac{c}{2a}+\frac{a^5f^2}{8b^4c^3},\\
 \dot c &=& \ds-1+\frac{c^2}{2a^2}+\frac{c^2}{2b^2}-\frac{3a^2f^2}{8b^4},
\end{array}
~~
\begin{array}{rcl}
 \dot b &=& \ds-\frac{c}{2b}-\frac{a^2(a^2-3c^2)f^2}{8b^3c^3},\\
 \dot f &=& \ds-\frac{a^4f^3}{4b^4c^3}.
\end{array}
\end{equation}
As initial conditions we put $a=c=0,~ b=1$ and $f=f_0$.
The regularity at $r=0$ requires
\begin{equation}
a=\frac{r}{2}+{\cal O}(r^2),~~~
c=-\frac{r}{2}+{\cal O}(r^2).
\end{equation}
If $f_0<1$, numerical solutions asymptote to $\mbox{(conifold)}\times S^1$,
\begin{equation}
 (a,b,c)\sim r\cdot(\sqrt{1/6},\sqrt{1/6},-\sqrt{1/9}),~~~
 f\sim {\rm const},~~~ g\sim 0,~~~ g_3\sim -1.
\end{equation}
These solutions were named $\MF$ in \cite{cvetic}.
At $f_0=1$ the $S^1$ decompactifies and the solution coincides
with the familiar asymptotically conical $G_2$ metric on the spin bundle over $S^3$.

Interestingly, the M-theory geometry has
$SU(2)\times SU(2)\times U(1)^M$ symmetry and no extra $U(1)^{\rm phase}$.
It seems that the $U(1)^{\rm phase}$ isometry corresponding to the phase rotation
$z_i\to e^{i\alpha}z_i$ of the resolved conifold is broken in the presence of flux.

One can increase the flux or introduce an orientifold action simply
by orbifolding the M-theory geometry.
The orbifold group can be either $A_{N-1}$ or $D_N$ groups 
acting on $(X,\tilde X)$ in the same way from the right, or it can be the
group $D'_N$ defined in (\ref{D'_N}). All these groups act freely.
The RR charge is $N$ for $A_{N-1}$ orbifolds, $N-2$ for $D_N$ orbifolds
and $2-N$ for $D'_N$ orbifolds.

\section{Quantum moduli space}
\label{analyze}

We are now in a position to study the quantum moduli space
of supersymmetric orientifold vacua for different values of RR charge $Q=N-2$.
We first analyze them through the behavior of membrane instanton
factors $\eta_i$ on various classical M-theory geometries, as in \cite{AW}.
We then study them using Vafa's exact superpotential.

From Table \ref{summary}, it appears that there are six
classes of cases.
$N-2>1$ and $N<0$ are ``regular'' and can be understood completely
using M-theory arguments or the Vafa superpotential.
$N=3$ is somewhat special, but the same analysis applies.
In all these cases, the moduli space turns out to be a copy of $\projective^1$ 
with some number of marked points.
When $N=0,1$ or $2$, the moduli space of vacua consists of
different branches and needs to be discussed separately.

\subsection{M-theory lift}
\label{topology}

Here we try to study the quantum moduli space within the M-theory
framework, through the behavior of various membrane instanton factors
as holomorphic functions on moduli space.
As preparation we need some basic results about the topology 
of the M-theory lifts.

We would like to find good chiral parameters to describe
the classical and quantum moduli space.
In an M-theory compactification on a $G_2$ manifold, $X$, the chiral 
parameters are of the form:
\begin{equation}
 u = \exp\left(\int_Q (k \Phi_3 + i C)\right) \, ,
\end{equation} 
where the integral is taken over a non-trivial three-cycle $Q \in
H_3(X,\mathbb{Z})$. Here, $\Phi_3$ is the $G_2$ three-form and $C$ is
the M-theory three-form potential. $k$ is a constant related to
the membrane tension and $u$ agrees with the exponential of the action 
of a membrane
instanton wrapping an associative three-cycle homologous to $Q$.

Following \cite{AW} we introduce a set of chiral parameters corresponding to
a basis of integral three-cycles in the boundary of 
the manifold $X$. The boundary of $X$ is
$Y_{\Gamma} \equiv (S^3 \times S^3)/\Gamma$ 
where $\Gamma$ is the relevant orbifold group which is
$D_N$ (when $N>2$) or $D'_{N'}$ (with $N':=4-N$ when $N<2$).
The space of integral 
three-cycles $H_3(Y,\zet)$ is thus
two-dimensional. On the various semi-classical branches, 
one three-cycle is `filled in'
so that $H_3(X,\zet)$ is one-dimensional. 
Classically, the chiral parameter 
corresponding to the filled in cycle takes 
the value $\exp(0)=1$. As explained 
in \cite{AW}, this is subject to quantum 
corrections, but the classical analysis 
(supplemented with knowledge of the
gauge theory) is reliable near limits of moduli 
space where the M-theory geometry 
is everywhere weakly curved (except for possible orbifold singularities). 
This gives information about the poles and zeros
of the holomorphic parameters which 
is sufficient to reconstruct the exact moduli space.

Our first task, then, is to understand the third homology group of the
boundary $Y_{\Gamma}$ for $\Gamma=D_N$ or $D'_{N'}$ ($N'=4-N$). 
To describe $H_3(Y_{\Gamma},\mathbb{Z})$, we follow \cite{AW}.
Before modding out by $\Gamma$, the boundary is $ Y=S^3\times S^3$.
This space can usefully be described in terms of
the three $SU(2)$ elements $g_1,g_2,g_3$ subject to the equivalence
relation: 
\begin{equation} 
\label{equiv}(g_1,g_2,g_3)= (g_1h,g_2h,g_3h).
\end{equation}
Let $\hat{E}_i \subset SU(2)^3$ be the $i^{th}$ copy of $SU(2)$ ---
so $\hat{E}_1$ is the set $(g,1,1), \, \, g\in SU(2)$. In $Y$, the
$\hat{E}_i$ project to cycles $E_i$ obeying 
\begin{equation} 
E_1 + E_2 + E_3 = 0 .
\label{relcover}
\end{equation} 

Under the orbifold projection $Y\to Y_{\Gamma}$ for $\Gamma=D_N$,
the $E_i$ are mapped to cycles in $Y_{\Gamma}$ which we label $E'_i$.
The map of $E_1$ to $E_1'$ is the $(4N-8)$-fold cover, 
$S^3\to S^3/D_N$, whilst $E_2$ and $E_3$
are mapped diffeomorphically to $E'_2$ and $E'_3$.
Thus we have
\begin{equation}
 (4N-8) E'_1 + E'_2 + E'_3 = 0. \eqlabel{homology}
\end{equation} 
$Y_{\Gamma}$ is simply the product $E_1'\times E_2'$ and
these cycles generate $H_3(Y_{\Gamma}, \zet)$. 

For $\Gamma = D'_{N'}$,
$E_2$ and $E_3$ are again mapped diffeomorphically to
$E_2'$ and $E_3'$ in $Y_{\Gamma}$.
To find another cycle, instead of $E_1$
we consider a little different three sphere in $Y$;
\begin{equation}
(g,g^{-1}\left(\begin{smallmatrix}i & 0 \\ 0 & -i 
\end{smallmatrix}\right) g,1),\,\,g\in SU(2).
\label{defE}
\end{equation}
Left multiplication by $D_{N'}$ elements on $g$ corresponds to
the $D'_{N'}$ action on $Y$. Thus, this $S^3$ defines a $(4N'-8)$-fold
cover over a cycle $E\cong S^3/D_{N'}$ in $Y_{\Gamma}$.
Note that the cover $S^3$ has the same homology class as
$E_1$.
 This is because the inverse $g\mapsto g^{-1}$
reverses the orientation of $SU(2)$ and therefore
the $g_2$ part in (\ref{defE}) does not contribute in homology. 
Using (\ref{relcover}),
we find the homology relation in $Y_{\Gamma}$,
$$
(4N'-8)E+E_2'+E_3'=0.
$$
Denoting the orientation reversal of $E$ by $E_1'$,
we find the same relation as \eqref{homology} with $N=4-N'$.
$Y_{\Gamma}$ is just the product $E_1'\times E_2'$
and these cycles generate $H_3(Y_{\Gamma},\zet)$.

Next we would like to describe the behavior of the holomorphic parameters 
\begin{equation}
 \eta_i = \exp\left(\int_{E'_i}(k\Phi_3 + i C)\right)\,
\end{equation} 
at the various semi-classical limits of moduli space.
The homology relation \eqref{homology} implies the following relation
of $\eta_i$'s
\begin{equation}
\eta_1^{4N-8} \eta_2 \eta_3 = 1.
\label{releta}
\end{equation}
As remarked in \cite{AW}, one must be careful about the definition
of $\exp(i\int_{E_i'}C)$ --- this must include the sign factor
of the fermionic determinant, and there is a potential sign error
in the right hand side of (\ref{releta}).
It is shown in \cite{AW} that the error is absent 
for $\Gamma=D_N$, and the same proof applies also
to $\Gamma=D'_{N'}$.
The reason is that, just as in $D_N$ case,
$Y_{\Gamma}$ is the product of spin manifolds
$Y_1=S^3/D_{N'}$ and $Y_2=S^3$,
and that $E_i'$ are all transverse to
a $Spin(3)$ sub-bundle of $TY_{\Gamma}\cong TY_1\oplus TY_2$
--- a slight rotation of $TY_1$
in the direction of $TY_2$.

Thus, the relation (\ref{releta}) exactly holds
at each of the semi-classical points, and therefore by holomorphy,
everywhere on the moduli space.
We shall determine the orders of the poles and
zeroes of the $\eta_i$ as functions on moduli space,
following the argument in \cite{AW}.
Part of this information is obtained classically by counting
the number of times which a boundary cycle $E'_i$ wraps
the minimal three-cycle of the geometry.
However, if the geometry has an orbifold singularity, the classical
analysis is modified by strong coupling effects at low energies.

Near the deformed classical points corresponding to
orbifolds of $\MD$ with large $S^3$, $E'_1$ is filled in
whilst $E'_2$ and $E'_3$ grow large with opposite orientation.
Thus 
\begin{equation} 
\begin{array}{lllll}
(N>2) &
\eta_1 \rightarrow 1  ~,&
\eta_2 \rightarrow 0^h      ~,&
\eta_3 \rightarrow \infty^h ~,&
h\equiv \check h_{SO(2N)}=2N-2, \\
&
\eta_1 \rightarrow -1  ~,&
\eta_2 \rightarrow 0^{2h'}      ~,&
\eta_3 \rightarrow \infty^{2h'} ~,&
h\equiv \check h_{Sp(N-4)}=N-3, \\
(N<2) &
\eta_1 \rightarrow 1  ~,&
\eta_2 \rightarrow \infty^{\tilde h} ~,&
\eta_3 \rightarrow 0^{\tilde h}      ~,&
\tilde h\equiv \check h_{SO(4-2N)}=2-2N, \\
&
\eta_1 \rightarrow -1  ~,&
\eta_2 \rightarrow \infty^{2\tilde h'} ~,&
\eta_3 \rightarrow 0^{2\tilde h'}      ~,&
\tilde h'\equiv \check h_{Sp(2-N)}=3-N. \\
\end{array}
\end{equation} 
Note that, as explained in \cite{AW}, the $\pm$ sign for $\eta_1$
corresponds to a choice of discrete flux which leads to $SO$
or $Sp$ gauge theory, and the order of zeroes or poles is related
to the degeneracy of vacua of the corresponding super Yang-Mills
theory\footnote{
The extra factor of 2 for the symplectic group was explained in \cite{AW}
for the $N>2$ cases with O6-plane at $S^3$, using an argument in which
D6-branes are deformed away from the orientifold plane.
In fact, we need this extra factor also for the $N<2$ cases
with free orientifold
with minimal $\RP^3$.
Derivation that applies to both cases is given in
Section~\ref{superpotential}.}.
Also, we have chosen the orientations of cycles $E'_i$ so that
the family of vacua preserves the same supersymmetry for all $N$,
in accordance with Table \ref{summary} in the introduction.
One can see that the orders of zeroes or poles depend linearly on
$N$ under this choice.

On the two resolved vacua described by orbifolds of $\MF$
we have either $E'_2$ or $E'_3$ filled in.
In the former case the minimal cycle is homologous to
$E'_1 = - \frac{1}{4N-8} E'_3$ and
\begin{equation} 
\eta_1 \rightarrow \infty \,\, , \,\,
\eta_2 \rightarrow 1 \,\, , \,\,
\eta_3 \rightarrow 0^{4N-8} \,\,,
\end{equation}  
whilst in the latter case one has
\begin{equation} 
\eta_1 \rightarrow 0              \,\, , \,\,
\eta_2 \rightarrow \infty^{4N-8}  \,\, , \,\,
\eta_3 \rightarrow 1              \,\, .
\end{equation}
The pole turn into zero and vice versa as $N-2$ flips sign, due to
our choice of the orientations of $E'_i$.

On the classical vacua corresponding to $\MO$ and its double
cover $\MOD$, we expect a similar behavior to the deformed classical
vacua except that the large $S^3$ should have opposite orientation.
In both cases, $E'_2$ and $E'_3$ wrap the minimal cycle
once and so the relevant poles and zeroes are simple.
\begin{equation} 
\eta_1 \rightarrow 1 \,\, , \,\,
\eta_2 \rightarrow 0 \,\, , \,\,
\eta_3 \rightarrow \infty \,\, .
\end{equation} 

Now, let us glue together the local behaviors of $\eta_i$ to get
the curve describing the moduli space.

\subsubsection{$N>3$}

\begin{figure}
\begin{center}
\scriptsize
 \psfrag{r+}{$P_2$: resolved}
 \psfrag{r-}{$P_3$: resolved}
 \psfrag{N>=4}{}
 \psfrag{d+}{$P_1':Sp(N-4)$}
 \psfrag{d-}{$P_1 :SO(2N)$}
 \epsfig{width=11cm,file=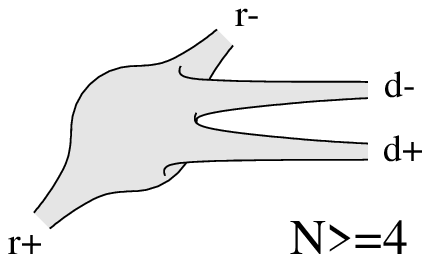}
\\ \vskip-12mm
 \psfrag{m+}{$\mu>0$}
 \psfrag{m-}{$\mu<0$}
 \psfrag{t+}{$t>0$}
 \psfrag{t-}{$t<0$}
 \epsfig{width=11cm,file=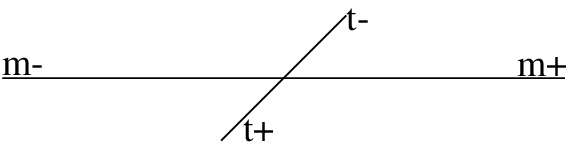}
\end{center}
\caption{\small The quantum moduli space for the cases $N>3$. Here
         $\mu$ and $t$ are the sizes of deformation and resolution.}
\label{fig:N>3} 
\end{figure}

The quantum moduli space of the orientifold plane system has been analyzed 
in \cite{AW} for the cases with $N > 3$ D6-branes. It consists of four 
classical geometries connected in the quantum moduli space (see Figure 
\ref{fig:N>3}) at the points $\eta_1= \pm 1, 0, \infty$.  The point $\eta_1 = 
1$ corresponds to an O6$^-$-plane and $N$ D6-branes wrapping 
the $S^3_>$ of the deformed conifold.
The gauge group at this point is $SO(2N)$. 
The point $\eta_1 = - 1$ corresponds to an O6$^+$-plane and $N-4$ 
D6-branes wrapping the $S^3_>$ of the deformed conifold that has a gauge group 
$Sp(N-4)$. The points $\eta_1= 0, \infty$ correspond to the two 
resolved conifolds with $N-2$ units of RR flux.

The quantum curve can be obtained by looking at the poles and zeros of the $\eta_i$ 
functions, as explained above:
\vspace{0.1cm}
\begin{center}
\begin{tabular}{|c||c|c|c|c|} \hline
 & $P_1$ & $P_1'$ & $P_2$ & $P_3$ \\ \hline \hline
$\eta_1$ & $1$ & $-1$ & 0 & $\infty$ \\ \hline
$\eta_2$ & $0^{2N-2}$ & $0^{2(N-3)}$ & $\infty^{4N-8}$ & $1$  \\ \hline 
$\eta_3$ & ${\infty}^{2N-2}$ & ${\infty}^{2(N-3)}$ & $1$ & $0^{4N-8}$ \\ \hline
\end{tabular}
\end{center}

From this table it is easy to deduce the quantum curve:
\begin{equation}
\begin{split}
\eta_2 & =  \eta_1^{-(4N-8)} (\eta_1 -1)^{(2N-2)} (\eta_1 + 1)^{(2N-6)},\\
\eta_3 & =   (\eta_1 -1)^{-(2N-2)} (\eta_1 + 1)^{-(2N-6)}.
\end{split}
\end{equation}

Notice that for this charge there is no classical configuration 
involving anti-D6-branes 
wrapped on $S^3_<$. They would preserve the same 
supersymmetry but the charge would be negative.

\subsubsection{$N=3$}

\begin{figure}
\begin{center}
\scriptsize
 \psfrag{r+}{$P_2$: resolved}
 \psfrag{r-}{$P_3$: resolved}
 \psfrag{N=3}{}
 \psfrag{d-}{$P_1 :SO(6)$}
 \epsfig{width=11cm,file=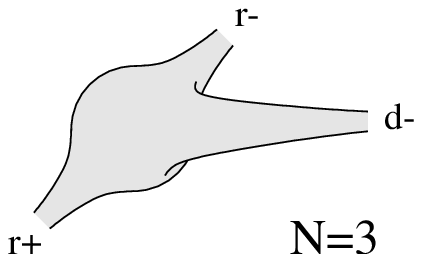}
\\ \vskip-12mm
 \psfrag{m+}{$\mu>0$}
 \psfrag{m-}{$\mu<0$}
 \psfrag{t+}{$t>0$}
 \psfrag{t-}{$t<0$}
 \epsfig{width=11cm,file=axes.eps}
\end{center}
\caption{\small The quantum moduli space for the case $N=3$.}
\label{fig:N=3} 
\end{figure}

The case $N=3$ with orientifold planes coincides with the $A_3$ case
(see Figure \ref{fig:N=3}).
The quantum moduli space has three points with classical descriptions
at $\eta_1=1, 0, \infty$.
The point $\eta_1 = 1$ corresponds to an O6$^-$-plane
and three D6-branes wrapping the $S^3_>$ of the deformed conifold.
The gauge group at this point is $SO(6) \sim SU(4)$, that has $h= 4$ vacua.
The points $\eta_1= 0, \infty$ correspond to the two
resolved conifolds  with $1$ unit of RR flux.
As expected the quantum curve is the same as in the $A_3$ case:
\begin{equation}
\begin{split}
\eta_2  = & \eta_1^{-4} (\eta_1 -1)^4, \\
\eta_3 = & (1 - \eta_1)^{-4}.
\end{split}
\end{equation}

The poles and the zeros for the $N=3$ case can be summarized
in the following table:
\vspace{0.1cm}
\begin{center}
\begin{tabular}{|c||c|c|c|} \hline
 & $P_1$ & $P_2$ & $P_3$ \\ \hline \hline
$\eta_1$ & 1 & 0 & $\infty$ \\ \hline
$\eta_2$ & $0^4$ & $\infty^4$ & 1  \\ \hline 
$\eta_3$ & $\infty^4$  & 1 & $0^4$  \\ \hline
\end{tabular}
\end{center}

As in the previous case there are no classical configurations involving
anti-D6-branes and preserving the same supersymmetries.

\subsubsection{$N=2$}

This case is truly exceptional, because the boundary of the relevant
M-theory geometry is not an orbifold of $S^3\times S^3$, but rather
\begin{equation}
 \left(\frac{S^2\times S^1}{\mathZ_2}\right)\times S^3.
\label{bXN=2}
\end{equation}
We denote the first factor by $E'_1$ and the second by $E'_2=-E'_3$,
and define the associated membrane instanton factors $\eta_i$ as
explained before.

The classical point $P_1$ corresponding to an O6$^-$-plane and two
D6-branes wrapped on $S^3_>$ supports an $SO(4)$ gauge theory.
Although $E'_2=-E'_3$ wraps the minimal cycle once, the corresponding
instanton factors $\eta_2, \eta_3$ develop double zero and pole
there due to the $SO(4)$ gauge dynamics.
Here we determined the degeneracy of vacua from
$\check h_{SO(N)}=2N-2=2$ for $N=2$,
though $SO(4)$ super Yang-Mills theory actually has four degenerate vacua.
At the two resolved classical points $P_{2,3}$ the $\eta_1$ has a simple
zero or pole, whereas $\eta_{2,3}$ remain finite.

In addition, we have geometries with large
$\mathR\mathP^3 \equiv S^3_</\mathZ_2$ corresponding to
free orientifold of deformed conifold.
We claim that there are two distinct classical points with
large $\mathR\mathP^3$.
The two will differ, from Type IIA viewpoint, in the action
on the Chan-Paton indices when one wraps some D6-branes on $\mathR\mathP^3$.
They should also be distinguished by the discrete torsion
of NSNS B-field or M-theory three-form potential.
Indeed, the third homology group $H_3(X,\mathZ)$ of the relevant
M-theory geometry
\[
 X~:~~\mathR^3\times S^1~\rightarrow~ \mathR\mathP^3
\]
is $\mathZ\oplus\mathZ_2$ and therefore has a torsion part,
since the third homology group of the boundary (\ref{bXN=2}) is
$\mathZ\oplus\mathZ$ and only {\it twice} the first generator
is trivial in $X$.
Therefore, one has classically the choice\footnote{
Alternatively, we may look at the twisted second homology of
the IIA reduction of $\tilde X$.
We again find the torsion $H_2(\tilde X,\tilde\mathZ)=\mathZ_2$
as $\tilde X$ contains an $\mathR\mathP^3$.
}
\begin{equation}
 \int_{E'_1}C ~=~ 0 ~~\mbox{or}~~\pi ~~~~(\mbox{mod}~~2\pi).
\end{equation}
We denote the corresponding two classical points by $P'_1$ and $P''_1$.

The table of singularities reads:
\vspace{0.1cm}
\begin{center}
\begin{tabular}{|c||c|c|c|c|c|c|} \hline
 & $P_1$ & $(P_1)$ & $P_1'$ & $P_1''$ & $P_2$ & $P_3$ \\ \hline \hline
$\eta_1$ & 1 & 1 & 1 & $-1$ & 0 & $\infty$ \\ \hline
$\eta_2$ & $0^2$ & $0^2$ & $\infty^2$ & $\infty^2$ & $1$ & 1  \\ \hline 
$\eta_3$ & $\infty^2$ & $\infty^2$ & $0^2$ & $0^2$ & 1 & $1$   \\ \hline
\end{tabular}
\end{center}
Here we included in the second column the contribution of
the two $SO(4)$ vacua that we missed by the counting based on
$\check h_{SO(4)}=2$.

\begin{figure}[htb]
\begin{center}
\scriptsize
 \psfrag{r+}{$P_2$: resolved}
 \psfrag{r-}{$P_3$: resolved}
 \psfrag{N=2}{}
 \psfrag{d-} {$P_1 :SO(4)$}
 \psfrag{d'-}{\hskip-3mm $P'_1 :SO(0)$}
 \psfrag{d'+}{\hskip-3mm $P''_1 :Sp(0)$}
 \epsfig{width=11cm,file=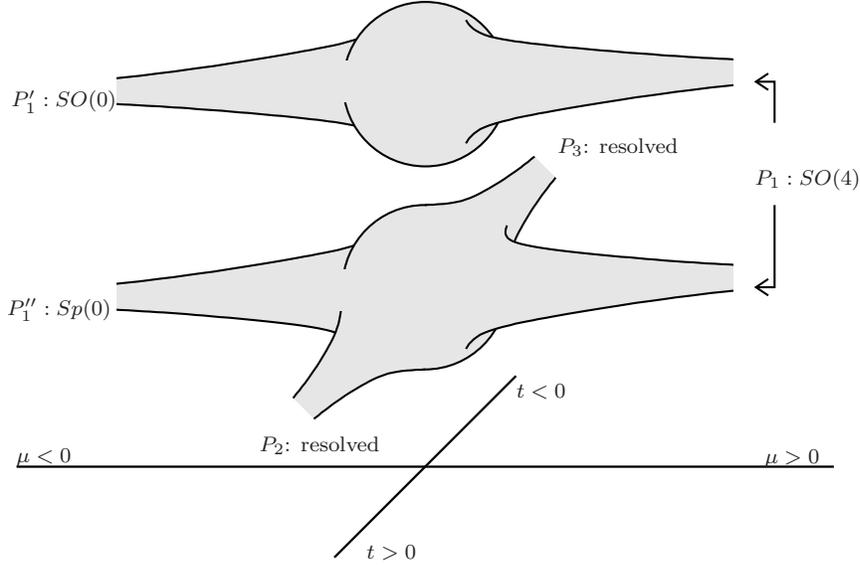}
\\ \vskip-12mm
 \psfrag{m+}{$\mu>0$}
 \psfrag{m-}{$\mu<0$}
 \psfrag{t+}{$t>0$}
 \psfrag{t-}{$t<0$}
 \epsfig{width=11cm,file=axes.eps}
\end{center}
\caption{\small The quantum moduli space for the case $N=2$.}
\label{fig:N=2} 
\end{figure}

From exact superpotential explained later, one can derive
the quantum curve:
\begin{equation}
\begin{split}
\eta_2  = & (\eta_1 -1)^{2} (\eta_1 + 1)^{-2},  \\
\eta_3  = &  (\eta_1 -1)^{-2} (\eta_1 + 1)^{2}. 
\end{split}
\end{equation}
This accounts for only four classical points $P_1,P''_1,P_2,P_3$.
It is thus expected that the moduli space consists of two branches,
one of which is the curve given above whereas the other contains
the two missing $SO(4)$ vacua as well as the classical point $P'_1$.
The latter branch is most likely a cylinder $\mathC^\times$.
If we parameterize it by $z$ the $\eta_i$'s on this branch will be given by
\begin{equation}
 \eta_1=1,~~~~
 \eta_2=z^2,~~~
 \eta_3=z^{-2}.
\end{equation}
The two branches meet at the classical point with an O6$^-$-plane
and two D6-branes on $S^3$.
The structure of quantum moduli space is summarized in Figure
\ref{fig:N=2}.

\subsubsection{$N=1$}
\label{N=1}

There are five classical points: the one corresponding to an
O6$^-$-plane and a D6-brane wrapping the $S^3_>$ which we denote
by $P_1$, the free orientifold plus two D6-branes wrapping the $S^3_<$
which are denoted by $P'_1$ or $P''_1$ according to the gauge group
being $SO(2)$ or $Sp(1)$, and the two resolved vacua
($P_2,\,P_3$).
The orders of zeroes or poles of $\eta_i$ is summarized as follows:
\vspace{0.1cm}
\begin{center}
\begin{tabular}{|c||c|c||c|c|c|} \hline
 & $P_1$ &  $P_1'$ & $P_1''$ & $P_2$ & $P_3$ \\ \hline \hline
$\eta_1$ & 1 & 1 & $-1$ & 0 & $\infty$ \\ \hline
$\eta_2$ & $0$ & $\infty$ & $\infty^4$ & $0^4$ & 1  \\ \hline 
$\eta_3$ & $\infty$ & $0$ & $0^4$ & 1 & $\infty^4$   \\ \hline
\end{tabular}
\end{center}
The first two have massless $U(1)$ gauge bosons at low energies
whereas the other three have a mass gap.
The moduli space should therefore consist of two smooth
components, one for vacua with $U(1)$ and the other for mass-gapped
vacua. From the analysis of zeroes and poles we expect that
the branch of vacua with $U(1)$ is a cylinder $\eta_2\eta_3=1$, and
the mass-gapped branch is given by the curve
\begin{equation}
\begin{array}{rcl}
 \eta_2 &=& \eta_1^4(\eta_1+1)^{-4},\\
 \eta_3 &=& (\eta_1+1)^4.
\end{array}
\label{curveN=1}
\end{equation}
\begin{figure}
\begin{center}
\scriptsize
 \psfrag{r+}{$P_2$: resolved}
 \psfrag{r-}{$P_3$: resolved}
 \psfrag{N=1}{}
 \psfrag{d-} {$P_1 :SO(2)$}
 \psfrag{d'-}{\hskip-3mm $P'_1 :SO(2)$}
 \psfrag{d'+}{\hskip-3mm $P''_1 :Sp(1)$}
 \epsfig{width=11cm,file=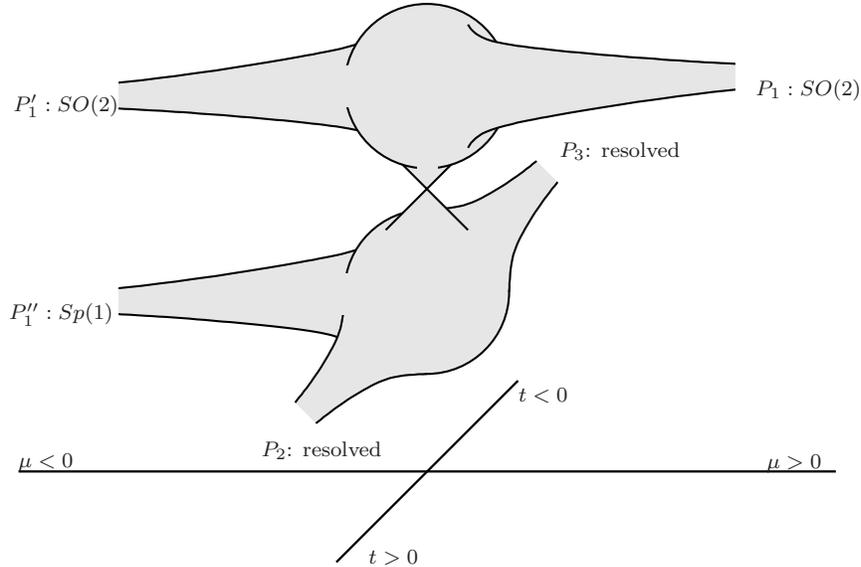}
\\ \vskip-12mm
 \psfrag{m+}{$\mu>0$}
 \psfrag{m-}{$\mu<0$}
 \psfrag{t+}{$t>0$}
 \psfrag{t-}{$t<0$}
 \epsfig{width=11cm,file=axes.eps}
\end{center}
\caption{\small The quantum moduli space for the case $N=1$.
         The two branches are connected at a phase transition point,
         as we will see in the next section.}
\label{fig:N=1} 
\end{figure}

We propose that the two branches meet at a phase transition point,
as shown in Figure \ref{fig:N=1}.
As will be discussed in detail in Section~\ref{branch}, one can get
the exact branch structure by following a long chain of dualities
to go to a ``mirror'' IIB theory.
Let us summarize here the main points.

The idea is to consider another Type IIA limit by reduction
on a different circle and then take its Type IIB mirror. This is
the route found in \cite{av} and the technique
is further developed in \cite{hp2005}
on which the present computation is based.
The Type IIB dual is the non-compact Calabi-Yau
$(\xi,\eta\in\mathC;~x,y\in\mathC^\times) $
$$
\xi\eta=F(x,y):=y^2-2s xy +x^3-2x^2+x,
$$
parameterized by $s^2$, together with a D5-brane located at
a line $\eta=0$, $\xi$ free.
The D5-brane position is parameterized by a point
$(x,y)=(x_{\bf D}, y_{\bf D})$ of the Riemann surface $F(x,y)=0$,
which is generically genus one
and has three punctures ${\bf A}$, ${\bf B}$, ${\bf C}$ 
at $(x,y)=(1,0), (0,0),(\infty,\infty)$.
The modulus $s$ is a normalizable dynamical variable
but ${\bf D}=(x_{\bf D}, y_{\bf D})$ is a coupling constant.
The presence of the D5-brane generates
a superpotential $W(x_{\bf D},s)$, and the extremization
$\partial_s W=0$ relates ${\bf D}$ and $s$ as follows:
When the curve $F(x,y)=0$ has genus one,
the D5-position ${\bf D}$ is determined by $s^2$. This is what we call
the $g=1$ branch. At $s^2=-4$, the curve degenerates to
genus zero and then ${\bf D}$ is free to move on this curve.
We call this the $g=0$ branch.

On the $g=1$ branch there is a $U(1)$ vector multiplet which is
an $\caln=2$ superpartner of the complex structure modulus $s$.
This corresponds to the upper branch in Figure~\ref{fig:N=1}.
The points $P_1$ (O6$^-$ with $D6$) and $P_1'$ (free orientifold
with $SO(2)$ D6's on $\RP^3$) correspond 
respectively to the large complex
structure limit $s^2=\infty$ and the ``orbifold limit'' $s^2=0$.
The $g=0$ branch has no massless vector and corresponds to
the lower branch in Figure~\ref{fig:N=1}. The point
$P_1''$ (free orientifold
with $Sp(1)$ D6's on $\RP^3$)
corresponds to the limit where ${\bf D}$ approaches the marked point
${\bf A}$ while the points $P_2$ and $P_3$ (two resolved conifolds)
correspond to ${\bf D}\longrightarrow {\bf B},{\bf C}$.
This branch structure is reminiscent of the result of
\cite{csw}, where the behavior of vacua of ${\cal N}=1$ gauge theory
with an adjoint matter was studied on the space of superpotential couplings.

Let us now focus on the transition point where the two branches meet.
From the $g=1$ side, this is the point $s^2=-4$ where a linear combination
of the $A$ cycle and $B$ cycle of the torus degenerates,
and Type IIB D3-brane wrapped on this vanishing cycle
becomes massless.
Such D3-brane states, which constitute a charged
hypermultiplet $(M,\tilde{M})$, must be included in the low energy
effective theory near the point $s^2=-4$.
The effective superpotential is then given by
\begin{equation}
W_{\it eff}=W(x_{\bf D},s)+(s^2+4)\tilde{M}M.
\label{Weff}
\end{equation}
Variation with respect to the normalizable variables
$s$, $M$ and $\tilde{M}$ yields
$$
\partial_s W(x_{\bf D},s)+2s \tilde{M}M=0,\quad
(s^2+4)\tilde{M}=0,\quad (s^2+4)M=0.
$$
The solutions with $M=\tilde{M}=0$ leave the $U(1)$ gauge symmetry
unbroken and lie in the $g=1$ branch.
There are other solutions with $s^2=-4$
in which $\tilde{M}M$ is determined by $x_{\bf D}$
through the first equation and its non-zero value higgses the
$U(1)$.
They constitute the $g=0$ branch.

In the original Type IIA or M-theory description,
what are the particles that become massless at the transition point?
Type IIB D3-branes wrapped on $A$ and $B$ cycles of the
curve (plus two other directions) near the large complex structure 
limit $s^2=\infty$ correspond
in M-theory on $\MOD$ to a membrane wrapped on the $S^2$ bolt of a
Dancer's fiber and a fivebrane wrapped on the $T^{1,1}$ bolt of
$\MOD$. 
The corresponding objects in
Type IIA orientifold with O6$^-$+D6 wrapped on $S^3$ (point $P_1$)
are essentially the non-BPS states discussed in \cite{sen}:
a membrane wrapped on the non-holomorphic $S^2$ bolt of Dancer's manifold
is a massive oscillation mode of the open string stretched
between a D6 and its orientifold image, while a
fivebrane wrapped on the bolt is a non-BPS threebrane
whose tension is $1/\ell_s^4$ in the strong coupling limit.
Thus, the charged particle responsible for the transition
to the confining branch is
the electrically charged massive open string mode on D6 or
the magnetic non-BPS threebrane wrapped on $S^3$, 
or some dyonic bound state.
Which one becomes massless at $s^2=-4$,
is an interesting question although somewhat ambiguous because of monodromies
around the other special points in moduli space.

\subsubsection{$N=0$}

There are at least five classical points: an O6$^-$-plane on $S^3_>$ ($P_1$),
the free orientifold plus two D6-branes on $S^3_<$ ($P'_1$ or $P''_1$
depending on the gauge groups $SO(4)$ or $Sp(2)$), and the two
of resolved vacua $P_{2,3}$.
The orders of zeroes and poles read
\vspace{0.1cm}
\begin{center}
\begin{tabular}{|c||c|c|c|c|c|c|} \hline
 & $P_1$ & $P_1'$ & $(P_1')$ & $P_1''$ & $P_2$ & $P_3$ \\ \hline \hline
$\eta_1$ & 1 & 1 & $1$ & $-1$ & $0$ & $\infty$ \\ \hline
$\eta_2$ & $0$ & $\infty^2$ & $\infty^2$ & $\infty^6$ & $0^8$ & 1  \\ \hline 
$\eta_3$ & $\infty$ & $0^2$ & $0^2$ & $0^6$ & 1 & $\infty^8$   \\ \hline
\end{tabular}
\end{center}
Here we included the contribution of two $SO(4)$ vacua at $P'_1$
that are missed by the counting based on $\check h_{SO(4)}=2$.
It follows that the four points $P'_1,P''_1,P_2,P_3$ can live on
a single Riemann sphere as described by the equations
\begin{equation}
\begin{array}{rcl}
 \eta_2&=&\eta_1^8(\eta_1-1)^{-2}(\eta_1+1)^{-6},\\
 \eta_3&=&(\eta_1-1)^{ 2}(\eta_1+1)^{ 6}.
\end{array}
\end{equation}
This curve actually follows also from Vafa's exact superpotential.
A natural guess then is that there is another branch of moduli
space containing all these missing vacua.

As a non-trivial check for this guess, let us consider the orders
of zeroes or poles of the functions $\eta_i$ on the new branch.
At the $SO(4)$ classical point on the new branch one should have
$\eta_2\sim 0^2,~\eta_3\sim \infty^2$ to account for the missing
vacua (third column of the table above).
On the other hand, at the vacuum $P_1$ corresponds to an M-theory
geometry $\MO$ which has a finite $S^3\simeq E'_2\simeq -E'_3$,
so that $\eta_2\sim\infty,~\eta_3\sim0$ at the corresponding classical
point.
Therefore, the numbers of poles and zeroes agree on the ``new'' branch
if there are two distinct classical points corresponding to an
O6$^-$-plane on $S^3_>$.
Remarkably, this is in agreement with the fact that the corresponding
M-theory geometry $\MO$ spontaneously breaks the $\mathZ_2'$ symmetry
of Section~\ref{symori}.

As we have seen in Section~\ref{symori}, the asymptotic symmetry of
M-theory geometry is $SO(4)\ltimes\widetilde\mathZ_2\times\mathZ'_2$.
In the interior of some solutions, a part of
$\tilde\mathZ_2\times Z'_2$ is broken.
The two resolved vacua are permuted under $\tilde\mathZ_2$ defined
in (\ref{flop}), while all the deformed vacua are invariant.
On the other hand, $\mathZ'_2$ is the centralizer of the $D'_4$ orbifold
group and is identified with a half-period shift along the M-theory
circle.
As such, it is broken in the solution $\MO$ corresponding to
an O6$^-$-plane on $S^3_>$ while all other classical solutions
are invariant.
Thus,
there should be two classical points corresponding to an O6$^-$-plane
on $S^3_>$, as claimed.

\begin{figure}[htb]
\begin{center}
\scriptsize
 \psfrag{r+}{$P_2$: resolved}
 \psfrag{r-}{$P_3$: resolved}
 \psfrag{N=0}{}
 \psfrag{d-} {$P_1 :SO(0)$}
 \psfrag{d'-}{\hskip-3mm $P'_1 :SO(4)$}
 \psfrag{d'+}{\hskip-3mm $P''_1 :Sp(2)$}
 \epsfig{width=11cm,file=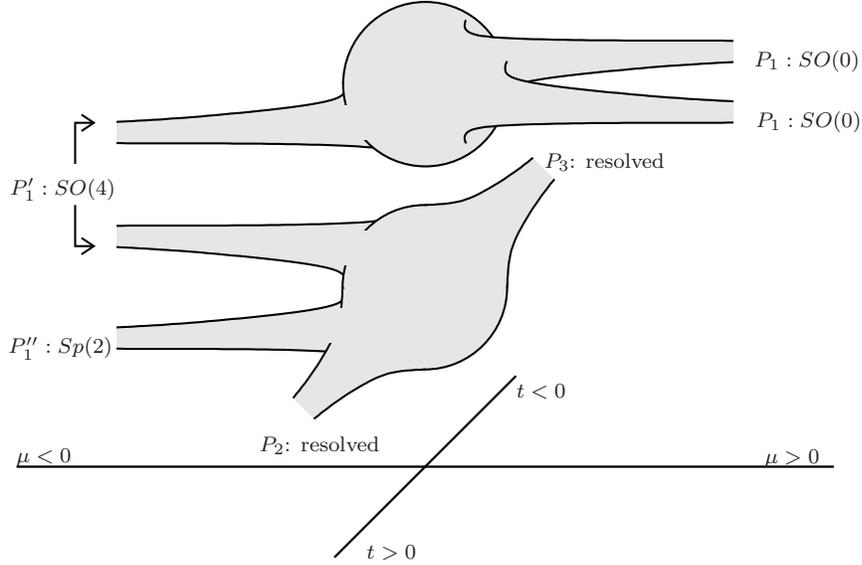}
\\ \vskip-12mm
 \psfrag{m+}{$\mu>0$}
 \psfrag{m-}{$\mu<0$}
 \psfrag{t+}{$t>0$}
 \psfrag{t-}{$t<0$}
 \epsfig{width=11cm,file=axes.eps}
\end{center}
\caption{\small The quantum moduli space for the case $N=0$.}
\label{fig:N=0} 
\end{figure}

Are the two branches connected? From what we discussed above it 
is clear that $\widetilde\mathZ_2$
acts non-trivially on the branch containing resolved vacua,
while $\mathZ'_2$ acts non-trivially on the other branch. Conversely, $\mathZ'_2$ should act trivially on the branch containing resolved vacua and $\widetilde\mathZ_2$ should act trivially on the other branch, under the assumption that both branches are $\projective^1$'s with various marked points, (since both $\mathZ_2$'s act holomorphically on moduli space and fix at least three points on the appropriate branches.)
The two branches can therefore be connected only through points
invariant under both $\mathZ_2$'s.
Since the classical points $\eta_1=\pm1$ are fixed by $\widetilde\mathZ_2$
and they are the only fixed points on the resolved branch, one can conclude immediately that the two branches are not connected
through interior points.
The moduli space is thus made of two disjoint branches as depicted
in Figure \ref{fig:N=0}.

One might have guessed that a phase transition as
in the $N=1$ case would connect the two branches
because the classical moduli space is connected.
Such a phase transition would be characterized by
the emergence of massless particles.
For the $N=1$ case, the relevant particle in the M-theory framework
is either a five-brane wrapped on the $T^{1,1}$ bolt of $\MOD$ or
a membrane wrapped on the $S^2$.
However, the corresponding cycles for the case $N=0$ are both
non-orientable, so there will be no massless particles when
they shrink to zero size.

\paragraph{$\zet_2'$ in Type IIA \protect\footnote{Discussions with S.\ 
Hellerman were instrumental in shaping the following arguments.}}

The presence of the discrete symmetry in $\zet_2'$ might at first sight 
appear puzzling from the Type IIA perspective: A half-period shift of
the M-theory circle descends in Type IIA to D0-brane charge modulo $2$,
which according to our discussion should be preserved in front of $N>0$ 
D6-branes wrapped on top of the O6$^-$, and broken precisely for $N=0$. 
Naively, this appears to be in conflict with the familiar classification 
of D-brane charge in string theory. A perturbative analysis in flat space
shows that the tachyonic ground state of the 0--0 strings is in the symmetric
representation. The tachyon is therefore not orientifolded out even for 
a single D0-brane, which should therefore not produce a conserved charge.
This situation is T-dual to the D3-brane in Type I, and in distinction
to the D($-1$)-brane there, for which the ground state is in the 
anti-symmetric representation. 

Wrapping on $S^3$ does not eliminate the tachyon, and one would therefore
not expect a stable D0-brane in our backgrounds, for any $N$. On the other 
hand, the M-theory analysis solidly establishes the existence of $\zet_2'$ 
for $N>0$, and its breaking for $N=0$.

To reconcile the two points of view, we note that the analog of $\zet_2'$
can already be seen in the context of O6$^-$/D6 systems in flat space
and their M-theory lifts, with the same breaking pattern. The discrete 
symmetry in this case can be nicely interpreted from the perspective of 
D2-brane probes \cite{seib,sewi}. (Its presence was also noticed, for example, 
in the D0-brane scattering analysis of \cite{imam}.)
The worldvolume theory of a D2-brane pair is a $3d$ $\caln=4$ supersymmetric
$Sp(1)$ gauge theory with $2N$ half-hypermultiplets in the
fundamental representation.
The hypers are from the $2$-$6$ strings and their masses
parameterize the position of the D6-branes.
Now, one may interpret the $\Z_2'$ as the global symmetry extending
$SO(2N)$ by $O(2N)$. It acts as the sign flip
of just one of the $2N$ half-hypermultiplets.
That this correspond to a half-period shift of the M-theory circle
follows from the fact that, in an odd monopole background,
the fermion zero mode measure is odd under this symmetry.
When some of the D6-brane pairs are on top of the O6-plane,
the sign flip of one of the corresponding massless half-hypermultiplets
is a symmetry of the system.
When all D6-branes are away from the O6,
all the masses are turned on, and the symmetry is broken.
Namely, $\zet_2'$ is a symmetry only when at least a single 
D6-brane is {\it exactly on top} of the O6-plane, and broken otherwise.
This is indeed the breaking pattern we have found in the curved
background.

Returning to the interpretation of $\zet_2'$ as ``D0-brane number
modulo 2'', we note that this is a good quantum number far away
from the O6-plane. Indeed, a D0-brane is then far away (in the
covering space) from its anti-D0-brane image, and the ground state 
of the open string between them is non-tachyonic.
(With more than one D0-$\overline{\rm D0}$ pair in the covering
space, a D0 and an anti-D0 from different pairs can approach each
other asymptotically, and annihilate.) In a process in which the
D0 approaches an O6$^-$-plane, the tachyon will develop and the
D0/$\overline{\rm D0}$ can annihilate. When $N=0$, this is fine
since $\zet_2'$ is broken. When $N>0$, in which case $\zet_2'$ 
is globally conserved, this charge is carried away by open strings 
on the D6-brane describing its separation from the O-plane. (These 
states are massive after wrapping on $S^3$, but the D0-brane is 
much heavier at weak string coupling.)

Let us also note that $\zet_2'$ is preserved in front of
an O6$^+$-plane (our case $N=4$). This is consistent with the
fact that the 0--0 tachyon lands in the anti-symmetric 
representation and is orientifolded out.

\subsubsection{$N<0$}

\begin{figure}
\begin{center}
\scriptsize
 \psfrag{r+}{$P_2$: resolved}
 \psfrag{r-}{$P_3$: resolved}
 \psfrag{N<0}{}
 \psfrag{d'-}{\hskip-12mm $P'_1 :SO(4-2N)$}
 \psfrag{d'+}{\hskip-12mm $P''_1 :Sp(2-N)$}
 \epsfig{width=11cm,file=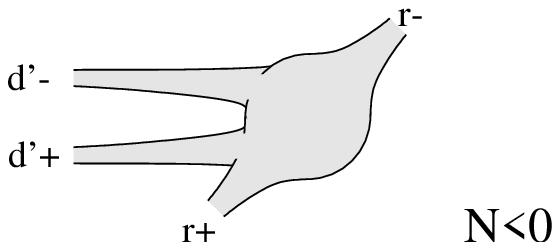}
\\ \vskip-12mm
 \psfrag{m+}{$\mu>0$}
 \psfrag{m-}{$\mu<0$}
 \psfrag{t+}{$t>0$}
 \psfrag{t-}{$t<0$}
 \epsfig{width=11cm,file=axes.eps}
\end{center}
\caption{\small The quantum moduli space for the case $N<0$.}
\label{fig:N<0} 
\end{figure}

Finally we come back to the regular case.
The singularities can be read from the following table:
\begin{center}
\begin{tabular}{|c||c|c|c|c|} \hline
 & $P_1'$ & $P_1''$ & $P_2$ & $P_3$ \\ \hline \hline
$\eta_1$ & 1 & $-1$ & 0 & $\infty$ \\ \hline
$\eta_2$ & ${\infty}^{2(\tilde{N} -1)}$ & ${\infty}^{2(\tilde{N} +1)}$ & $0^{4\tilde{N}}$ & 1  \\ \hline 
$\eta_3$ & $0^{2(\tilde{N} -1)}$ & ${0}^{2(\tilde{N} +1)}$ & 1 & $\infty^{4\tilde{N}}$   \\ \hline
\end{tabular}
\end{center}
where $\tilde N=2-N>2$.
And the quantum curve is:
\begin{equation}
\begin{split}
\eta_2  = & \eta_1^{4 \tilde{N}} 
(\eta_1 -1)^{-2 (\tilde{N}-1)} (\eta_1 + 1)^{-2 (\tilde{N}+1)},\\
\eta_3  = &(\eta_1 -1)^{2 (\tilde{N}-1)} (\eta_1 + 1)^{2 (\tilde{N}+1)}.
\end{split}
\end{equation}
The moduli space consists of a single smooth component
as in Figure \ref{fig:N<0}.

\subsection{Using superpotential}
\label{superpotential}

A part of the results of the previous subsection can be
obtained also by studying the 
superpotential proposed by Vafa \cite{VafaSup}
and the relevant computations in \cite{Wafa}.

\newcommand{\e}{e}

The superpotential is computed on the branch of
the resolved conifold with flux through $\reals\projective^2$.
It consists of three parts, coming from the four-form flux,
the two-form flux and worldsheet instantons:
$$
W=W_{4\,{\rm flux}}+W_{2\,\,{\rm flux}}+W_{\rm crosscap}.
$$
Let $t$ be the complexified K\"ahler class of the base $\projective^1$
of the resolved conifold.
In the present orientifold, the periodicity of the parameter is
doubled,
\begin{equation}
t\equiv t+4\pi i.
\label{doubled}
\end{equation}
 The reason is that
there exist crosscap diagrams
associated with the odd degree maps $S^2\to \projective^1$
which are equivariant
with respect to the involution $\Omega:w\to -1/\bar w$ on the domain
and the anti-holomorphic involution (\ref{inv1})
on the target. For example, the identity map is such a map and has degree 1.
The path-integral weight of
such a diagram include odd powers of
$e^{-i{\rm Im}(t)/2}$.
Thus, $z=e^{-t/2}$ is the single valued parameter
of the theory.
Now, let us describe each term of the superpotential.
 The contributions from the RR two-form flux through the $\RP^2$
and worldsheet instantons are \cite{VafaSup,Wafa}
$$
W_{2\,\,{\rm flux}}=(N-2){\partial F_0\over\partial t}
=-(N-2)Li_2(z^2),
$$
$$
W_{\rm crosscap}=-4\sum_{m:{\rm odd}\geq 1}{e^{-mt/2}\over m^2}
=-2(Li_2(z)-Li_2(-z)),
$$
where $Li_2$ is the dilogarithm function. 
For convenience, some of its properties are collected
in Appendix \ref{lithium}.
The contribution from the four-form flux
is given by
$$
W_{4\,{\rm flux}} = -\int F_4 \wedge \omega,
$$
where $\omega$ is the 
complexified K\"ahler form of the resolved conifold. 
According to 
\cite{VafaSup}, $F_4$ has an imaginary part corresponding to 
the RR four-form and a real NSNS part coming from the failure of 
the metric to be Calabi-Yau. On a non-compact space it is  natural to 
interpret this formula as follows:
$$
W_{4\,{\rm flux}} = -\int_M d(\Re(\hat{\Omega}) + i C_3) \wedge \omega = 
-\int_{\partial M} (\Re(\hat{\Omega}) + i C_3) \wedge \omega ,
$$
where $\hat{\Omega}$ is a suitably normalized form of the holomorphic 
three-form which is the superpartner of $C_3$.
Evaluating this for the $\Z_2$ quotient of
the conifold we obtain
$$
W_{4\,{\rm flux}}=-Yt/2,
$$
where $Y$ is the holomorphic volume of the boundary $S^3$:
\begin{equation}
Y:=\int_{S^3_{\infty}}({\rm Re}(\hat{\Omega})+iC_3).
\end{equation}
Summing up the three terms, we obtain the total superpotential
\begin{equation}
W= -Yt/2-(N-2) Li_2(z^2) -2 (Li_2(z) - Li_2(-z)).
\label{W(z)}
\end{equation}
Note that the parameters $t$ and $Y$ introduced here are related to
the coordinates that we used in the M-theory 
description of the moduli space as
\begin{equation}
z=\exp(-t/2) = \eta_1, \qquad \exp(Y) = \eta_3 \, .
\label{zY}
\end{equation}

Following \cite{VafaSup,av} we can use this superpotential 
to find the exact form of the moduli space.
In order to have a supersymmetric background we should vary the
superpotential with respect to $t$ and find a stationary point. This gives 
\begin{equation}
\partial_t W = 0  \Rightarrow
Y=\log{\left( (z -1)^{-(2N-2)} (z + 1)^{-(2N-6)}\right) }.
\label{Yeq} 
\end{equation}
Using the relation (\ref{zY}), we see that this is nothing
but the equation describing 
the component of the moduli space 
which is smoothly connected to the resolved geometries.

\subsection*{\it Comparison to 4d gauge theory}

\begin{figure}
\begin{center}
\input{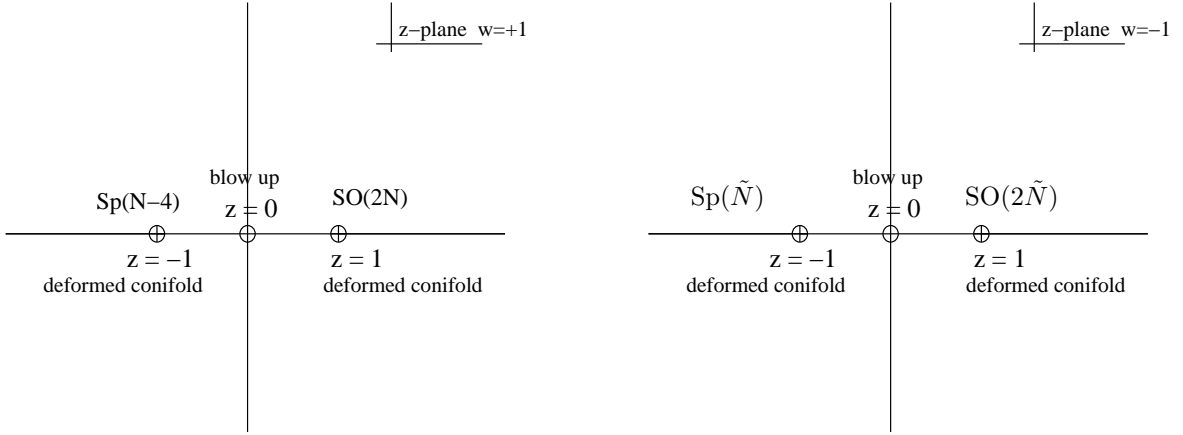}
\end{center}
\caption{\small The superpotential for the orientifold system has two branch cuts 
at $z=\pm 1$ labeling different gauge groups depending on the orientifold action. 
On the left hand side, the orientifold action with fixed points (orientifold planes).
On the right hand side the free orientifold.}
\label{Wafa} 
\end{figure}

The superpotential (\ref{W(z)})
has two branch cuts (see Figure \ref{Wafa}),
starting at $z = \pm 1$. For $z=\pm e^{-\varepsilon}$
with small $\varepsilon$, we have
\begin{equation}
W= -Y\varepsilon -b\varepsilon\log \varepsilon+\cdots,
\qquad
b=\left\{\begin{array}{ll}
2N-2&\mbox{at $z=1$},\\
2N-6&\mbox{at $z=-1$},
\end{array}\right.
\label{beha}
\end{equation}
where $+\cdots$ is a power series in $\varepsilon$.
When $b$ is positive ({\it resp.} negative), by the relation (\ref{Yeq}),
${\rm Re}(Y)$ diverges to $+\infty$ ({\it resp.} $-\infty$)
as $\varepsilon\to 0$.
This is the classical limit where 
we have a large minimal three sphere $S^3_{>}$ ({\it resp.} $S^3_{<}$)
on which a certain number of D6-branes ({\it resp.} anti-D6-branes)
are wrapped.
Below we compare this behavior of the superpotential with
what we expect from the gauge theory on the sixbranes.

For this purpose, one needs to understand the precise relation
of the parameter $Y$, which can be regarded as the membrane
instanton action on a cycle homologous to $S^3_{\infty}$,
 and the holomorphic gauge coupling constant
${8\pi^2\over g^2}-i\theta$ of the $4d$ gauge theory
on the (anti-)D6-branes wrapped on the minimal three sphere.
This was discussed in \cite{AW} for the case with O6-plane at $S^3$.
Here we present another argument, using the embedding of
$SO(2n)$ or $Sp(n)$ into $U(2n)$ defined by the orientifold projection,
which is applicable to the more general systems we
are studying. We first note that the instanton number 
of $4d$ Yang-Mills theory of a simple gauge group $G$ is defined
as 
$$
k=-{1\over 8\pi^2}\int_{\mathR^4}{\rm Tr}(F_A\wedge F_A),
$$
where ${\rm Tr}$ is such that the long root has length squared 2,
or the trace in the adjoint representation is given by
${\rm tr}_{\rm adjoint}(XY)=2\check{h}{\rm Tr}(XY)$, with $\check{h}$ being
the dual Coxeter number of $G$.
For the groups $SO(2n)$ and $Sp(n)$, it is related to the trace
of the fundamental representation of $U(2n)$ by
$$
{\rm Tr}(F_A\wedge F_A)
=\sigma\times {\rm tr}_{\rm fund}(F_{\tilde A}\wedge F_{\tilde A}),
\qquad
\sigma=\left\{\begin{array}{cl}
{1\over 2}&\mbox{for $SO(2n)$}\\
1&\mbox{for $Sp(n)$}
\end{array}\right.
$$
where $\tilde{A}$ is the $U(2n)$ gauge field which is obtained from 
$A$ by the embedding of $SO(2n)$ or $Sp(n)$ into $U(2n)$.
Thus, {\it \underline{one} 
$SO(2n)$ instanton corresponds to \underline{two} $U(2n)$ instantons,
while \underline{one} $Sp(n)$ instanton corresponds to \underline{one}
$U(2n)$ instanton.}

First, we consider the case ${\rm Re}(Y)\gg 0$ for which
the Type IIA geometry is the deformed conifold with an O6-plane
at the large minimal three sphere $S^3_>$.
Note that this $S^3_>$ is homologous to $S^3_{\infty}$, and hence
$Y$ is the action for one D2-brane wrapped on $S^3_{>}$.
Before orientifold, this D2-brane is wrapped twice on $S^3$ and
corresponds to {\it two} $U(2n)$ instantons on $2n$ D6-branes wrapped on
$S^3$. By the remark above,
after orientifold, it corresponds to {\it one} $SO(2n)$ instanton
or {\it two} $Sp(n)$ instantons.
This is indeed the claim in \cite{AW}.
Thus, the relation of $Y$ and the holomorphic gauge coupling is
\begin{equation}
Y=2\sigma \left({8\pi^2\over g^2}-i\theta\right).
\label{Yg}
\end{equation}
Note that the Chan-Paton factor of the D2-branes on $S^3_>$ is symplectic
for O6$^-$ ($SO(2n)$ on $n$ D6), while it is
orthogonal for O6$^+$ ($Sp(n)$ on $n$ D6).
Thus, the number of D2-branes for O6$^-$ must be even in the double
cover.
For O6$^+$, on the other hand, a ``half'' D2-brane (one in the cover)
is allowed and corresponds to {\it one}
$Sp(n)$ Yang-Mills instanton with instanton factor $\e^{-Y/2}$.
In terms of $S=2\sigma \varepsilon$, the superpotential
(\ref{beha}) can be written as
\begin{equation}
W= -\left({8\pi^2\over g^2}-i\theta\right)S
-{b\over 2\sigma}S\log S+\mbox{power series in $S$}.
\label{VY}
\end{equation}
If we identify $z=1$ as the large $S^3_>$ limit
with O6$^-$-plane and $z=-1$ as the large $S^3_>$ limit
with O6$^+$-plane, the coefficient of the $-S\log S$ term is
$$
{b\over 2\sigma}=\left\{\begin{array}{ll}
2N-2&\mbox{at $z=1$}\\
N-3&\mbox{at $z=-1$},
\end{array}\right.
$$
which are the dual Coxeter number of the groups $SO(2N)$ and
$Sp(N-4)$ respectively.
Then, (\ref{VY}) is exactly the Veneziano-Yankielowicz superpotential,
up to a power series in $S$, for the gauge group
$SO(2N)$ for $z=1$ and $Sp(N-4)$ for $z=-1$.

Let us next consider the case ${\rm Re}(Y)\ll 0$
which corresponds to the free orientifold of the deformed conifold
with large minimal three sphere $S^3_<$.
The cycle $S^3_{<}$ is also homologous to $S^3_{\infty}$, and
$Y$ corresponds to the action for D2-brane wrapped
twice on $\RP^3=S^3_</\zet_2$. This again corresponds
to {\it two} $U(2n)$ instantons on $2n$ D6-branes wrapped on $S^3_<$
before orientifold, and thus to
{\it one} $SO(2n)$ or {\it two} $Sp(n)$ instanton after orientifold.
On anti-D6-branes, instantons with positive instanton numbers
correspond to {\it anti}-D2-branes.
Thus the relation of $Y$ and the holomorphic gauge
coupling on the anti-D6-branes is
\begin{equation}
Y=-2\sigma \left({8\pi^2\over g^2}-i\theta\right).
\label{Ygneg}
\end{equation}
As in the ${\rm Re}(Y)\gg 0$ case, 
the minimal instanton factor is $e^{Y}$ for
$SO(2n)$ and $e^{Y/2}$ for $Sp(n)$.
In terms of $S=-2\sigma \varepsilon$, the superpotential is
$$
W= -\left({8\pi^2\over g^2}-i\theta\right)S
+{b\over 2\sigma}S\log S+\mbox{power series in $S$}.
$$
This is exactly the expected Veneziano-Yankielowicz superpotential
if we identify $z=1$ ({\it resp.} $z=-1$)
as the large $S^3_<$ limit
with $SO$ ({\it resp.} $Sp$) gauge group.
Indeed, under this identification,
the coefficient of the $-S\log S$ term is
$$
-{b\over 2\sigma}=\left\{\begin{array}{ll}
-2N+2=2\tilde{N}-2&\mbox{at $z=1$}\\
-N+3=\tilde{N}+1&\mbox{at $z=-1$},
\end{array}\right.
$$
which are the dual Coxeter number of the groups $SO(2\tilde{N})$ and
$Sp(\tilde{N})$ respectively, where $\tilde{N}=2-N$ is
half the number of anti-D6-branes.

%%%%%%%%%%%%%%%%%%%%%%%%
%%%%%%%%%%%%%%%%%%%%%%%%%

\subsection*{\it Summary on the component including the resolved conifolds}

By the above comparison with $4d$ gauge theory
along with the information about the 
expected classical gauge groups in the various limits, 
we arrive at the following
consistent set of rules for reading the classical configurations 
from the superpotential analysis depending on the RR charge and the 
particular point in the moduli space:

\newenvironment{rrrlist}{%
  \begin{list}{}{%
      \setlength{\labelwidth}{.5cm}%
      \setlength{\leftmargin}{.5cm}%
      \setlength{\rightmargin}{0cm}%
      \setlength{\listparindent}{0cm}%
      \setlength{\parskip}{0cm}%
      \setlength{\topsep}{0cm}
      \setlength{\itemsep}{-0.1cm}%
      }%
    }%
  {\end{list}}

\begin{rrlist}

\item[A)] at the point $z=-1$ there is an $Sp$ group, the number of vacua (or better said the 
instanton counting) is $2N-6$ . The gauge group will 
depend on the number of vacua:

\begin{rrrlist}
\item[A.1)] if $2N-6 > 0$ the gauge group is $Sp(N-4)$. The classical description is 
an O6$^+$ with $N-4$ D6-branes. 

\item[A.2)] if $2N-6 = 0$ there is no classical limit at $z=-1$.

\item[A.3)] if $2N-6 < 0$ the gauge group is $Sp(2-N)$. The classical description is 
a free orientifold with $4-2N$ anti-D6-branes. 
\end{rrrlist}

\item[B)] At the point $z=1$ the structure is similar:

\begin{rrrlist}

\item[B.1)] if $2N-2 > 0$ the gauge group is $SO(2N)$. The classical description is
an O6$^-$ with $N$ D6-branes. 

\item[B.2)] if $2N-2 = 0$ there is no classical limit at $z=1$.

\item[B.3)] if $2N-2 < 0$ the gauge group is $SO(4-2N)$. The classical description
is the free orientifold with $4-2N$ anti-D6-branes.
\end{rrrlist}

\end{rrlist}

These rules allow us to classify the different possibilities depending on the 
RR charge:

\begin{rrlist} 

\item[i)] For $N > 3$ there are four classical points: 
        \begin{center}
	O6$^-$ and $N$ D6, gauge group $SO(2N)$ \\
        O6$^+$ and $N-4$ D6, gauge group $Sp(N-4)$ \\
        two free orientifolds of the resolved conifold
	\end{center}

\item[ii)] For $N=3$ there are three classical points (the curve is the same as 
the $SU(4)$ curve):
	\begin{center}
        O6$^-$ and $N$ D6, gauge group $SO(6) \sim SU(4)$ \\
        two free orientifolds of the resolved conifold
	\end{center}

\item[iii)] For $N=2$ there are four classical points
        (and others sitting on the other branch):
	\begin{center}
        O6$^-$ and $N$ D6, gauge group $SO(4)$\\
        free orientifold of deformed conifold, gauge group $Sp(0)$ (= nothing) \\
        two free orientifolds of the resolved conifold
	\end{center}

\item[iv)] For $N=1$ there are 3 classical points (two others
	sitting on the branch with an infrared $U(1)$):
	\begin{center}
        free orientifold and $2$ anti-D6 on $\reals\projective^3$, gauge group $Sp(1)=SU(2)$ \\
        two free orientifolds of the resolved conifold
	\end{center}

\item[v)] For $N=0$ there are 4 classical points
        (and a few others on the other branch):
	\begin{center}
        free orientifold and $4$ anti-D6 on $\reals\projective^3$, gauge group $SO(4)$ \\
        free orientifold and $4$ anti-D6 on $\reals\projective^3$, gauge group $Sp(2)$ \\
        two free orientifolds of the resolved conifold
	\end{center}
 
\item[vi)] For $N < 0$ there are four classical points: 
	\begin{center}
        free orientifold and $4-2N$ anti-D6 on $\reals\projective^3$, gauge group $SO(2(2-N))$ \\
        free orientifold and $4-2N$ anti-D6 on $\reals\projective^3$, gauge group $Sp(2-N)$ \\
        two free orientifolds of the resolved conifold
	\end{center}

\end{rrlist}

\section{Exact branch structure for $N=1$}
\label{branch}

In this section we wish to present a detailed analysis
of the branch structure for the $N=1$ case of the previous section
from various dual pictures.
As advertised, an exact description of the quantum moduli space
is obtained by moving to a mirror Type IIB theory.

We consider the strong coupling limit of the original Type IIA,
for which the relevant M-theory geometry is asymptotically a cone
over $(S^3\times S^3)/D'_3$.
Here the group $D'_3\simeq \mathZ_4$ is introduced in
Section~\ref{symori} and acts on the triplet of $SU(2)$ matrices as
\[
 (g_1,g_2,g_3) \to (i\tau_2 g_1,-g_2,g_3).
\]
Dimensional reduction along the orbit of the $U(1)$ action
$(g_1,g_2,g_3)\to(e^{i\alpha\tau_3}g_1,g_2,g_3)$ brings the system
back to the original Type IIA.
If we reduce instead along a diagonal $S^1$,
\[
 (g_1,g_2,g_3)\to
 (e^{i\alpha\tau_2}g_1,e^{i\alpha\tau_2}g_2,e^{i\alpha\tau_3}g_2),
\]
the resulting Type IIA configuration is a partially blown-up
orbifold $\mathC^3/D'_3$ with a D6-brane of topology $\mathR^2\times S^1$.
Here the generator of $D'_3$ acts on the coordinates of $\mathC^3$ as
\begin{equation}
 (z_1,z_2,z_3)~\to~ (-z_1,iz_2,iz_3).
\end{equation}
The whole system admits a GLSM description.

The main advantage of this framework is that one can read off
the quantum moduli space rather directly by moving to the mirror
IIB description.
This chain of dualities was used in \cite{av} to study the
geometric transition of D6-branes wrapped on the $S^3$ of deformed
conifold, and also applied in \cite{hp2005} to study its $\mathZ_2$
orbifold.

\subsubsection*{GLSM description}

\begin{figure}[htb]
\centerline{
{\small
\psfrag{v1}{$v_1=(1,0,0)$}
\psfrag{v2}{$v_2=(0,1,0)$}
\psfrag{v3}{$v_3=(0,0,1)$}
\psfrag{v4}{$v_4$}
\psfrag{v5}{$v_5$}
\epsfig{width=4cm,file=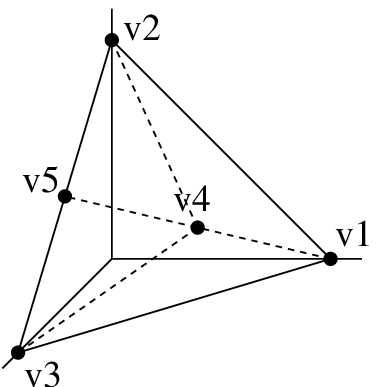}}
\hskip3cm
{\small
\psfrag{z1}{$z_1$}
\psfrag{z2}{$z_2$}
\psfrag{z3}{$z_3$}
\psfrag{z4}{$z_4$}
\psfrag{z5}{$z_5$}
\psfrag{A}{A}
\psfrag{A'}{A'}
\psfrag{B}{B}
\psfrag{C}{C}
\epsfig{width=4cm,file=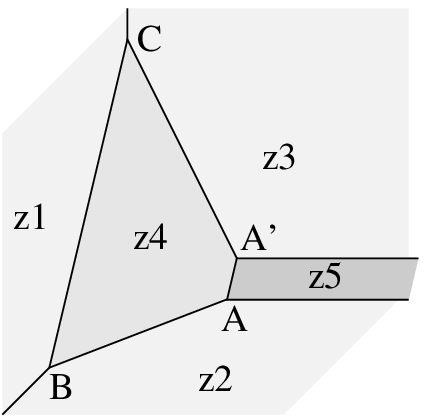}}
}
\caption{(Left) toric fan for the orbifold $\mathC^3/\mathZ_4$,
         (Right) skeleton for the fully blown-up phase.}
\label{fig:fan}
\end{figure}

Consider the edge vectors for the toric fan of $\mathC^3$,
\[
 v_1=(1,0,0),~~~
 v_2=(0,1,0),~~~
 v_3=(0,0,1),
\]
generating the lattice $\mathZ^3$ of torus actions.
The orbifolding makes the lattice finer by the inclusion of an extra
generator $\rho=(-\frac12,\frac14,\frac14)$.
The fan for Calabi-Yau resolution of orbifold singularity is given by
including the lattice points $v_4=(\frac12,\frac14,\frac14)$ and
$v_5=(0,\frac12,\frac12)$ as new edge vectors and subdividing the fan.
See Figure \ref{fig:fan}.

The GLSM consists of five chiral fields $z_{1,\cdots,5}$ and
$U(1)^2$ gauge symmetry.
The $U(1)$ charges which span the Mori cone are given by
\begin{equation}
 Q^1=(1,0,0,-2,1),~~~
 Q^2=(0,1,1,0,-2),
\end{equation}
so that the fully blown-up phase is given by the D-term equations
\begin{equation}
 |z_1|^2 -2|z_4|^2  +|z_5|^2 ~=~ r^1 >0,~~~~
 |z_2|^2 + |z_3|^2 -2|z_5|^2 ~=~ r^2 >0.
\end{equation}
In the orbifold phase with negative FI parameters the fields $z_4,z_5$
acquire vev and break the gauge group down to $\mathZ_4$, which acts
$(z_1,z_2,z_3)$ to $(-z_1,iz_2,iz_3)$.
It is useful to draw the toric skeleton diagram describing the
base polytope of $T^3$ fibration.

In addition to the K\"ahler parameters, one has to specify the
location of D6-brane which projects to a half-line
ending on a one-dimensional face of the toric polytope.
As was discussed in \cite{hp2005} for related models,
one cannot choose the K\"ahler parameters and the location of
the D6-brane independently, because of a superpotential
generated by the D6-brane.
Also, by looking into the asymptotics one finds that the resolution
mode corresponding to $z_5$ should not be turned on.
In other words, in the skeleton diagram the points A, A' should
coincide.

\subsubsection*{Three-dimensional Five-brane Web}

\begin{figure}[htb]
\centerline{
{\footnotesize
\psfrag{A}{$(0,2)$}
\psfrag{B}{$(1,-1)$}
\psfrag{C}{\hskip-6mm$(1,1)$}
\psfrag{D}{$(1,0)$}
\psfrag{E}{\hskip-3mm$(2,1)$}
\psfrag{F}{$(2,-1)$}
\epsfig{width=6cm,file=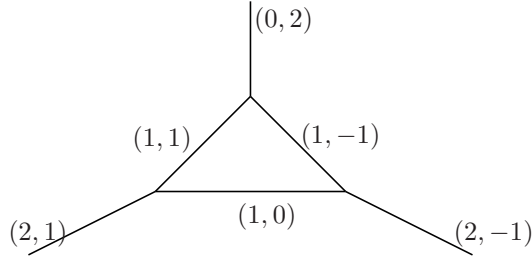}}}
\caption{the fivebrane web which is dual to a partial
         resolution of the orbifold $\mathC^3/\mathZ_4$.}
\label{fig:web}
\end{figure}

One can understand the effect of superpotential semiclassically
by relating our GLSM picture to a Type IIB five-brane web.
Under a suitable choice of basis for the charge, the partial
blowup of our orbifold $\mathC^3/\mathZ_4$ is mapped to the
two-dimensional web of Figure \ref{fig:web}.
The D6-brane in GLSM picture turns into another five-brane leg
carrying a new kind of charge ending on a leg of the web.
Addition of such a five-brane makes the web three-dimensional.
The supersymmetry condition for the resulting web
constrains the allowed locations of the D6-brane endpoint
for each choice of K\"ahler parameter.

The charges of each leg of the three-dimensional web can be
obtained by regarding the $G_2$ holonomy manifolds as $T^3$
fibrations, and analyzing the locus of degenerate fiber in the
base.
Since these charges are relevant in calculating the superpotential,
let us go back to M-theory and calculate them explicitly.
As the $T^3$ we take the orbit of $U(1)^3$ action
\begin{equation}
 [\alpha_1,\alpha_2,\alpha_3]~:~
 (g_1,g_2,g_3)~\mapsto~
 (e^{2\pi i\alpha_1\tau_2}g_1,
  e^{2\pi i\alpha_2\tau_2}g_2,
  e^{2\pi i\alpha_3\tau_2}g_3),
\end{equation}
modulo identification
\begin{eqnarray}
 [\alpha_1,\alpha_2,\alpha_3]
      &\sim& [\alpha_1,\alpha_2,\alpha_3] +\ts[-\frac14,-\frac12,0]
\nn\\ &\sim& [\alpha_1,\alpha_2,\alpha_3] +\ts[\frac12,0,0]
\nn\\ &\sim& [\alpha_1,\alpha_2,\alpha_3] +\ts[\frac12,\frac12,\frac12].
\label{aperiod}
\end{eqnarray}
In the following we will also use the coordinate
$(X,\tilde X)\equiv (g_2g_1^{-1},g_3g_1^{-1})$ of $S^3\times S^3$.

The classical moduli space consists of the following branches.
A free orientifold of deformed conifold with two D6-branes
corresponds to $g_1$ filled in, and the orientifold of resolved conifold
with flux correspond to either $g_2$ or $g_3$ filled in.
The O6$^-$+D6 configuration corresponds to the geometry with
the following $S^1$ shrinking at $r=0$,
\begin{equation}
  (X,\tilde X) \mapsto
  (Xe^{2\pi i\alpha\tau_2}, \tilde Xe^{-2\pi i\alpha\tau_2}).
\label{S1}
\end{equation}

The cross-section of the seven-manifold at any finite $r$ is
$S^3\times S^3$.
It can be viewed as a $T^3$ fibration over $S^3$, and a section is
given by
\[
 (X,\tilde X)=(e^{i\theta\tau_1/2}e^{i\varphi\tau_2},
               e^{i\tilde\theta\tau_1/2}),
~~~~~~
(0\le\theta\le\pi,~~
 0\le\tilde\theta\le\pi,~~
 0\le\varphi\le 2\pi).
\]
At four special points on the base, the fiber $T^3$ has a vanishing
one-cycle labeled by the ratio of $[\alpha_1,\alpha_2,\alpha_3]$,
\begin{equation}
\begin{array}{lcl}
(\theta,\tilde\theta)=(0  ,0  ) &\cdots&
 [\alpha_1,\alpha_2,\alpha_3]=[+\frac12,+\frac12,+\frac12],\\
(\theta,\tilde\theta)=(0  ,\pi) &\cdots&
 [\alpha_1,\alpha_2,\alpha_3]=[-\frac12,-\frac12,+\frac12],\\
(\theta,\tilde\theta)=(\pi,0  ) &\cdots&
 [\alpha_1,\alpha_2,\alpha_3]=[-\frac12,+\frac12,-\frac12],\\
(\theta,\tilde\theta)=(\pi,\pi) &\cdots&
 [\alpha_1,\alpha_2,\alpha_3]=[+\frac12,-\frac12,-\frac12].
\end{array}
\label{loci1}
\end{equation}
Thus we find four half-lines of degenerate fiber.
There are additional loci of degenerate fiber at $r=0$.
For the classical branch with $g_1$ filled in one finds a line
segment of degenerate $[1,0,0]$ one-cycle at $r=0$, and similarly
for the other two branches where $g_2$ or $g_3$ are filled in.
For the last branch, one finds that the vanishing $S^1$ given in
(\ref{S1}) lies along the $T^3$ fiber when
\begin{equation}
\begin{array}{lcl}
\theta=0   &\cdots&
 [\alpha_1,\alpha_2,\alpha_3]=[+\frac14,+\frac12,0],\\
\theta=\pi &\cdots&
 [\alpha_1,\alpha_2,\alpha_3]=[+\frac14,-\frac12,0],\\
\tilde\theta=0   &\cdots&
 [\alpha_1,\alpha_2,\alpha_3]=[-\frac14,0,-\frac12],\\
\tilde\theta=\pi &\cdots&
 [\alpha_1,\alpha_2,\alpha_3]=[-\frac14,0,+\frac12].
\end{array}
\end{equation}

\begin{figure}[htb]
\centerline{\scriptsize
{\psfrag{A}{\hskip-6mm$(0,2,-1)$}
 \psfrag{B}{\hskip-6mm$(-2,-1,-1)$}
 \psfrag{C}{\hskip-4mm$(2,-1,1)$}
 \psfrag{D}{\hskip-4mm$(0,0,1)$}
 \psfrag{I1}{\hskip-3mm$(0,2,0)$}
 \epsfig{width=3cm,file=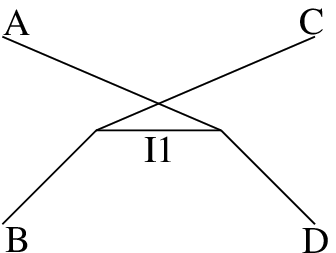}} ~~~
{\psfrag{A}{}
 \psfrag{B}{}
 \psfrag{C}{}
 \psfrag{D}{}
 \psfrag{I2}{\hskip-1mm$(-2,-1,0)$}
 \epsfig{width=3cm,file=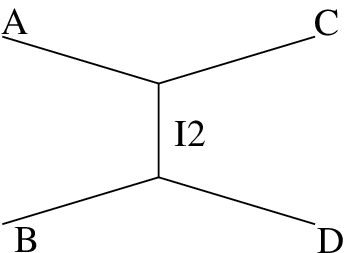}} ~~~
{\psfrag{A}{}
 \psfrag{B}{}
 \psfrag{C}{}
 \psfrag{D}{}
 \psfrag{I3}{\hskip-5mm$(2,-1,2)$}
 \epsfig{width=3cm,file=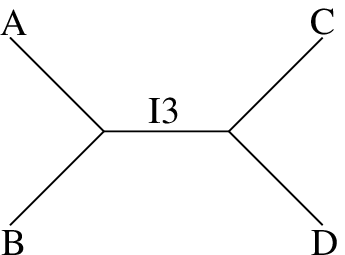}} ~~~
{\psfrag{A}{}
 \psfrag{B}{}
 \psfrag{C}{}
 \psfrag{D}{}
 \psfrag{J1}{\hskip-4mm$(1,-1,1)$}
 \psfrag{J2}{\hskip-0mm$(-1,0,0)$}
 \psfrag{J3}{\hskip-4mm$(-1,0,-1)$}
 \psfrag{J4}{\hskip-7mm$(1,1,0)$}
 \epsfig{width=3cm,file=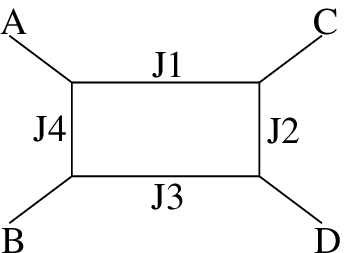}}}
\caption{the fivebrane web corresponding to the four families of
         $G_2$ holonomy spaces.}
\label{fig:3dweb}
\end{figure}

By re-labeling the shrinking one-cycle in
terms of the fundamental periods given in (\ref{aperiod}),
one finds that the four half-lines of (\ref{loci1}) are labeled by
\begin{equation}
 (0,0,1),~~~(2,-1,1),~~~(-2,-1,-1),~~~(0,2,-1).
\end{equation}
We can identify them with the charges of semi-infinite five-brane legs
forming three-dimensional webs.
Indeed, the first leg corresponds to the D6-brane while the other three
project to the legs of two-dimensional web of Figure \ref{fig:web}.
The four families of $G_2$ holonomy spaces thus correspond to the
three-dimensional webs summarized in Figure \ref{fig:3dweb}.
All these four webs are rigid apart from overall rescaling.

\subsubsection*{IIB mirror}
The mirror of this GLSM is given by
a LG model with superpotential
\begin{equation}
 F(x,y) = y^2 + e^{\frac{t_1}{2}+\frac{t_2}{4}} xy + x^3
        + e^{\frac{t_2}{2}} x^2 + x.
\end{equation}
Here $(x,y)$ are $\mathC^\times$-valued chiral superfields,
and $t_1,t_2$ are the K\"ahler parameters in the GLSM.
The spacetime theory is a Type IIB superstring on a local Calabi-Yau manifold
\begin{equation}
\xi\eta ~=~ F(x,y),
\end{equation}
with a D5-brane wrapping a holomorphic curve $\eta=F(x,y)=0$.
We denote by $(x_{\bf D},y_{\bf D})$ its position on the curve
$\Sigma:F(x,y)=0$.

The parameter $t_2$ is fixed by requiring the curve $\Sigma$
to have only three punctures.
There seems to be two choices $e^{\frac{t_2}{2}}=\pm2$, but they should
lead to the same structure for the moduli space so we choose
$e^{\frac{t_2}{2}}=-2$.
The remaining normalizable parameter
$s\equiv -ie^{\frac{t_1}{2}}/\sqrt2$ in the curve,
\begin{equation}
 \Sigma~:~~
 y^2-2sxy+x^3-2x^2+x ~=~ 0,
\end{equation}
is stabilized by the superpotential generated by the D5-brane.
\begin{equation}
W(s)= \int_{\bf A}^{\bf D}+\int_{\bf C}^{\bf B} \log y\cdot d\log x.
~~~~
\left(
{\bf A}(1,0),~~
{\bf B}(0,0),~~
{\bf C}(\infty,\infty)
\right).
\end{equation}
This integration contour is chosen from the observation that
the four semi-infinite five-brane legs all have non-zero ``third''
charge, which we interpret as the presence of D5 or $\overline{\rm D5}$-branes
at three punctures {\bf A, B, C} of the curve.

Note that the moduli space of curves $\Sigma$ is the complex $s$-plane
modulo identification $s\sim -s$, since $s\sim e^{\frac{t_1}{2}}$.
Indeed, the sign flip of $s$ can be absorbed by the sign flip of $y$.
The good modulus of the curve is therefore $s^2$.

For generic $s$ the curve $\Sigma$ is of genus one.
The degeneration to genus zero occurs at $s=0$ or $s=\pm2i$.
For genus one curve, the modulus $s$ is related
to the position of {\bf D} because of the superpotential.
Using $\hat y=y-sx$ we rewrite the equation for the curve as
\begin{equation}
 \hat y^2+x^3-(2+s^2)x^2+x ~=~ 0.
\end{equation}
The F-term condition reads
\begin{equation}
 \int_{\overline{\rm D5}={\bf A,C}}^{{\rm D5}={\bf B,D}}\frac{dx}{\hat y}=0.
\end{equation}
One can solve this equation by fixing $s$ in such a way that
there is a meromorphic function on $\Sigma$ with simple zeroes at
{\bf B, D} and simple poles at ${\bf A, C}$.
\begin{equation}
\left|\begin{array}{rrr}
 1 & x_{\bf A} & -\hat y_{\bf A} \\
 1 & x_{\bf B} & \hat y_{\bf B} \\
 1 & x_{\bf D} & \hat y_{\bf D} \\
\end{array}\right| ~=~
\left|\begin{array}{ccc}
 1 & 1         & s \\
 1 & 0         & 0 \\
 1 & x_{\bf D} & \hat y_{\bf D} \\
\end{array}\right| ~=~
sx_{\bf D}-\hat y_{\bf D} ~=~0
~~~~\Rightarrow~~~~
(x_{\bf D},y_{\bf D})=(1,2s).
\end{equation}

The moduli space for the case $N=1$ is made of $g=0$ and $g=1$ branches.
The $g=1$ branch is the moduli space of genus one curves with three
punctures, and is the complex $s^2$-plane as explained above.
It has special points $s^2=0,-4$ and infinity.
At $s=2i$ the curve degenerates to genus zero,
\begin{equation}
 \Sigma~:~ y^2 - 4ixy +x^3-2x^2+x~=~0.
\label{g=0curve}
\end{equation}
This curve itself is regarded as the $g=0$ branch.
Various semi-classical points on moduli space are identified with
those in the conifold picture as follows:
\begin{center}
\begin{tabular}{|l|l|l|l|}
\hline
branch & point & gauge group & description \\
\hline
$g=0$ & {\bf A  } & $Sp(1)$ & 
 deformed$_{(\mu<0)}$, with 2D6 on $\mathR\mathP^3$ \\
$g=0$ & {\bf B,C} & none    &
 resolved, with ($-1$) flux \\
$g=1$ & $s^2=0$     &  $O(2)$ &
 deformed$_{(\mu<0)}$, with 2D6 on $\mathR\mathP^3$ \\
$g=1$ & $s^2=\infty$&  $O(2)$ &
 deformed$_{(\mu>0)}$, with O6$^-$+D6 on $S^3$ \\
\hline
\end{tabular}
\end{center}

By a suitable identification of the variables $\eta_i$ with $(x,y)$
one can identify the $z$-plane of Section~\ref{superpotential}
with the $g=0$ branch here, and also see the expected behavior
of $\eta_i$'s on the other branch which can be identified with
the $s^2$-plane.
The key fact is that the membrane instantons wrapping on three-cycles
$\{2E'_1,E'_2,E'_3\}$ of Section~\ref{topology} turn into disc instantons
bounded by the D6-brane.
Matching of their volumes gives
\begin{equation}
 \eta_1^2 ~\sim~ e^{(|z_2|^2-|z_3|^2)/2} ~\sim~ x,~~~
 \eta_2   ~\sim~ e^{ |z_1|^2-|z_3|^2   } ~\sim~ x^3y^{-2},~~~
 \eta_3   ~\sim~ e^{ |z_2|^2-|z_1|^2   } ~\sim~ x^{-1}y^2,
\label{matching}
\end{equation}
where we used the standard mirror identification
\begin{equation}
 e^{-|z_1|^2}:e^{-|z_2|^2}:e^{-|z_3|^2}:e^{-|z_4|^2}:e^{-|z_5|^2}
~\sim~
 y^2:x^3:x:xy:x^2,
\end{equation}
and ``$\sim$'' expresses the identification up to phase.
Under this identification, the $g=0$ curve (\ref{g=0curve})
precisely agrees with the relation (\ref{curveN=1})
among $\eta_i$ on the branch of mass-gapped vacua obtained
in the previous section.
Note that the $\eta_1$ defined as
\begin{equation}
 \eta_1 = i\left(\frac{y-2ix}{1+x}\right)
\label{matching2}
\end{equation}
is single valued on $\Sigma$ and it indeed squares to $x$ on $\Sigma$.
Note also that, although the $g=0$ curve (\ref{g=0curve}) has
a double point, the singularity is blown up on generic points
of $g=0$ branch and the $\eta_i$ indeed take two different values
there.
The functional form of $\eta_i$'s on the $g=1$ branch is obtained
simply by substituting $(x, y)=(1,2s)$ into (\ref{matching}):
\begin{equation}
  \eta_1=1,~~~~
  \eta_2=\frac{1}{4s^2},~~~~
  \eta_3=4s^2.
\end{equation}
This agrees with the expectation in Section~\ref{N=1} that the
branch of vacua with infrared $U(1)$ is a cylinder $\eta_2\eta_3=1$.

\section{Other cases}
\label{other}

We wish to briefly comment on the possibility of $\mu$-transitions
at the conifold when the orientifold action is in one of the other
classes discussed in Section~\ref{oricon}. It is clear that it will
be much harder to find the associated $G_2$ holonomy metrics.
Nevertheless, qualitative considerations similar to the ones we
sometimes used above give good indications when we should expect 
a $\mu$-transition.

\subsection{Case (1)$\leftrightarrow$(3)}

The first thing to notice in cases (1) and (3) is that the O-plane 
intersects the compact three-cycle of the deformed 
conifold. This creates flux that cannot escape to infinity, and we should 
cancel the flux by wrapping a fixed number of D6-branes. A simple class 
of cycles to wrap branes around is the fixed point set of some 
anti-holomorphic involution which can be different from the one used 
to define the orientifold, but must be in the same class to preserve
supersymmetry.

In case (1), the O6-plane is the fixed point locus of the involution
$z\mapsto M_O \bar z$, where $M_O$ is the orthogonal matrix ${\rm diag} 
(-1,1,1,1)$. The O6-plane is topologically $S^2\times\reals$ and the
intersection number with the $S^3_>$ is two. If this is a standard
O6$^-$-plane with negative twice the charge of a D6-brane (as measured 
in the quotient space), this means that we need to wrap a D6-brane 
configuration that intersects the compact cycle exactly four times 
(namely, an invariant configuration in the covering space intersecting
$S^3_>$ in eight points). We know of supersymmetric cycles intersecting 
the $S^3_>$ twice: The fixed point locus of $z\mapsto M \bar z$, where 
$M$ is another orthogonal matrix with eigenvalues $(-1,1,1,1)$ (generally 
distinct from $M_O$). We can write $M = UM_O U^T$, where $U$ is an element 
of $SO(4)$, and two $U$'s give the same $M$ if they differ by an element 
of $SO(3)$. In the covering space, the possible $M$'s live in 
$SO(4)/SO(3)\simeq S^3$, and we need two (pairs of) such cycles to 
cancel the charge. Orientifolding maps the brane associated with $U$ 
to the brane associated with $M_O U M_O$, which corresponds to acting 
on $S^3$ as the element ${\rm diag}(1,-1,-1,-1)$ of $SO(4)$. In the 
orientifold, the space of possible brane wrappings is therefore the 
symmetric product $\calm^{(1)}_-\simeq\cals^2(\call)$, where 
$\call=S^3/\zet_2$ with the given action of $\zet_2$.

Note that if the O-plane is an O6$^-$, we cannot wrap branes 
on fixed point loci of involutions with three negative eigenvalues, 
since those would preserve the opposite supersymmetry, and we cannot 
wrap anti-branes if we are to cancel the charge.

On the other hand, if the orientifold plane is an O6$^+$ with
positive twice the charge of a D6-brane, we need anti-branes
to cancel the charge, and those are most conveniently wrapped on
fixed point loci of $z\mapsto M \bar z$, where $M$ is an orthogonal
matrix with eigenvalues $(-1,-1,-1,1)$. As we recall, this gives a 
cycle with two disconnected components, each of which is a copy of
$\reals^3$ and intersects the $S^3_>$ once. In that case, we
have the choice of four such cycles, each of which is again 
parameterized by the choice of $\call\simeq S^3/\zet_2\simeq 
\bigl(SO(4)/SO(3)\bigr)/\zet_2$. Thus, $\calm^{(1)}_+\simeq\cals^4(\call)$.

In case (3), the situation is reversed: If the O-plane on $\reals^3
\cup\reals^3$ is an O6$^-$, we can wrap four D6-branes on four 
different copies of $\reals^3$, while with an O6$^+$, we 
can wrap two anti-D6-branes on $S^2\times \reals$. We have 
$\calm^{(3)}_-\simeq\cals^4(\call)$, and $\calm^{(3)}_+\simeq
\cals^2(\call)$, where, importantly, $\call\simeq S^3/\zet_2$ 
is the same quotient as before.

It is worthwhile to point out that the parameters associated with
the brane wrappings are not on the same footing as the parameter
associated with the size of the $S^3$. The latter parameter,
although not normalizable in the non-compact geometry, is still 
localized and will survive embedding in a compact model. The data
parameterizing the positions of the branes, on the other hand,
is completely fixed at infinity. Not even their complexification
can be determined before embedding in a compact model.

As a consequence, in addition to fixing the sign of the O-plane,
we should fix a point $(M_1,M_2)\in \calm^{(1)}_-$ when we attempt 
to take case (1) through a $\mu$ transition to case (3). In so 
doing, we can see from the matching of the D-brane configuration 
at infinity that we end up with a configuration in $\calm^{(3)}_-$ 
of the special form $(M_1,M_1,M_2,M_2)$. Conversely, starting
from $(M_1,M_2,M_3,M_4)\in\calm^{(1)}_+$, we can match with a
point in $\calm^{(3)}_-$ only for $M_1=M_3$, $M_2=M_4$. 

These constraints allow us to predict a smooth $\mu$-transition 
only for this subset of D-brane configurations. We are not able
to determine the fate of the other configurations as $\mu$ goes 
to zero.

Another point that remains unclear is whether there will be a
point of enhanced gauge symmetry in the moduli space. As we have
explained before, the orientifolds in the classes (1) and (3)
do not admit the resolved conifold. Naively, we can say that the 
orientifold is projecting out the scalars of the $\caln=2$ 
vectormultiplet that was associated with the $\projective^1$ 
of the resolved conifold (whereas in cases (0), (2) and (4) we 
are projecting out the $\caln=1$ vector). One possible conclusion 
is that the $\caln=1$ vector half of the $\caln=2$ vectormultiplet, 
which is broken by the hypermultiplet vev at a generic point, 
reappears at the singular conifold. However, it is not clear that 
such a naive argument will survive a more careful treatment.

\subsection{Case (2)}

Now the O-plane locus is $S^1\times\reals^2$ and has zero intersection 
with the compact three-cycle. We can then wrap supersymmetrically
D6-branes on the compact three-cycle and get a dynamical gauge theory
in four dimensions. This is very similar to the case (0)/(4) that
we have focused on before. We can not find an M-theory lift, but
since the small resolution of the conifold is allowed, we can apply
the Vafa superpotential method to predict the structure of 
a part of the quantum parameter space.

Assume that we want to wrap $2N$ D6-branes on $S^3_>$. By the
computation of the period integrals in Section~\ref{Ohomology},
this will preserve the same supersymmetry as an anti-O6-plane
wrapped on the noncompact cycle $S^1\times\reals^2$. If this 
O-plane is an anti-O6$^-$ (with positive charge), the gauge 
group is $Sp(N)$. This gauge group type follows because the 
relative codimension between O-plane and D-brane is $4$.

A configuration with the same supersymmetry and charge at infinity is 
obtained by wrapping $2N+4$ branes, or rather $2 \tilde N=-2N-4$ 
anti-D6-branes, on $S^3_<$. Here, the $4$ comes again from the jump 
\eqref{delta2} in the class of the fixed point locus, multiplied by
the charge of the anti-O6$^-$-plane. Note again that this is 
supersymmetric and gives the gauge group $Sp(\tilde N)$ in front
of the anti-O6$^-$-plane because of the lower-dimensional 
intersection.

We see that $N$ and $\tilde N$ are never both non-negative at the
same time, so that we do not expect a $\mu$-transition to be possible
for any $N$. In fact, when $N=-1$, also $\tilde N=-1$, so this
value of the flux does not admit a deformed conifold for the fixed
supersymmetry.

Meanwhile, if the O-plane is an anti-O6$^+$, and we wrap $2N$ 
D6-branes on $S^3_>$, the gauge group will be $SO(2N)$. Going
through the $\mu$-transition, we could also wrap $2N-4$ D6-branes,
or rather $2 \tilde N=-2N+4$ anti-D6-branes on $S^3_<$, with gauge
group $SO(2\tilde N)$. The jump by $-4$ is the value familiar from 
the case (0)/(4), and as in that case, we expect a $\mu$-transition 
to be possible for $N=0,1,2$, but no other values of $N$.

Let us now figure out the moduli space using the superpotential.

\subsection*{\it The moduli space}

To start with, we discuss the right parameter of the moduli space
near the resolved conifold points.
Let $t$ be as before the complexified K\"ahler class of the $\projective^1$
of the resolved conifold. Unlike in the cases (0) and (4) considered
in Section~\ref{superpotential} (see (\ref{doubled})),
the periodicity in the present case is the same as the one before
the orientifold:
\begin{equation}
t\equiv t+2\pi i,
\label{notdoubled}
\end{equation}
so that the single valued coordinate of the parameter space is
$e^{-t}$.
This is because
the worldsheet diagram must always have even powers of
$\exp(i{\rm Im}(t)/2)$.
Namely, there is no odd degree
smooth map of the worldsheet $S^2$ to the target
$\projective^1$ compatible with
the involution $\Omega:w\to -1/\bar w$ on the domain
and the involution $\tau$ given by (\ref{accompa}) on the target.
This can be shown as follows. Let $X:S^2\to\projective^1$ be such
a map, where the compatibility means $\tau\circ X\circ\Omega=X$.
Choosing a K\"ahler form $\omega$ of $\projective^1$
 of volume $1$, the degree is defined as $d=\int_{S^2}X^*\omega$.
Let us decompose $S^2$ as a union of the upper and lower hemi-spheres
$S^2=H_+\cup H_-$ which are oriented such that 
$[S^2]=H_++H_-$ and $\Omega(H_+)=-H_-$.
Using $\tau^*\omega=-\omega$, one can express the degree as
\begin{eqnarray}
d&=&\int_{H_+}X^*\omega+\int_{H_-}X^*\omega
=\int_{H_+}X^*\omega-\int_{H_+}\Omega^*X^*\omega
\nn\\
&=&\int_{H_+}(X^*\omega+\Omega^*X^*\tau^*\omega)=2\int_{H_+}X^*\omega.
\nn
\end{eqnarray}
The idea is to show that $\int_{H_+}X^*\omega$ is an integer.
Note that $d$ is an integer and thus
$\int_{H_+}X^*\omega$ is deformation invariant, as long as
$X:H_+\to \projective^1$ extends to an equivariant map of $S^2$ to 
$\projective^1$.
Extension to $S^2$ is possible if and only if the restriction to
the boundary $X:\partial H_+\to\projective^1$ is equivariant.
It is always possible to shrink this loop to a constant map to
a $\tau$-fixed point, keeping the equivariance all the way.
Once this is done, we obtain a map
$X:H_+/\partial H_+\to\projective^1$, which is a map between 
 two spheres.
$\int_{H_+}X^*\omega$ is its degree and thus is an integer.

We first consider the case of an anti-O6$^-$ and $2N\ge 0$ D6-branes 
on $S^3_>$ or $2\tilde N=-2N-4\geq 0$ anti-D6-branes on $S^3_<$.
The computation is done in the orientifold of
the resolved conifold with certain RR two-form flux through $\RP^2$.
The flux is $N$ since the O-plane does not contribute to
the two-form flux through $\RP^2$ (the O-plane is still there on 
the resolved side). The superpotential again has three terms corresponding to
the three origins: four-form flux, two-form flux and worldsheet instantons.
The result is
\begin{eqnarray}
W &=& -\int F_4\wedge\omega +N {\partial F_0\over
\partial t} -4\sum_{m:\,{\rm even}>0}{e^{-mt/2}
\over m^2}
\nn\\
&=&
-Yt/2-N Li_2(e^{-t})-2(Li_2(e^{-t/2})+Li_2(-e^{-t/2}))
\nn\\
&=& -Yt/2-(N+1) Li_2(\e^{-t}).
\eqlabel{otherW}
\end{eqnarray}
The parameter $Y$ is again defined by
$$
Y:=\int_{S^3_{\infty}}({\rm Re}\hat{\Omega}+iC_3).
$$
The sum over $m$ in the crosscap part is over even integers
because only even degree maps are compatible
with the present orientifold, as remarked above.
Solving $\partial_t W=0$, we find the equation determining the moduli
space
\begin{equation}
e^{-Y}=(1-e^{-t})^{2N+2}.
\label{otherC}
\end{equation}
It is a copy of the Riemann sphere with three marked points
--- $(e^{-Y},e^{-t})=(1,0)$, $(\infty,\infty)$, $(0,1)$ for $N\geq 0$,
while $(e^{-Y},e^{-t})=(1,0)$, $(0,\infty)$, $(\infty,1)$ for $N\leq -2$.
In either case, the superpotential has a branch cut at the last marked point,
$e^{-t}=1$, at which it behaves as follows:
\begin{equation}
W=-Yt/2-(N+1)t\log t+\cdots.
\label{case2Sup}
\end{equation}

If $N\geq 0$, the point $(e^{-Y},e^{-t})=(0,1)$
corresponds to
a large minimal three sphere $S^3_>$ with $2N$ D6-branes supporting
an $Sp(N)$ gauge field.
The relation of the parameter $Y$ and the holomorphic gauge coupling
$({8\pi^2\over g^2}-i\theta)$ on the D6-brane
can be determined following the argument given in
Section~\ref{superpotential}.
Notice that it is similar to the case (4) in that
the orientifold acts non-trivially on the three sphere.
Noting that we expect symplectic gauge group, we find
\begin{equation}
Y=2\left({8\pi^2\over g^2}-i\theta\right).
\label{otherY}
\end{equation}
Then, with the identification of $t$ as the glueball field $S$,
 (\ref{case2Sup}) is indeed the Veneziano-Yankielowicz superpotential
for the gauge group $Sp(N)$, up to a power series in $S$.

If $N\leq -2$, at the point $(e^{-Y},e^{-t})=(\infty,1)$
we have a large $S^3_<$ with $2\tilde N$ anti-D6-branes supporting
an $Sp(\tilde N)$ gauge field.
The relation of the parameter $Y$ and the holomorphic gauge coupling
is the same as (\ref{otherY}) up to sign.
With the identification $S=-t$,
(\ref{case2Sup}) agrees with the Veneziano-Yankielowicz
superpotential for the $Sp(\tilde N)$ super Yang-Mills,
since $-(N+1)=\tilde N+1$ is the dual Coxeter number of $Sp(\tilde N)$.

One may also consider the case $N=-1$. In the resolved side,
the superpotential from the flux is exactly canceled by
the contribution form the crosscap instantons. The superpotential is simply
$W=-Yt/2$ and $\partial_tW=0$ requires $Y=0$.
Indeed, we do not have any candidate classical limit 
with deformed conifold for this value of
the charge, as we have seen.
What is most interesting is that the resolved branch
has no singularity in the interior.
By the combined effect of
the flux and worldsheet instantons,
the singularity, which was present in $\caln=2$ systems, is completely
washed out!

Let us next consider the case of an anti-O6$^+$-plane and
$2N>4$ D6-branes on $S^3_>$ or $2\tilde N=-2N+4>4$ anti-D6-branes
on $S^3_<$.
In this case, the crosscap
contribution in (\ref{otherW}) changes sign, and the equation for
the moduli space is
\begin{equation}
e^{-Y}=(1-e^{-t})^{2N-2}.
\label{otherCp}
\end{equation}
It is again a complex plane with three marked points.
For $N>2$, the holomorphic gauge coupling is related to the parameter $Y$
by
\begin{equation}
Y={8\pi^2\over g^2}-i\theta,
\label{otherYp}
\end{equation}
and we have the expected behavior of the superpotential for $S=t/2$
near the point $(e^{-Y},e^{-t})=(0,1)$,
as $2(N-1)$ is the dual Coxeter number of the gauge group $SO(2N)$.
For $N<0$, the relation of $Y$ and the gauge coupling is opposite,
but again the superpotential behaves as it should.

\begin{figure}[htb]
\def\axes{\tiny
\psfrag{m+}{$\mu>0$}
\psfrag{m-}{$\mu<0$}
\psfrag{t+}{$t>0$}
\psfrag{t-}{$t<0$}
\epsfig{width=7cm,file=axes.eps}   }
\def\res{\scriptsize
\psfrag{r+}{(res)}
\psfrag{r-}{(res)}   }
\begin{center}
\begin{tabular}{cc}
\res
\psfrag{N=2}{\fbox{\large $\sf N=2$}}
\psfrag{?}  {\sf\large ?}
\psfrag{d-} {\scriptsize $SO(4)$}
\psfrag{d'-}{\hskip-3mm\scriptsize $SO(0)$}
\epsfig{width=7cm,file=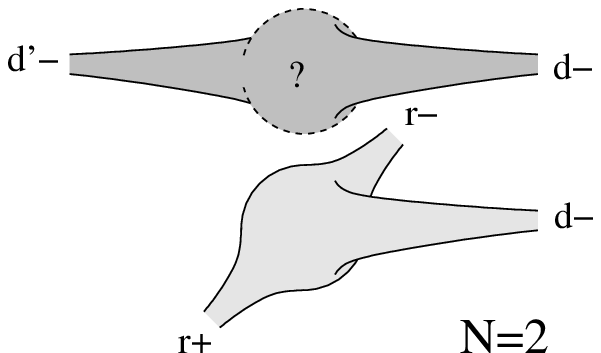}
~~&~~
\res
\psfrag{N=1}{\fbox{\large $\sf N=1$}}
\psfrag{?}  {\sf\large ?}
\psfrag{phase}{\parbox{1cm}{phase\\ transition?}}
\psfrag{d-} {\scriptsize $SO(2)$}
\psfrag{d'-}{\hskip-3mm\scriptsize $SO(2)$}
\epsfig{width=7cm,file=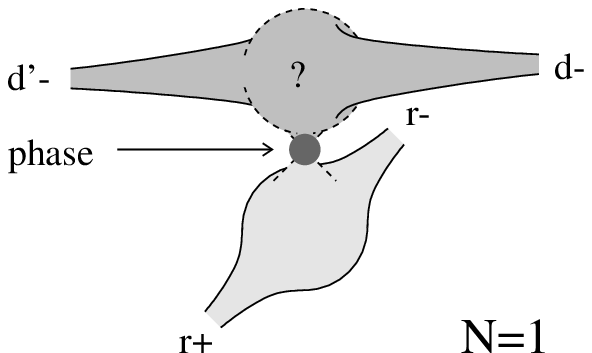}
\\
\axes~~&~~\axes
\end{tabular}
\end{center}
\caption{Expected structure of moduli space for case (2) with $N=2$ and $1$.
         The branch with darker shade contains vacua that are not
         accounted for by the superpotential.}
\label{fig:case(2)}
\end{figure}

All these cases having been checked quite nicely, we briefly
comment on the exceptional cases $N=0,1,2$.
Here we expect a $\mu$-transition between vacua with $SO(2N)$ and
$SO(4-2N)$ super Yang-Mills theories to be possible, but this will occur on a
branch of moduli space that is not described by the exact superpotential.
Though we do not have very powerful tools of analysis for the new
branch, we can at least make a guess and check the consistency by
examining the order of poles of holomorphic parameters at classical points.
The expected branch structure is summarized in Figure \ref{fig:case(2)}.

For the case $N=2$ (related to the case $N=0$ by sign flip of $Y$),
the superpotential accounts for only two vacua of the super Yang-Mills theory
on large $S^3_>$.
Other vacua should sit on a ``new'' branch which we expect to contain
also the deformed vacuum with negative $\mu$.
The holomorphic parameter $e^{-Y}$ has double zero at $\mu\to +\infty$,
whereas the complex volume of minimal three-cycle at $\mu\to -\infty$
is minus $Y/2$ so that $e^{-Y/2}$ has a simple pole there.
The new branch will therefore be a cylinder parametrized by
$e^{-Y/2}$.

For $N=1$, the flux and crosscap terms in the superpotential cancel
out in the same way as the $Sp$ case with $N=-1$.
A branch of the moduli space is therefore a
cylinder interpolating two resolved classical points.
We expect another branch of vacua with low energy $U(1)$ gauge
dynamics that contains deformed classical points
with either sign of $\mu$.
It would be interesting to study how the two branches are connected.

\section{Conclusions}
\label{conclusions}

We have discussed supersymmetric quantum transitions between various 
orientifolds of the conifold. We have constructed the possible 
orientifolds of the deformed and resolved conifold in Type IIA 
string theory. In the primary case where this is possible, we have
answered our basic question by considering the M-theory lift of
the various IIA orientifold configurations. We identified the 
corresponding $G_2$ holonomy manifolds, and studied the quantum moduli 
space connecting different configurations through their topology and 
also the IIA exact superpotential.

Our main results are valid for the orientifold $z_i\to \bar z_i$, of
deformed conifold $\sum_iz_i^2=\mu$. With $\mu$ real, this has two
phases. Depending on whether $\mu$ is positive or negative, 
the orientifold fixes the $S^3$, or acts freely so that the minimal
cycle is an $\mathR\mathP^3$. The transition between positive and 
negative $\mu$ is possible only for special values of RR charge.
Once we fix the supersymmetry, one may either consider the
O$6^\pm$-plane ($\mu>0$) with charge $\pm2$ and increase the charge by
adding D6-branes, or start with free orientifolds ($\mu<0$) with zero
charge and decrease the charge by adding anti-D6-branes.
Therefore, $\mu$ can flip the sign only when the total charge $(N-2)$
equals $0,-1$ or $-2$.

\begin{figure}[htb]
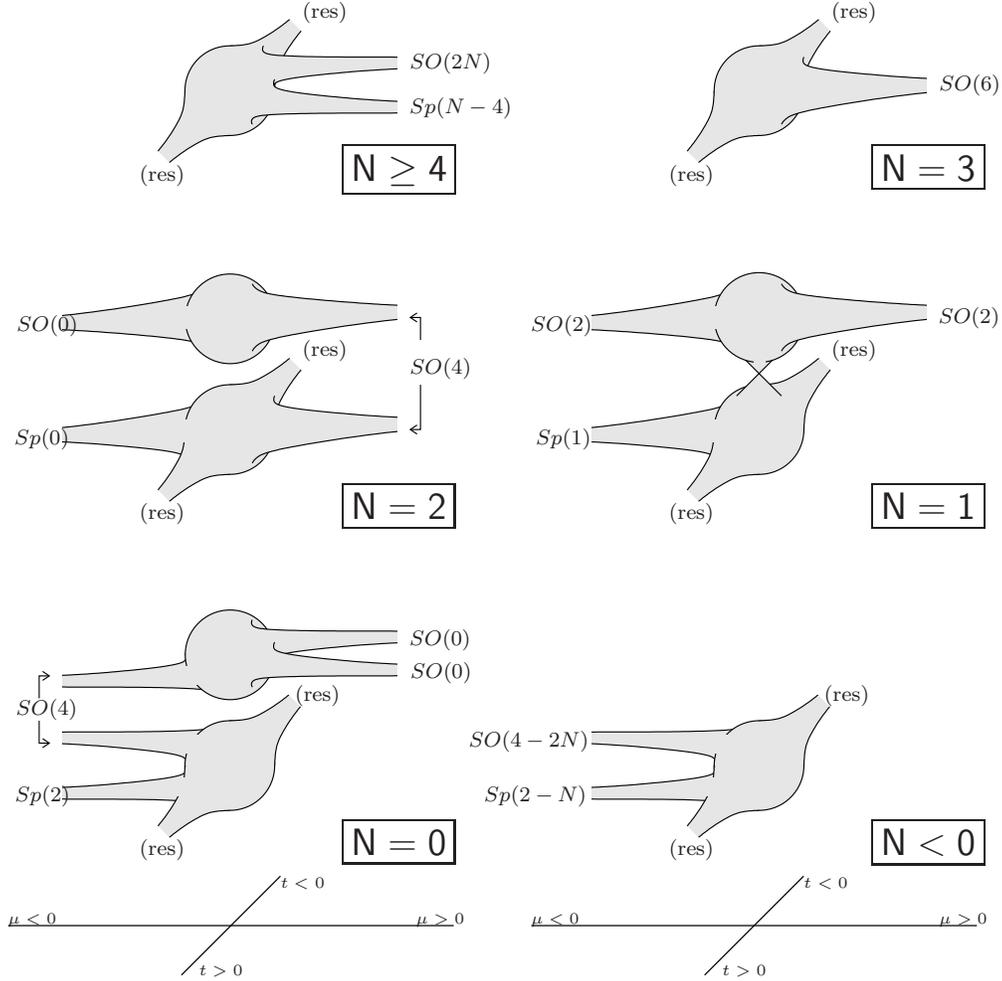

\def\axes{\tiny
\psfrag{m+}{$\mu>0$}
\psfrag{m-}{$\mu<0$}
\psfrag{t+}{$t>0$}
\psfrag{t-}{$t<0$}
\epsfig{width=6cm,file=axes.eps}   }
\def\res{\scriptsize
\psfrag{r+}{(res)}
\psfrag{r-}{(res)}   }
\begin{center}
\begin{tabular}{cc}
\res
\psfrag{N>=4}{\fbox{\large $\sf N\ge4$}}
\psfrag{d+}{\scriptsize $Sp(N-4)$}
\psfrag{d-}{\scriptsize $SO(2N)$}
\epsfig{width=6cm,file=Nge4.eps}
~~&~~
\res
\psfrag{N=3}{\fbox{\large $\sf N=3$}}
\psfrag{d-}{\scriptsize $SO(6)$}
\epsfig{width=6cm,file=Neq3.eps}
\\[1cm]
\res
\psfrag{N=2}{\fbox{\large $\sf N=2$}}
\psfrag{d-} {\scriptsize $SO(4)$}
\psfrag{d'+}{\scriptsize $Sp(0)$}
\psfrag{d'-}{\scriptsize $SO(0)$}
\epsfig{width=6cm,file=Neq2.eps}
~~&~~
\res
\psfrag{N=1}{\fbox{\large $\sf N=1$}}
\psfrag{d-} {\scriptsize $SO(2)$}
\psfrag{d'+}{\hskip-1mm\scriptsize $Sp(1)$}
\psfrag{d'-}{\hskip-2mm\scriptsize $SO(2)$}
\epsfig{width=6cm,file=Neq1.eps}
\\[1cm]
\res
\psfrag{N=0}{\fbox{\large $\sf N=0$}}
\psfrag{d-} {\scriptsize $SO(0)$}
\psfrag{d'+}{\scriptsize $Sp(2)$}
\psfrag{d'-}{\scriptsize $SO(4)$}
\epsfig{width=6cm,file=Neq0.eps}
~~&~~
\res
\psfrag{N<0}{\fbox{\large $\sf N<0$}}
\psfrag{d'+}{\hskip -8mm\scriptsize $Sp(2-N)$}
\psfrag{d'-}{\hskip-10mm\scriptsize $SO(4-2N)$}
\epsfig{width=6cm,file=Nlt0.eps}
\\
\axes~~&~~\axes
\end{tabular}
\end{center}
\caption{Moduli space of vacua for various total RR charge $(N-2)$.}
\label{fig:cartoon}
\end{figure}

On the other hand, some deformed or resolved geometries uplift
to $D_N$-type orbifolds of the $G_2$ holonomy spaces $\MD$ or $\MF$,
both of which are topologically $\mathR^4\times S^3$.
Remarkably, we found that this is true also for negatively large
RR charges ($N-2\le -3$), though the action of the dihedral group turned
out to be non-standard.
A careful analysis of the behavior of membrane instanton factors
allows us to determine the structure of quantum moduli space
unambiguously.
Also, the exact IIA superpotential tells us how the resolved vacua
are connected smoothly to other vacua.

We concluded that when $N=0,1,2$ the moduli space
consists of two branches.
For $N=0$ and $N=2$, the two branches meet at infinity where
there is a weakly coupled $SO(4)$ super Yang-Mills theory,
and each branch contains two of the four vacua.
For $N=1$, the branch of mass-gapped vacua meets the branch of
vacua with infrared $U(1)$ at a phase transition point, and
we found a precise description of the transition via a mirror
Type IIB picture.

The quantum moduli space of IIA orientifolded conifolds thus 
depends on the RR charge in an interesting way. We summarize these
main results of our paper in Figure \ref{fig:cartoon}. We found
similar results in other cases of orientifolded
conifolds as discussed in Section \ref{other}.

\begin{acknowledgments}
We would like to thank B.\ Acharya, J.\ Gomis, S.\ Hellerman, M.\ Kleban, 
J.\ Maldacena and E.\ Witten for very useful discussions.
K.H., K.H. and D.P. thank
Research Institute for Mathematical Sciences and Yukawa Institute for
Theoretical Physics, Kyoto University, for the support and hospitality
during their visits where a part of this work was done.
  R.R.\ and J.W.\ 
acknowledge the hospitality of the Fields and Perimeter Institutes where 
some of this work was realized. 
The work of K.H. and K.H. was supported by
NSERC and Alfred P. Sloan Foundation.
The work of D.P. is supported by PREA.
The work of R.R.\ and J.W.\ was supported 
by the DOE under grant number DE-FG02-90ER40542.
\end{acknowledgments}

\newpage

\appendix

\noindent {\Large\bf Appendix}

\section{Conifold}

Ricci-flat K\"ahler metrics for conifold or its small
deformation or resolution were obtained in \cite{cande}.
The singular conifold is defined by $\sum_{i=1}^4z_i^2=0$ or equivalently
by ${\rm det}W=0$ with 
\begin{equation}
 W=\frac1{\sqrt2}
  \left(\begin{array}{cc}z_1+iz_2 & -z_3+iz_4 \\ z_3+iz_4 & z_1-iz_2
	\end{array}\right).
\end{equation}
Regarding $z_i$ as holomorphic coordinates and putting
$K=\rho^{2/3}$ with $\rho\equiv {\rm Tr}WW^\dag$, one obtains a
Ricci-flat K\"ahler metric which is symmetric under
$SU(2)_L\times SU(2)_R$ acting on $W$ as $W\to LWR^\dag$.
To see the symmetry of the metric, we use the coordinates (which
is slightly different from the one conventionally used)
\begin{equation}
 W = X(\theta,\phi,\psi)\cdot W_0\cdot
     \tilde X(\tilde\theta,\tilde\phi,\tilde\psi)^\dag,~~~
     W_0 = r^{3/2}\left(\begin{array}{cc}1 & 0\\ 0 & 0 \end{array}\right)
\end{equation}
or more explicitly
\begin{equation}
 W = r^{3/2}\left(\begin{array}{rr}
	  \cos\frac\theta2\cos\frac{\tilde\theta}2
	  e^{\frac{i}{2}(\psi-\tilde\psi+\phi-\tilde\phi)} &
	  \cos\frac\theta2\sin\frac{\tilde\theta}2
	  e^{\frac{i}{2}(\psi-\tilde\psi+\phi+\tilde\phi)} \\
	  \sin\frac\theta2\cos\frac{\tilde\theta}2
	  e^{\frac{i}{2}(\psi-\tilde\psi-\phi-\tilde\phi)} &
	  \sin\frac\theta2\sin\frac{\tilde\theta}2
	  e^{\frac{i}{2}(\psi-\tilde\psi-\phi+\tilde\phi)}
      \end{array}\right)
\end{equation}
and get
\begin{eqnarray}
 ds^2 &=& dr^2 + r^2ds_{T^{1,1}}^2,\nn\\
 ds_{T^{1,1}}^2 &=&
  \frac16(\sigma_1^2+\sigma_2^2+\tilde\sigma_1^2+\tilde\sigma_2^2)
 +\frac19(\sigma_3-\tilde\sigma_3)^2 \nn\\
 &=& \frac16(d\theta^2+\sin^2\theta d\phi^2
           + d\tilde\theta^2 +\sin^2\tilde\theta d\tilde\phi^2)
\nn\\ && \hskip1cm
    +\frac19( d(\psi-\tilde\psi) + \cos\theta d\phi
             -  \cos\tilde\theta d\tilde\phi)^2.
\label{T11}
\end{eqnarray}
The coordinate $\check\psi \equiv \psi-\tilde\psi$ has the period $4\pi$
and defines an $S^1$ which is fibered over $S^2\times S^2$ labeled by
$(\theta,\phi)$ and $(\tilde\theta,\tilde\phi)$.
This metric has an $O(4)\times U(1)$ symmetry:
$SO(4)\simeq (SU(2)_L\times SU(2)_R)/\mathZ_2$ acts on $W$ as explained
and in particular $z_i$ are transformed as a 4-vector.
A parity transform in $O(4)$ exchanges the two $S^2$'s, and the $U(1)$
shifts $\psi$ or acts on $z_i$ as phase rotation.

Small resolution is an ${\cal O}(-1)\oplus{\cal O}(-1)$ bundle
over $\mathC\mathP^1$.
To write down the K\"ahler metric, write the matrix $W$ with
vanishing determinant as
\begin{equation}
 W = \left(\begin{array}{cc} -u\lambda & u \\ -y\lambda & y\end{array}\right)
   = \left(\begin{array}{cc} x & -x\mu \\ v & -v\mu\end{array}\right).
\end{equation}
$(\lambda,\mu)$ with $\lambda\mu=1$ are regarded as coordinates on
$\mathC\mathP^1$, and from the relation between $(u,y)$ and $(x,v)$
one finds they are coordinates on the fiber.
A natural ansatz for the K\"ahler potential of the resolved conifold is
\begin{equation}
  K = K(\rho) + 4a^2\ln(1+|\lambda|^2),
\end{equation}
the second term yielding a Fubini-Study metric on $\mathC\mathP^1$.
From the Ricci-flatness one finds that $r^2\equiv \rho\frac{dK}{d\rho}$
has to satisfy
\begin{equation}
  r^4(r^2+6a^2) ~=~ c\rho^2 + c'
\end{equation}
for some constants $c,c'$.
Setting $c=1,c'=0$ one obtains
\begin{equation}
  ds^2 \propto k^{-1}(r)dr^2 + \frac{r^2}{6}(\sigma_1^2+\sigma_2^2)
  + \frac{r^2+4a^2}{6}(\tilde\sigma_1^2+\tilde\sigma_2^2)
  +\frac{r^2k(r)}{9}(\sigma_3-\tilde\sigma_3)^2,
\end{equation}
with $k(r) = \frac{r^2+6a^2}{r^2+4a^2}$.
The metric is invariant under $SU(2)\times SU(2)\times U(1)$,
but the $\mathZ_2$ symmetry of singular conifold is lost.

Small deformation is defined by $\sum_iz_i^2=\epsilon^2$ or
${\rm det}W =\frac{\epsilon^2}{2}$.
Hereafter we assume $\epsilon$ to be real positive, as the Ricci-flat
metric will depend only on the modulus $|\epsilon|^2$.
The $SU(2)_L\times SU(2)_R$ invariant metric can be found by
assuming the K\"ahler potential to be a function $K(\rho)$ of
$\rho={\rm Tr}WW^\dag$.
The Ricci-flatness can be solved easily by introducing
$\rho = \epsilon^2\cosh\tau$ and putting
\begin{equation}
  W_0 = \frac{\epsilon}{\sqrt2}
   \left(\begin{array}{cc}e^{\tau/2}&0\\0&e^{-\tau/2}\end{array}\right).
\end{equation}
The Ricci-flatness amounts to
$\frac{dK}{d\rho}=\frac{(\sinh2\tau-2\tau)^{1/3}}{\sinh\tau}\equiv k(\tau)$,
and one obtains the following metric
\begin{equation}
 ds^2 \propto k(\tau)\left\{
     \frac{4}{3k(\tau)^3}(d\tau^2+(\sigma_3-\tilde\sigma_3)^2)
    +\cosh\tau(\sigma_1^2+\tilde\sigma_1^2+\sigma_2^2+\tilde\sigma_2^2)
    -2(\sigma_1\tilde\sigma_1+\sigma_2\tilde\sigma_2)\right\} .
\label{defcon}
\end{equation}
The metric is invariant under $O(4)$ but not under $U(1)$.

\section{Properties of $Li_2(z)$ }
\label{lithium}
 
We can define the Euler dilogarithm function\footnote{Defined by Euler 
in 1768.} in the disk $|z| <1$ as a convergent power series \footnote{For 
more properties of this function see \cite{dilog}}:

\begin{equation}
Li_2(z) = \sum_{n=1}^\infty \frac{z^n}{n^2} .
\end{equation}

The function can be extended to the whole complex plane as a multi valued 
analytical function. There is a branch cut from $z=1$ to $z= \infty$. 
Alternatively it can be defined as an integral:

\begin{equation}
Li_2(z) = - \int_0^z \frac{\log{(1-z)}}{z} dz  .
\end{equation}

A related function often found in literature is the Rogers L-function:

\begin{equation}
L(z) = Li_2(z) + \frac{1}{2} \log{z} \log{(1-z)} .
\end{equation}

Some functional equations that we will use in the main text are:

\begin{itemize}
\item  The Euler identity provides an expansion around the branch point $z=1$:
\begin{equation}
Li_2(1-z) = - Li_2(z) + \frac{\pi^2}{6} - \log{z} \log{(1-z)}
\end{equation}

Or in terms of the Rogers L-function:

\begin{equation}
L(z) + L(1-z) = L(1)
\end{equation}

\item The expansion around $z= \infty$

\begin{equation}
Li_2(1/z) = -Li_2(z) - \frac{\pi^2}{6} + \frac{1}{2} (\log{(- z)})^2
\end{equation}

\item A simple relation between the value of $Li_2(z)$ and $Li_2(z^2)$

\begin{equation}
Li_2(z) + Li_2(-z) = \frac{1}{2} Li_2(z^2)
\end{equation}

\item The above relations can be obtained in terms of the Abel identity 

\begin{multline}
Li_2(x) + Li_2(y) = Li_2(x y) + Li_2 \left( \frac{x(1-y)}{1-xy} \right) + \\
Li_2 \left( \frac{y(1-x)}{1-xy} \right) + \log{\left( \frac{1-x}{1-xy} \right)} 
\log{\left( \frac{1-y}{1-xy} \right)}
\end{multline}

Or in terms of the Rogers L-function:

\begin{equation}
L(x) + L(y) = L(x y) + L \left( \frac{x(1-y)}{1-xy} \right) + 
L \left( \frac{y(1-x)}{1-xy} \right)
\end{equation}

\end{itemize}

And some particular values of the dilogarithm:

\begin{equation}
Li_2(1)  ~=~   \frac{\pi^2}{6},~~~
Li_2(-1) ~=~ - \frac{\pi^2}{12},~~~
Li_2(1/2)~=~   \frac{\pi^2}{12} - \frac{1}{2} (\log{(2)})^2,
\end{equation}
\begin{equation}
L(1)  ~=~  \frac{\pi^2}{6},~~~
L(0)  ~=~ 0.
\end{equation}

\end{document}